\documentclass[a4paper,11pt]{article}

\usepackage[T1]{fontenc}
\usepackage[utf8]{inputenc}
\usepackage{lmodern}

\usepackage{tikz} 
\usepackage{xcolor}
\usepackage{graphicx}
\graphicspath{{Figures/}}

\usepackage[left=2.5cm,right=2.5cm,bottom=2.5cm,top=2.5cm]{geometry}

\usepackage[english]{babel}

\usepackage{microtype}

\usepackage{amsmath,amssymb}

\usepackage{float}

\usepackage{bm}

\usepackage{enumitem}

\usepackage[hang,flushmargin]{footmisc}

\usepackage{relsize,exscale}

\usepackage{fancyhdr}

\usepackage{caption}
\usepackage{subcaption}

\usepackage{changepage}

\usepackage[most]{tcolorbox}
\tcbset{colback=yellow!10!white, colframe=red!50!black, 
	highlight math style= {enhanced, 
		colframe=red,colback=red!10!white,boxsep=0pt}}

\usepackage{empheq}

\usepackage{gensymb}

\usepackage{dsfont}

\usepackage[only,llbracket,rrbracket]{stmaryrd}

\usepackage{mathtools}

\usepackage{slashed}

\usepackage{adjustbox}

\DeclareMathAlphabet{\mathdutchcal}{U}{dutchcal}{m}{n}
\SetMathAlphabet{\mathdutchcal}{bold}{U}{dutchcal}{b}{n}
\DeclareMathAlphabet{\mathdutchbcal}{U}{dutchcal}{b}{n}

\DeclareFontFamily{OT1}{pzc}{}
\DeclareFontShape{OT1}{pzc}{m}{it}{<-> s * [1.10] pzcmi7t}{}
\DeclareMathAlphabet{\mathpzc}{OT1}{pzc}{m}{it}

\usepackage{cite}

\makeatletter
\let\oldthebibliography\thebibliography
\renewcommand\thebibliography[1]{%
  \oldthebibliography{#1}%
  \setlength{\itemsep}{5pt}%
  \setlength{\parskip}{0pt}%
}
\makeatother

\usepackage{hyperref}

\hypersetup{
	colorlinks=true,
	linktoc=section,   
	citecolor=blue, 
	filecolor=blue,
	linkcolor=blue,
	urlcolor=blue,
	pdfborder={0 0 0.5},
	pdfmenubar=false,
	pdftoolbar=false
}
\numberwithin{equation}{section}

\providecommand{\delimsize}{\relax}

\DeclareRobustCommand{\rbig}[1]{\mathopen{\big#1}}

\DeclareRobustCommand{\rbigg}[1]{\mathopen{\bigg#1}}
\DeclareRobustCommand{\rBigg}[1]{\mathopen{\Bigg#1}}

\DeclarePairedDelimiterX{\bra}[1]{\delimsize\langle}{\delimsize\rvert}{#1}
\DeclarePairedDelimiterX{\ket}[1]{\delimsize\lvert}{\delimsize\rangle}{#1}
\DeclarePairedDelimiterX{\makebraket}[1]{\delimsize\langle}{\delimsize\rangle}{#1}

\makeatletter
\newcommand{\braketbar}{%
	\, \delimsize\vert\@ifnextchar|{\!}{\,\!\!\:}%
}
\makeatother
\newcommand{\activatebraketbar}{%
	\begingroup\lccode`~=`|\lowercase{\endgroup\let~}\braketbar
	\mathcode`|="8000
}

\NewDocumentCommand{\braket}{som}{%
	\mathord{%
		\begingroup
		\activatebraketbar
		\IfBooleanTF{#1}
		{\makebraket*{#3}}
		{\IfNoValueTF{#2}{\makebraket{#3}}{\makebraket[#2]{#3}}}%
		\endgroup
	}%
}

\newcommand{\ols}[1]{\mskip.5\thinmuskip\overline{\mskip-.5\thinmuskip {#1} \mskip-.5\thinmuskip}\mskip.5\thinmuskip} 
\newcommand{\olsi}[1]{\,\overline{\!{#1}}} 
\makeatletter
\newcommand\closure[1]{
	\tctestifnum{\count@stringtoks{#1}>1} 
	{\ols{#1}} 
	{\olsi{#1}} 
}
\long\def\count@stringtoks#1{\tc@earg\count@toks{\string#1}}
\long\def\count@toks#1{\the\numexpr-1\count@@toks#1.\tc@endcnt}
\long\def\count@@toks#1#2\tc@endcnt{+1\tc@ifempty{#2}{\relax}{\count@@toks#2\tc@endcnt}}
\def\tc@ifempty#1{\tc@testxifx{\expandafter\relax\detokenize{#1}\relax}}
\long\def\tc@earg#1#2{\expandafter#1\expandafter{#2}}
\long\def\tctestifnum#1{\tctestifcon{\ifnum#1\relax}}
\long\def\tctestifcon#1{#1\expandafter\tc@exfirst\else\expandafter\tc@exsecond\fi}
\long\def\tc@testxifx{\tc@earg\tctestifx}
\long\def\tctestifx#1{\tctestifcon{\ifx#1}}
\long\def\tc@exfirst#1#2{#1}
\long\def\tc@exsecond#1#2{#2}
\makeatother

\makeatletter
\newlength\xvec@height%
\newlength\xvec@depth%
\newlength\xvec@width%
\newcommand{\xvec}[2][]{%
	\ifmmode%
	\settoheight{\xvec@height}{$#2$}%
	\settodepth{\xvec@depth}{$#2$}%
	\settowidth{\xvec@width}{$#2$}%
	\else%
	\settoheight{\xvec@height}{#2}%
	\settodepth{\xvec@depth}{#2}%
	\settowidth{\xvec@width}{#2}%
	\fi%
	\def\xvec@arg{#1}%
	\def\xvec@dd{:}%
	\def\xvec@d{.}%
	\raisebox{.2ex}{\raisebox{\xvec@height}{\rlap{%
				\kern.05em
				\begin{tikzpicture}[scale=1]
					\pgfsetroundcap
					\draw (.05em,0)--(\xvec@width-.05em,0);
					\draw (\xvec@width-.05em,0)--(\xvec@width-.15em, .1em);
					\draw (\xvec@width-.05em,0)--(\xvec@width-.15em,-.1em);
					\ifx\xvec@arg\xvec@d%
					\fill(\xvec@width*.45,.5ex) circle (.5pt);%
					\else\ifx\xvec@arg\xvec@dd%
					\fill(\xvec@width*.30,.5ex) circle (.5pt);%
					\fill(\xvec@width*.65,.5ex) circle (.5pt);%
					\fi\fi%
				\end{tikzpicture}%
	}}}%
	#2%
}
\makeatother

\renewcommand{\vec}[1]{\xvec[]{#1}}

\usepackage{contour}
\usepackage[normalem]{ulem}

\contourlength{0.8pt}

\newcommand{\myuline}[1]{%
	\uline{\phantom{#1}}%
	\llap{\contour{white}{#1}}%
}

\makeatletter
\g@addto@macro\bfseries{\boldmath}
\makeatother

\newcommand\scaleddot{\scalebox{.89}{.}}

\makeatletter
\renewcommand{\dddot}[1]{%
	{\mathop{\kern\z@#1}\limits^{\makebox[0pt][c]{\vbox to-2.2\ex@{\kern-\tw@\ex@
					\hbox{\normalfont\scaleddot\kern-0.5pt\scaleddot\kern-0.5pt\scaleddot}\vss}}}}}
\renewcommand{\ddddot}[1]{%
	{\mathop{\kern\z@#1}\limits^{\makebox[0pt][c]{\vbox to-2.2\ex@{\kern-\tw@\ex@
					\hbox{\normalfont\scaleddot\kern-0.5pt\scaleddot\kern-0.5pt\scaleddot\kern-0.5pt\scaleddot}\vss}}}}}
\makeatother

\newcounter{daggerfootnote}
\newcommand*{\daggerfootnote}[1]{%
	\setcounter{daggerfootnote}{\value{footnote}}%
	\renewcommand*{\thefootnote}{\fnsymbol{footnote}}%
	\footnote[2]{#1}%
	\setcounter{footnote}{\value{daggerfootnote}}%
	\renewcommand*{\thefootnote}{\arabic{footnote}}%
}

\makeatletter
\DeclareRobustCommand{\cbig}[1]{\mathopen{\setbox0=\hbox{$\m@th\big#1$}\scalebox{1.2}{\copy0}}}
\makeatother
\makeatletter
\DeclareRobustCommand{\cBig}[1]{\mathopen{\setbox0=\hbox{$\m@th\Big#1$}\scalebox{1.15}{\copy0}}}
\makeatother
\makeatletter
\DeclareRobustCommand{\cbigg}[1]{\mathopen{\setbox0=\hbox{$\m@th\bigg#1$}\scalebox{1.15}{\copy0}}}
\makeatother
\makeatletter
\DeclareRobustCommand{\cBigg}[1]{\mathopen{\setbox0=\hbox{$\m@th\Bigg#1$}\scalebox{1.15}{\copy0}}}
\makeatother

\newcommand\sbullet[1][.5]{\mathbin{\vcenter{\hbox{\scalebox{#1}{$\bullet$}}}}}

\newcommand\normordbig{\raisebox{-0.7pt}{$\scalebox{1.5}{:}$}}

\newcommand{\Englob}{\smash{E\raisebox{-3pt}{\scalebox{0.7}{$\ell$}}\!\!\:\!\:\!\raisebox{5pt}{\scalebox{0.65}{$(\text{global})$}}}} 
\newcommand{\Enloc}{\smash{E\raisebox{-3pt}{\scalebox{0.7}{$\ell$}}\!\!\:\!\:\!\raisebox{5pt}{\scalebox{0.65}{$(\text{local})$}}}} 
\newcommand{\Enres}{\smash{E\raisebox{-3pt}{\scalebox{0.7}{$\ell$}}\!\!\:\!\:\!\raisebox{5pt}{\scalebox{0.65}{$(\text{resonant})$}}}}
\newcommand{\Eglob}{E\raisebox{-2.5pt}{\scalebox{0.65}{$0$}}\!\!\!\:\raisebox{4.5pt}{\scalebox{0.65}{$(\text{global})$}}} 
 
\newcommand{\Eres}{E\raisebox{-2.5pt}{\scalebox{0.65}{$0$}}\!\!\!\:\raisebox{4.5pt}{\scalebox{0.65}{$(\text{resonant})$}}} 

\begin{document}
	\renewcommand{\headrulewidth}{0pt}
	\pagestyle{fancy}
	\pagenumbering{Roman}
	\fancyhead{}
	\cfoot{$-\;\!$\thepage $\;\!-$}
	
	\begin{flushright}
		{\small
			\textcolor{black}{TUM--HEP--1567/25}\\
			\textcolor{black}{MPP--2025--150}
		}
	\end{flushright}
	\vspace{0.5cm}
	
	\begin{center}
        
        \begin{adjustwidth}{1.5cm}{1.5cm} 
        {\LARGE\bf 
        \begin{center}
        Path integral analysis of Schrödinger-type eigenvalue problems in the complex plane:
        \end{center}}
        \end{adjustwidth} \vskip0.45cm
        \begin{adjustwidth}{0.25cm}{0.25cm} 
        {\Large\bf Establishing the relation between instantons and resonant states} \\[1.25cm]
        \end{adjustwidth}
        
		\textsc{Björn Garbrecht}$^{1}$ and \textsc{Nils Wagner}$^{1,2,\dagger}$

		\vspace{0.5cm}
		
		{\it ${}^1$ Physik--Department T70, Technische Universität München, James--Franck--Stra{\ss}e 1, \\ D--85748 Garching, Germany \\[0.15cm]}
		{\it ${}^2$ Max--Planck--Institut für Physik (Werner--Heisenberg--Institut), Boltzmannstra{\ss}e 8, \\ D--85748 Garching, Germany\\}
		
		\vspace{0.5cm}
		\emph{E--mail:} \href{mailto:garbrecht@tum.de}{{\tt garbrecht@tum.de}}, 
		\href{mailto:nils.wagner@tum.de}{{\tt nils.wagner@tum.de}}\daggerfootnote{Corresponding author}

		\medskip
		
	\end{center}
	
	\vspace{1cm}
	
	\begin{abstract}
		\vspace{0.2cm}
		
		\noindent
        Schrödinger-type eigenvalue problems are ubiquitous in theoretical physics, with quantum-mechanical applications typically confined to cases for which the eigenfunctions are required to be normalizable on the real axis. However, seeking the spectrum of resonant states for metastable potentials or comprehending $\mathcal{PT}$-symmetric scenarios requires the broader study of eigenvalue problems for which the boundary conditions are provided in specific angular sectors of the complex plane. We generalize the conventional path integral treatment to such nonstandard boundary value problems, allowing the extraction of spectral information using functional methods. We find that the arising functional integrals are naturally defined on a complexified integration contour, encapsulating the demanded sectorial boundary conditions of the associated eigenvalue problem. The attained results are applied to the analysis of resonant ground-state energies, through which we identify the previously elusive one-to-one correspondence between decay rates derived from real-time quantum tunneling dynamics and those obtained via the Euclidean instanton method.
	\end{abstract}

    \vfill
    \noindent 
	 
\newpage

\tableofcontents

\newpage

\renewcommand{\headrulewidth}{0pt}
\pagestyle{fancy}
\pagenumbering{arabic}
\fancyhead{}
\cfoot{$-\;\!$\thepage $\;\!-$}

\section{Introduction}
\label{sec:1_Introduction}

Even almost a hundred years after the pioneering work of Gamow~\cite{GamowAlphaDecay}, quantum tunneling still remains a topic of considerable interest, bearing great significance in a plethora of physical phenomena. Originally tackled mostly utilizing WKB techniques~\cite{JWKB_combined,GamowAlphaDecay,Gurney_Decay,BerryWKBinWaveMechanics}, the later rise of quantum field theories demanded the use of more sophisticated tools, as most traditional methods are faced with profound challenges in such infinite-dimensional settings~\cite{BenderMultiDimensionalWKB,GervaisMultiDimWKB,BitarWKBFieldTheory}. Since the early contributions of Langer~\cite{LangerCondensationPoint,LangerMetastability}, it was found that a functional formulation, employing Euclidean path integrals, enables an effective treatment of nonperturbative aspects of field theories~\cite{GelfandFunctionalIntegration,DashenNonperturbativeMethods,BRSTInstantons,tHooftInstantons}, being especially suited to account for quantum-mechanical barrier penetration~\cite{LangerCondensationPoint,LangerMetastability,KobzarevMetastableVacuum,GildenerInstantonMethod,ColemanFateOfFalseVac1,CallanColemanFateOfFalseVac2}. The emerging instanton method, popularized by the seminal work of Callan \& Coleman~\cite{CallanColemanFateOfFalseVac2}, has since then manifested itself to be the go-to method for studying tunneling processes in QFT, particularly in the context of false vacuum decay~\cite{SherVacuumStabilitySM,Espinosa_VacSStability,IsidoriMetaStabilitySM,ButtazzoSMLifetime,SchwartzPrecisionDecayRate,SchwartzScaleInvInstantons}. Although the formalism successfully retrieves the expected result, conceptual issues have remained, as the justification for associating the computed quantity with the decay rate is not immediately evident. Recent mathematical insights into the asymptotic approximation of path integrals and their analytic continuation, drawing on concepts of Picard--Lefschetz theory~\cite{PhamPicardLefschetz,HowlsPicardLefschetzTheory,WittenAnalyticContinuation,TanizakiLefschetz,UnsalTowardsPicLefschetz,GarbrechtFunctionalMethods}, have therefore sparked renewed interest in clarifying the correspondence between the Euclidean instanton method and real-time tunneling computations, even within purely quantum-mechanical settings~\cite{BenderTunnelingAsAnomaly,TurokRealTimeTunneling,UnsalRealTimeInstantons,TanizakiLefschetz,LehnersComplexTimePaths,SchwartzDirectMethod,GarbrechtFunctionalMethods,MouRealTimeTunneling,BradenVacDecay,HertzbergRealTimeTunneling,MatsuiRealTimePI,NishimuraRealTimeTunneling,BlumRealTimeTunneling,SteingasserFiniteTemp,SteingasserRealTimeInstantons,LawrenceRealTimeTunneling,FeldbruggeRealTimeTunneling,WagnerExcitedStateTunneling}. Despite these efforts, it has not been conclusively demonstrated that the procedure's agreement with calculations based on solutions to the Schrödinger equation is more than a mere coincidence.\\

\noindent 
The traditional instanton approach of extracting decay rates is based on a semiclassical evaluation of the Euclidean transition amplitude from the false vacuum back to itself at large times, capturing contributions from both the constant false vacuum trajectory and a nontrivial configuration known as the ``bounce''. Since the fluctuation operator pertaining to the bounce saddle admits a negative mode, the associated steepest-descent thimble necessarily extends into the complexified function space. To retain the desired bounce contribution, one must therefore explain why the functional integration contour passes into the complex domain. While heuristically motivated with a formal deformation of the potential, this procedure, going back to Callan and Coleman, has yet to be placed on firm mathematical footing. Our work addresses this persistent conceptual shortcoming of the instanton method by elucidating how the functional integration contour, dictating the relevant saddle points contributing to the path integral, is linked to the notion of resonant states, thereby naturally encapsulating the Gamow--Siegert boundary conditions essential for describing decay dynamics. Starting directly from a boundary value problem whose eigenvalues amount to the desired resonant energies inside a metastable potential, we develop a generalized path integral formula for which the integration contour is naturally deformed according to the prescription proposed by Callan \& Coleman. The developed methods possess yet broader applicability, accommodating general Schrödinger-type eigenvalue problems with boundary conditions specified in certain angular sectors of the complex plane, as e.g. pursued by the community surrounding $\mathcal{PT}$-symmetric quantum mechanics~\cite{BenderPT1,BenderPT2,BenderPTReview,DoreyODEIM}.\\

\noindent 
Our exposition is structured as follows: At the onset of section~\ref{sec:2_Resonant_State_Introduction}, we first explain why resonant states, characterized by outgoing Gamow--Siegert boundary conditions, constitute the natural framework for analyzing decay processes, thereby motivating the subsequent developments. In the following subsections~\ref{sec:2_1_Eigenvalue_Problems_ComplexPlane} and~\ref{sec:2_2_AnalyticContinuation_EigenvalueProblems}, we introduce how these sought-after states can be captured as solutions to generalized eigenvalue problems whose boundary conditions are specified in angular sectors of the complex plane. To this end, we briefly review some properties of solutions to Schrödinger-type ODEs featuring a polynomial potential, along with the arising notion of Stokes sectors, in section~\ref{sec:2_1_Eigenvalue_Problems_ComplexPlane}. We then wield the introduced tools to relate resonant states to such boundary value problems in subsection~\ref{sec:2_2_AnalyticContinuation_EigenvalueProblems}, motivating this connection through a simple example that hinges on analytic continuation. In section~\ref{sec:3_QM_on_ComplexContour}, it is shown that one can associate a (quantum-mechanical) propagator to the desired class of eigenvalue problems by formally restricting the discussion to a complex contour terminating in the specified regions of subdominance of the eigenfunctions. Both the arising spectral and path integral representations for the so-defined propagator, computed respectively in subsections~\ref{sec:3_1_Spectral_Representation} and~\ref{sec:3_2_PathIntegral_Representation}, are found to naturally extend the conventional notions. Equating both representations in subsection~\ref{sec:3_3_Equating_Representations}, we arrive at the pivotal master formulas~\eqref{eq:Final_Relation_Propagator} and~\eqref{eq:Final_Relation_Trace}, constituting the central result of this work. In section~\ref{sec:4_Quantum_Tunneling_Revisited}, we apply these previously attained results to quantum tunneling. Briefly reviewing the traditional potential-deformation argument for justifying the instanton method in section~\ref{sec:4_1_PotentialDeformationMethod}, we subsequently compile some of the open questions raised by this approach in subsection~\ref{sec:4_2_OpenQuestions_ColemanApproach}. The following section~\ref{sec:4_3_FormalizingColemansIdeas} then formalizes the initial ideas by Callan \& Coleman, thereby elucidating most of the glaring questions posed previously. Section~\ref{sec:5_Assessment_Earlier_Attempts} addresses earlier attempts of finding a one-to-one correspondence between decay rate calculations hinging on real-time computations of the probability loss inside the FV region and the instanton method. We particularly devote subsections~\ref{sec:5_1_Flaws_DirectMethod} and~\ref{sec:5_2_Flaws_Steadyons} to a critical assessment of the ``direct method'' introduced by Andreassen et al.~\cite{SchwartzDirectMethod} and the ``steadyon'' interpretation proposed by Steingasser \& Kaiser~\cite{SteingasserRealTimeInstantons}, arguing that both prominent attempts fall short of their proclaimed goal of providing a first-principles account of the decay rate in terms of functional integrals in order to illuminate the instanton method. Subsection~\ref{sec:5_3_Other_Approaches} composes additional brief comments on several other recent approaches that also have striven toward a real-time understanding of instanton equivalents. We conclude with a summary of our findings in section~\ref{sec:6_Conclusion}. Additional supplementary material and auxiliary computations, which would burden the main exposition with excessive detail, are deferred to appendices~\ref{sec:A_PathIntegralTechniques}--\ref{sec:D_LargeOrderBehavior_Eigenvalues}. In anticipation of working primarily with complex variables, we consistently denote the position variable by $z$ throughout most of this work.

\section{Motivation: Extracting decay rates via resonant states}
\label{sec:2_Resonant_State_Introduction}

\begin{figure}[H]
\centering
\includegraphics[width=0.97\textwidth]{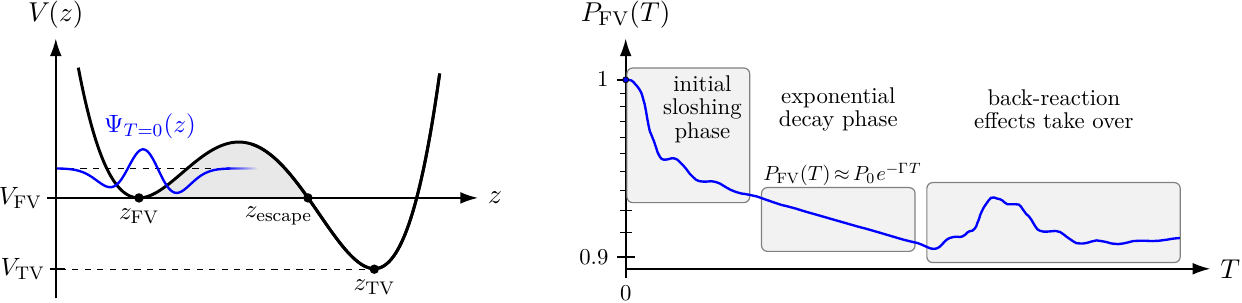}
\caption{(Left) Prototypical potential exhibiting decay from the false vacuum (FV) region toward the global minimum $z_\mathrm{TV}$, subsequently referred to as the true vacuum (TV). The shaded barrier region is assumed to be sufficiently high (and broad) in order for the decay dynamics to be slow compared to the oscillation periods in the FV and TV region, constituting the natural time scales in such a potential. (Right) Given an exemplary initial wave function $\Psi_{T=0}(z)$ supported solely inside the FV basin, the schematic time-dependence of the survival probability $\smash{P_\mathrm{FV}(T)=\displaystyle{\scalebox{1.3}{$\int$}}_{\!\!\mathrm{FV}}\big\lvert \Psi(z,\!\!\:T) \big\rvert^2 \mathrm{d}z}$ under the full time evolution is shown, albeit to a steeper potential and a simpler Gaussian initial state; data taken from Andreassen et al.~\cite[see figure 1]{SchwartzPrecisionDecayRate}. For sufficiently simple cases, one can distinguish three phases of the evolution, with only the intermediate temporal regime closely following an exponential decay law.}
\label{fig:GenericMetastablePot+TimeEvolution}
\end{figure}

Serving as the motivation for all subsequent developments, we will briefly introduce the physical reasoning behind the use of so-called \emph{resonant states} when investigating problems related to quantum-mechanical barrier penetration and their relation to eigenvalue problems in the complex plane. For simplicity, let us restrict our view to single-dimensional problems, with a typical potential allowing for quantum tunneling depicted in the left panel of figure~\ref{fig:GenericMetastablePot+TimeEvolution}.

\noindent 
The quantum-mechanical decay of a particle initially confined to the false vacuum (FV) region constitutes an inherently time-dependent phenomenon, rendering in-depth theoretical studies solely by analytic means practically infeasible. Although numerically solving the time-dependent Schrödinger equation allows access to the full time evolution for any given initial state, thereby providing precise estimates for the desired survival probability $P_\mathrm{FV}(T)$ within the FV region for arbitrary times, purely theoretical endeavors are limited to certain regimes of the time evolution~\cite{KhalfinLateTimeBehavior,PetzoldExponentialgesetz,NewtonUnstableSystems,TimeEvolutionQuantumSystem,FondaDecayTheory,PeresDecayLaw,SchwartzPrecisionDecayRate}. In simple cases, the survival probability $P_\mathrm{FV}(T)$ exhibits three distinct phases depending on the time scales investigated, which is schematically illustrated in the right panel of figure~\ref{fig:GenericMetastablePot+TimeEvolution}. For short times, the evolution is dominated by the initial wave function ``sloshing'' around inside the FV region, roughly ending once the highly excited modes in the metastable FV region have decayed and the FV ground state dominates~\cite{SchwartzPrecisionDecayRate}.\footnote{In the general case, the FV ground state prevails the longest, dominating the intermediate time regime. However, initializing the wave function in an approximate energy eigenstate of the FV basin effectively bypasses the initial sloshing phase. The resulting exponential decay behavior then directly probes the decay rate of the prepared state, which may correspond either to the ground state or to an excited state.} Additional deviations for small time scales arise from the fact that unitarity of the quantum-mechanical time evolution disallows a linear behavior of the survival probability at the onset of the decay, leading to the quantum Zeno effect~\cite{TimeEvolutionQuantumSystem,PeresDecayLaw}. Once the system has settled into an approximate eigenstate inside the FV region, the decay dynamics becomes uniform, exhibiting the well-known exponential behavior~\cite{NewtonUnstableSystems,PeresDecayLaw}. For very late times, the wave function inside the FV region becomes sufficiently depleted such that tunneling in the reverse direction becomes non-negligible. Apart from such back-reaction phenomena, quantum interference effects would generally lead to the probability decay asymptotically approaching a power-law behavior for late times~\cite{KhalfinLateTimeBehavior,PeresDecayLaw}.\footnote{Beware that this either requires the potential to become asymptotically flat or an additional coupling to an external system such that there exists a continuous energy band around the true vacuum energy, leading to these effects arising from the low-energy tail of the energy spectrum around $E_\mathrm{TV}$. In the simplistic case depicted in figure~\ref{fig:GenericMetastablePot+TimeEvolution}, possessing a fully discretized spectrum, the time evolution would instead be quasi-periodic, with interference effects playing no considerable role.} Summarizing, only at intermediate times does the decrease in probability of the wave function inside the FV region closely follow an exponential decay law, allowing for the introduction of a well-defined decay rate $\Gamma$. This intermediate phase usually lasts for an elongated period of time, which, together with the uniformity of the decay during this time frame, allows for a deeper analytic understanding. \\ 

\noindent 
One of the most significant insights attributed to Gamow is the realization that when decay occurs sufficiently slowly, the initially time-dependent problem can be effectively approximated by a steady-state one, significantly simplifying the analysis. During the previously introduced intermediate time regime, the quantum state is seen to possess a tiny, but practically constant flux directed toward the TV region. This enables us to model the arising wave function at these time scales, resembling an outgoing wave, as a (quasi-)stationary solution to the time-\myuline{in}dependent Schrödinger equation. To this end, instead of demanding the eigenfunctions of the Hamiltonian $\widehat{H}$ to be normalizable on $\mathbb{R}$, one requires the resonant states to possess a purely outward-directed flux in the spatial directions for which tunneling is allowed~\cite{GamowAlphaDecay,SiegertRadiativeStates}. These so-called (radiating) \emph{Gamow--Siegert boundary conditions} explicitly break the Hermiticity of the Hamiltonian $\widehat{H}$ due to the arising boundary contributions when integrating by parts, leading to complex energy eigenvalues. In the single-dimensional case illustrated in figure~\ref{fig:GenericMetastablePot+TimeEvolution}, any stationary solution obeying the time-independent Schrödinger equation $\widehat{H}\Psi=E\Psi$ satisfies
\begin{align}
 	\mathrm{Im}(E) = - \:\!\frac{\hbar}{2}\, \scalebox{1.1}{\bigg\{}\!\int_\mathrm{FV} \big\lvert\:\! \Psi(z)\:\!\big\rvert^2 \, \mathrm{d}z\scalebox{1.1}{\bigg\}}^{\!-1}\, J_\mathrm{outward} \, ,
 	\label{eq:Relation_ImaginaryEnergy_ProbabilityCurrent}
\end{align}
indicating that the imaginary part of the so-found energy eigenvalue is directly proportional to the (constant) flux $J_\mathrm{outward}$ leaving the FV region, leading to the desired exponential decay law with the associated rate given by $\Gamma=-2\,\mathrm{Im}(E)/\hbar$. \\

\noindent 
The above discussion will suffice for our needs to grasp the broader context of why resonant states with outgoing Gamow--Siegert boundary conditions are relevant when computing physical decay rates; for a more in-depth treatment of resonant states, consult e.g.~\cite{HislopSpectralTheory,IntroductionGamovVectors,ResonancesIntroduction}. Before returning to address how these outgoing boundary conditions can be implemented concisely, let us briefly digress into the more general theory of Schrödinger-type eigenvalue problems in section~\ref{sec:2_1_Eigenvalue_Problems_ComplexPlane}, which will equip us with the necessary tools to further illuminate the previously introduced resonant states in the subsequent section~\ref{sec:2_2_AnalyticContinuation_EigenvalueProblems}.


\subsection{Schrödinger-type eigenvalue problems in the complex plane}
\label{sec:2_1_Eigenvalue_Problems_ComplexPlane}

\begin{figure}[H]
    \centering
    \begin{subfigure}[c]{0.42\textwidth}
    \includegraphics[width=\textwidth]{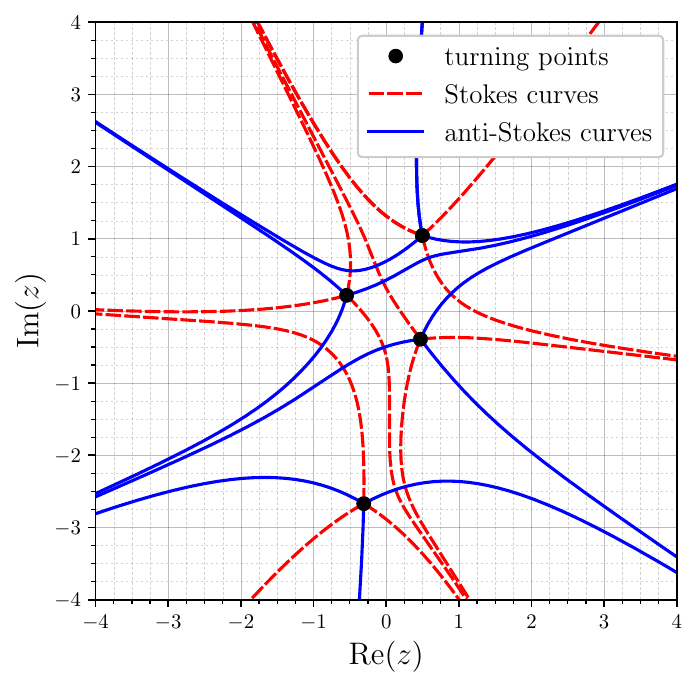}
    \end{subfigure}
    \begin{subfigure}[c]{0.12\textwidth}
    \begin{tikzpicture}
        \draw (-3,3) node{$\xrightarrow{\,\text{``zoom out''}\,}$};
    \end{tikzpicture}
    \end{subfigure}
    \begin{subfigure}[c]{0.42\textwidth}
    \includegraphics[width=\textwidth]{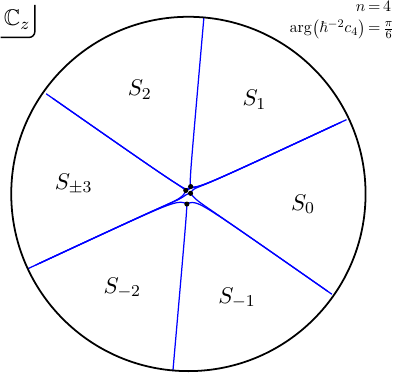}
    \end{subfigure}
    \caption{(Left) Stokes graph for the polynomial $Q(z)=\big(\sqrt{3}+i\big)z^4-\big(2-3i\big)z^3+5z^2-\big(1-2i\big)$ together with the associated anti-Stokes lines. (Right) Partitioning of the complex plane into the Stokes wedges $S_k$ arising from the asymptotic behavior of the anti-Stokes curves, illustrated for the identical polynomial $Q(z)$. The bounding circle represents complex infinity.}
    \label{fig:StokesGeometryQuarticPot}
\end{figure}
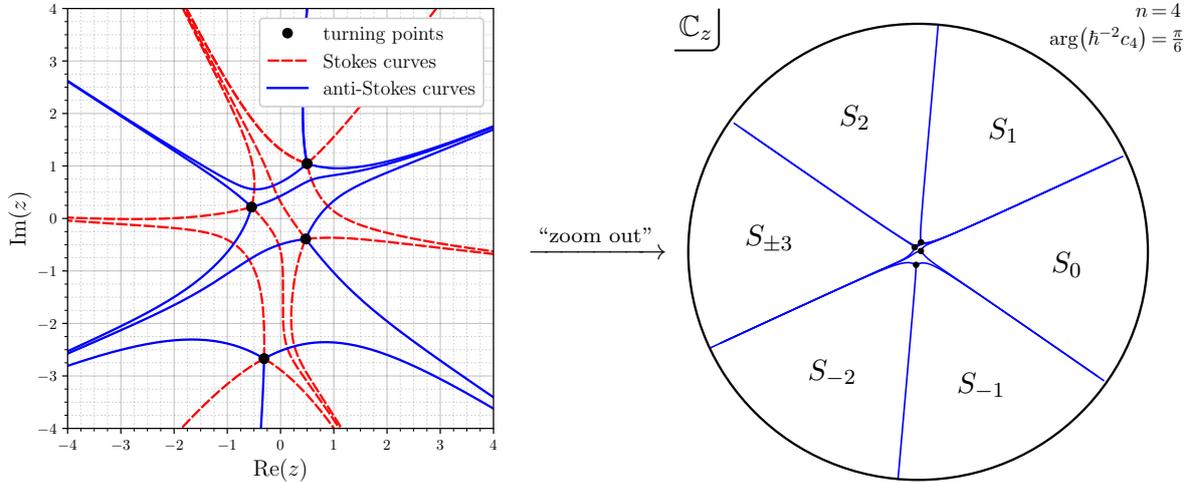

\begin{figure}[H]
    \centering
    \begin{subfigure}[c]{0.32\textwidth}
    \includegraphics[width=\textwidth]{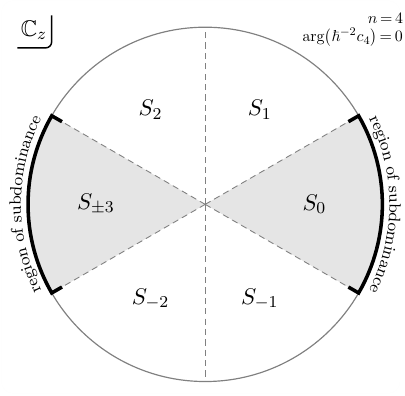}
    \end{subfigure}
    \hfill
    \begin{subfigure}[c]{0.32\textwidth}
    \includegraphics[width=\textwidth]{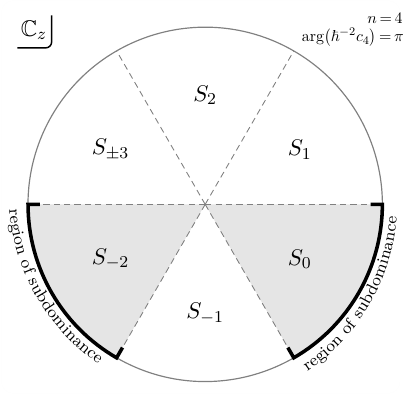}
    \end{subfigure}
    \hfill
    \begin{subfigure}[c]{0.32\textwidth}
    \includegraphics[width=\textwidth]{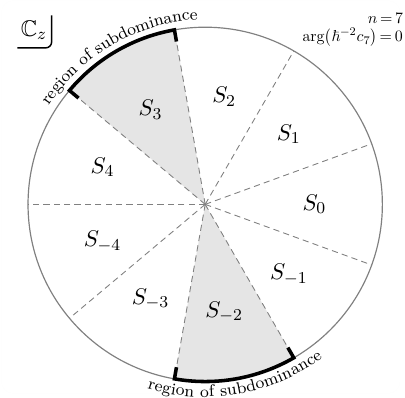}
    \end{subfigure}
    \caption{Examples of different admissible eigenvalue problems admitting a purely discrete spectrum---the regions of subdominance $S_\pm$ specifying the boundary conditions have been shaded. (Left) A conventional eigenvalue problem that can be defined on the real axis, being fully entailed in the Stokes sectors where the eigenfunctions are mandated to be subdominant. In that case, one encounters the typical requirement of the eigenfunctions to be square-integrable on $\mathbb{R}$. (Center) Example of a $\mathcal{PT}$-symmetric problem, assuming the condition $V(z)=\;\!\!\closure{\:\!V\:\!}\;\!\!(-z)$ has been met, for which subdominance is mandated in Stokes sectors that are symmetric under reflection along the imaginary axis. (Right) A more exotic eigenvalue problem.}
    \label{fig:StokesWedgeExamples}
\end{figure}

As with many mathematical problems, broadening the overly restrictive real perspective to the complex domain reveals a much richer structure. This approach turns out to be equally valuable in our analysis of solutions to the time-independent Schrödinger equation. For this purpose, let us briefly review some exact statements on Schrödinger-type ODEs of the form
\begin{align}
    \bigg\{\!\!\!\:-\!\!\:\frac{\mathrm{d}^2}{\mathrm{d}z^2}+\hbar^{-2}Q(z)\bigg\}\,\Psi(z)=0\, ,
    \label{eq:SchrödingerTypeODE}
\end{align}
with $Q(z)$ for now denoting a meromorphic function---we will restrict our view to polynomial potentials momentarily. For proofs and further in-depth analysis of the subsequent statements, consult the literature on the \emph{exact WKB} method, see e.g. the lucid monograph by Kawai \& Takei~\cite{KawaiExactWKB} and the references therein. It is well-known that the two linearly independent solutions $\Psi_\pm(z)$ of equation~\eqref{eq:SchrödingerTypeODE} attain the asymptotic form
\begin{align}
    \Psi_{\pm}(z) \sim \exp\cbigg\{\pm\:\!\frac{1}{\hbar}\mathlarger{\int}_{z_0}^z \sqrt{Q(z')}\:\mathrm{d}z'\cbigg\}\;\mathlarger{\sum}_{\ell=0}^\infty\: \Psi_{n,\pm}(z) \:\!\hbar^{n+\frac{1}{2}}\, ,
    \label{eq:ExactWKBAnsatz}
\end{align}
with the asymptotic series being Borel-summable in disjoint regions of the complex plane, failing precisely on so-called \emph{Stokes curves}, implicitly defined by
\begin{align}
    \mathrm{Im}\cbigg\{\frac{1}{\hbar}\mathlarger{\int}_{\scalebox{0.75}{$z_\mathrm{turn}$}}^{\scalebox{0.75}{$z$}} \sqrt{Q(z')}\:\mathrm{d}z'\cbigg\}=0\, .
\end{align}
Hereby, $z_\mathrm{turn}$ denotes a \emph{turning point} of $Q(z)$, defined as a zero of $Q(z)$, satisfying $Q(z_\mathrm{turn})=0$. Our primary interest moving forward will however lie in \emph{anti-Stokes curves}, which are similarly given by
\begin{align}
    \mathrm{Re}\cbigg\{\frac{1}{\hbar}\mathlarger{\int}_{\scalebox{0.75}{$z_\mathrm{turn}$}}^{\scalebox{0.75}{$z$}} \sqrt{Q(z')}\:\mathrm{d}z'\cbigg\}=0\, .
    \label{eq:AntiStokesCurveDefinition}
\end{align}
These partition the complex plane by delineating the strips where the dominant behavior between the two independent solutions $\Psi_\pm$ switches. Let us focus on the especially illustrative case of $Q(z)= \sum_{\ell=0}^n \:\!c_\ell \:\!z^\ell$ being a polynomial of degree $n$ with complex coefficients $c_\ell$, which has been extensively studied by Sibuya~\cite{SibuyaEigenvalueProblems}. When zooming out to large magnitudes of the complex variable $z$, the anti-Stokes curves partition the complex plane into $n+2$ wedge-shaped regions dubbed \emph{Stokes sectors} (also referred to as \emph{Stokes wedges}), defined by\footnote{The given angular sectors emerge when solving the defining relation~\eqref{eq:AntiStokesCurveDefinition} asymptotically for $\lvert z\rvert\to \infty$, for which one arrives at the leading-order result $\smash{\mathrm{Re}\big(\hbar^{-1}c_n^{1/2}z^{(n+2)/2}\big)=0}$. Note that the term Stokes sector is sometimes utilized to refer to the individual regions of the complex plane separated by Stokes lines. Keep in mind that this usage differs from the definition adopted here.}
\begin{align}
	S_k\coloneqq\Bigg\{z\in\mathbb{C}: \cbigg\lvert\, \mathrm{arg}(z)-\frac{2\pi k}{n+2}+\frac{\mathrm{arg}\big(\hbar^{-2}c_n\big)}{n+2}\cbigg\lvert <\frac{\pi}{n+2}\Bigg\}\, ,\qquad k\in \mathbb{Z}_{n+2}\, .
    \label{eq:DefinitionStokesSectors}
\end{align}
An illustration of an exemplary \emph{Stokes graph}, defined as the union of all Stokes curves, together with the important division of the complex plane into Stokes sectors, is provided in figure~\ref{fig:StokesGeometryQuarticPot}. To simplify the illustrations of the relevant anti-Stokes curves in all subsequent figures, we only show their asymptotic behavior, formally zooming out far enough such that the turning points virtually cluster around the origin, as is illustrated in the right panel of figure~\ref{fig:StokesGeometryQuarticPot}. 
\noindent
The significance of the so-defined Stokes sectors lies in the fact that one can prove the following assertions: 
\begin{itemize}
    \item Every non-constant (and necessarily holomorphic) solution $\Psi(z)$ of the ODE~\eqref{eq:SchrödingerTypeODE} with polynomial $Q(z)$ either decays or grows exponentially in each Stokes sector $S_k$ for $\lvert z\rvert\to\infty$, a behavior arising from the exponential contribution in equation~\eqref{eq:ExactWKBAnsatz}. Members of the one-dimensional space of decaying solutions in the Stokes sector $S_k$ are said to be \emph{subdominant in $S_k$}, satisfying $\Psi(\eta z)\rightarrow 0$ for $\eta\to \infty$ $(\eta>0)$ and arbitrary $z\in S_k$.
    \item In case a non-constant solution asymptotically decays in $S_k$, it necessarily grows in the adjacent wedges $S_{k\pm 1}$. The relation however does not hold the other way around, as a solution exponentially blowing up in $S_k$ is not required to be subdominant in either $S_{k-1}$ or $S_{k+1}$. Intuitively, this suggests that the occurrence of a solution being subdominant within multiple sectors is quite rare, with the generic behavior constituting exponential growth for $\lvert z\rvert \to\infty$. This realization subsequently leads to the important observation stated below.
    \item Provided $n\geq 2$, solutions $\Psi(z)$ of the ODE~\eqref{eq:SchrödingerTypeODE} that are demanded to be subdominant in two non-adjacent Stokes sectors $S_+$ and $S_-$ only exist for a discrete set of choices of the constant term $c_0$ of the polynomial $Q(z)$. To be more precise: The eigenvalue problem
    \begin{align}
        \bigg\{\!\!\!\:-\!\!\:\frac{\mathrm{d}^2}{\mathrm{d}z^2}+\hbar^{-2}Q(z)\bigg\}\:\!\Psi_\lambda(z)=\lambda \Psi_\lambda(z)\, , \qquad \Psi_\lambda(\eta z)\xrightarrow{\,\eta\to\infty \, (\eta>0)\,} 0 \;\;\;\; \mathrm{for} \;z\in S_\pm
    \label{eq:SchrödingerTypeEigenvalueProblem}
    \end{align}
    is well-defined and possesses a discrete set of simple eigenvalues $\big\{\lambda_\ell\big\}_{\ell\in\mathbb{N}_0}$ whose moduli accumulate at infinity~\cite{SibuyaEigenvalueProblems,FedoryukAsymptoticAnalysis,EremenkoODE}.
\end{itemize}
This general characteristic allows for a unified approach to various classes of eigenvalue problems, a small selection being illustrated in figure~\ref{fig:StokesWedgeExamples}. As displayed, depending on the symmetry properties of the chosen regions of subdominance $S_\pm$, this includes many regular Sturm--Liouville problems defined on the real axis, along with a wide range of $\mathcal{PT}$-symmetric eigenvalue problems. For other notable works and reviews employing the portrayed methods, consult e.g.~\cite{BenderAnharmonicOscillator,BenderAnalyticylContinuation,VorosPolynomial,ShinPTEigenvalues,ShinODE,DoreyODEIM,BenderPTReview}.
\noindent
Having introduced the notion of Stokes sectors and their importance for defining eigenvalue problems in the complex plane (considering polynomial potentials), we can proceed by explaining the connection between this machinery and the Gamow--Siegert boundary conditions introduced previously. This is most easily seen by providing a simple example hinging on analytic continuation, setting the stage for a broader understanding of the resonant states under investigation.

\subsection{Resonant states as solutions to generalized eigenvalue problems}
\label{sec:2_2_AnalyticContinuation_EigenvalueProblems}

\begin{figure}[H]
    \centering
    \begin{subfigure}[c]{0.32\textwidth}
    \includegraphics[width=\textwidth]{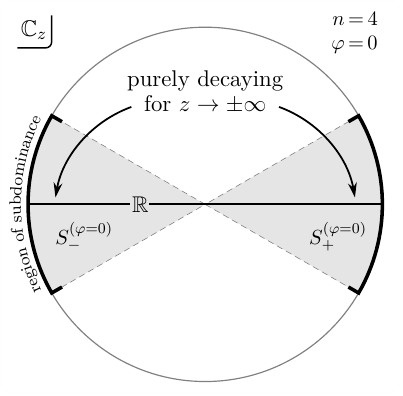}
	\includegraphics[width=\textwidth]{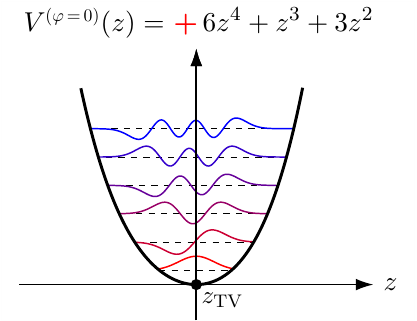}
    \end{subfigure}
    \hfill
    \begin{subfigure}[c]{0.32\textwidth}
    \includegraphics[width=\textwidth]{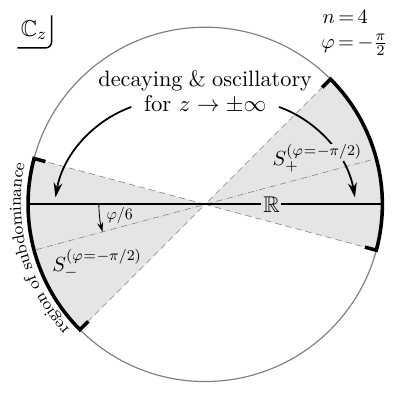}
	\includegraphics[width=\textwidth]{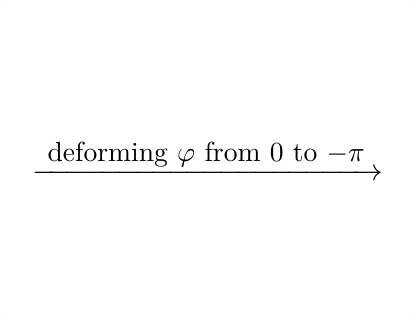}
    \end{subfigure}
    \hfill
    \begin{subfigure}[c]{0.32\textwidth}
    \includegraphics[width=\textwidth]{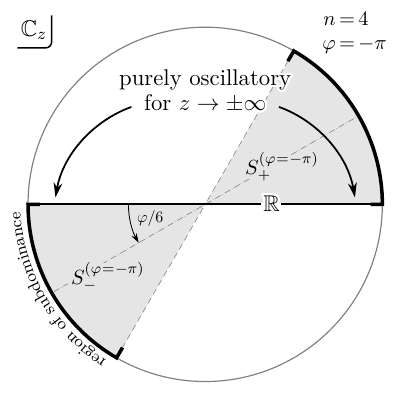}
	\includegraphics[width=\textwidth]{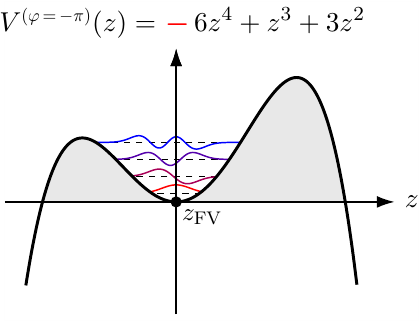}
    \end{subfigure}
    \caption{Process of analytic continuation of the eigenvalue problem~\eqref{eq:TimeIndependentSchrödingerEquation} given the exemplary quartic potential $\smash{V^{(\varphi)}(z)=6e^{i\varphi}z^4+z^3+3z^2}$ from $\varphi=0$ to $\varphi=-\pi$, rendering the initially stable potential unstable in the process. Whereas the boundary value problem associated to $\smash{V^{(\varphi\,=\,0)}(z)}$ can be defined on the real axis and admits purely real energy eigenvalues, the analytically continued eigenvalue problem obeys strictly oscillatory boundary conditions on $\mathbb{R}$ due to the Stokes sectors $\smash{S_\pm^{(\varphi)}}$ rotating counter-clockwise for decreasing $\varphi$. The schematically depicted bound states in the stable potential turn into the desired resonant states upon analytic continuation.}
    \label{fig:AnalyticContinuationQuarticPotential}
\end{figure}

We now wield the tools to make the notion of the previously characterized resonant states, possessing radiative outgoing boundary conditions, more explicit. To this end, we relate such solutions to the time-independent Schrödinger equation to a proper eigenvalue problem in the complex plane. These insights go back to the seminal work by Bender \& Wu on the anharmonic oscillator~\cite{BenderAnharmonicOscillator}, which were later generalized by Bender \& Turbiner~\cite{BenderAnalyticylContinuation}.\footnote{Similar ideas are utilized by the \emph{method of complex dilatation}, also dubbed \emph{complex scaling method}, going back to works by Aguilar, Balslev \&
Combes~\cite{AguilarComplexDilatation,BalslevComplexDilatation}, Simon~\cite{SimonComplexDilatation} and Yaris et al.~\cite{BenderComplexDilatation}; for dedicated reviews on this approach consult e.g.~\cite{ReinhardReviewComlexDilatation,MoiseyevComplexDilatation,HislopSpectralTheory} or the lucid textbook by Mariño~\cite{MarinoAdvancedQM}.} It was noticed that analytic continuation in the highest-order monomial coefficient $c_n$ goes hand in hand with a rotation of the Stokes sectors $S_\pm$ in which one demands subdominance, effectively changing the boundary conditions of the pertaining eigenvalue problem. This is best seen in the relevant case of interest for which a stable potential is rendered unstable upon analytic continuation, allowing us to view the resonant spectrum in the unstable potential as the analytic continuation of the spectrum of bound states in the associated stabilized potential. To see this explicitly, let us investigate the eigenfunctions of the time-independent Schrödinger equation 
\begin{align}
	\widehat{H}^{(\varphi)}\Psi_{\ell}^{(\varphi)}(z)=\scalebox{1.1}{\bigg\{}\!-\frac{\hbar^2}{2m}\frac{\mathrm{d}^2}{\mathrm{d}z^2} + V^{(\varphi)}(z)\scalebox{1.1}{\bigg\}} \,\Psi_\ell^{(\varphi)}(z) = E_{\ell}^{(\varphi)} \,\Psi_{\ell}^{(\varphi)}(z)\, ,
	\label{eq:TimeIndependentSchrödingerEquation}
\end{align}
given the polynomial potential
\begin{align}
	V^{(\varphi)}(z)=W(z)+e^{i\varphi} \lvert c_n\rvert z^n\, .
\end{align}
Hereby $W(x)$ shall denote a polynomial of order $n-1$ with purely real coefficients, whereas $n$ constitutes a positive, even integer.
\begin{itemize}
    \item For $\varphi=0$, the potential is real and stable, such that when demanding the eigenfunctions to be normalizable on the real axis, the boundary value problem is of regular Sturm--Liouville type. This yields a discretized set of real eigenenergies $\smash{E_{\ell}^{(\varphi\,=\,0)}}$, for which the associated eigenfunctions $\smash{\Psi_{\ell}^{(\varphi\,=\,0)}(z)}$ describe confined bound states, possessing a purely decaying behavior for $z\to \pm\infty$.
    \item With the sign of the highest-order monomial term flipped for $\varphi=\pm\pi$, the potential is rendered unstable and does not allow for normalizable eigenstates on the real axis anymore, i.e. it admits no bound states. Instead, as illustrated in figure~\ref{fig:AnalyticContinuationQuarticPotential}, the analytic continuation is accompanied by a rotation of the regions of subdominance $\smash{S_{\pm}^{(\varphi)}}$. While the sought-after eigenfunctions $\smash{\Psi_{\ell}^{(\varphi\,=\,\pm\:\!\pi)}(z)}$ are exponentially decaying when approaching complex infinity from within the tilted Stokes wedges $\smash{S_{\pm}^{(\varphi\,=\,\pm\:\!\pi)}}$, when restricting our view to their asymptotic behavior on $\mathbb{R}$, one will find a purely oscillatory character due to the sectors $\smash{S_{\pm}^{(\varphi\,=\,\pm\:\!\pi)}}$ precisely bordering the real axis. Consequently, the resulting eigenfunctions $\smash{\Psi_{\ell}^{(\varphi\,=\,\pm\:\!\pi)}(z)}$ precisely capture the desired behavior mandated by (anti-)resonant states.
\end{itemize}
\noindent 
Let us make the statements provided in the previous bullet point explicit: One finds that an increase $\Delta \varphi$ of the argument $\varphi$ of the coefficient $c_n$ multiplying the highest monomial $z^n$ is accompanied by a clockwise rotation of the Stokes sectors by an angle $\Delta\varphi/(n+2)$, see equation~\eqref{eq:DefinitionStokesSectors}. With the real axis dividing the initial regions of subdominance $\smash{S_\pm^{(\varphi\,=\,0)}}$ with width $2\pi/(n+2)$ precisely in half, a full rotation by an angle $\pi$ makes it such that the real axis is not entailed in the Stokes wedges $\smash{S_\pm^{(\varphi\,=\,\pm\:\!\pi)}}$ anymore. Instead, the real axis lies precisely on the boundary of the Stokes sectors defining the eigenvalue problem, leading to a purely oscillatory behavior of the eigenfunctions when viewing them as functions on $\mathbb{R}$. With the asymptotic behavior of the wave functions inside the sectors $\smash{S_\pm^{(\varphi)}}$ given by 
\begin{align}
    \Psi^{(\varphi)}_\ell(z)\sim \exp\!\!\:\Bigg\{\!-\!\!\:\frac{2e^{i\varphi/2}\sqrt{\lvert c_n\rvert}}{(n+2)\hbar}\, (\pm z)^{\frac{n+2}{2}}\Bigg\} \qquad \text{for } \lvert z \rvert \rightarrow \infty \text{ and } z\in S_\pm^{(\varphi)}\, ,
\end{align}
for $\varphi\to -\pi$ the behavior on the real axis is found to be
\begin{align}
    \Psi^{(\varphi\,=\,-\pi)}_\ell(z)\sim \exp\!\!\:\Bigg\{\frac{2i\sqrt{\lvert c_n\rvert}}{(n+2)\hbar}\, \lvert z\rvert^{\frac{n+2}{2}}\Bigg\} \qquad\qquad\!\!\!\,\!\!\: \text{for } z\to \pm\infty\, . \qquad\qquad\quad\;\,
\end{align}
It is apparent that these states describe a purely outgoing wave in both spatial directions, which is easily seen when computing the probability flux 
\begin{align}
    J(z)=\frac{\hbar}{m}\: \mathrm{Im}\Bigg\{\closure{\Psi^{(\varphi\,=\,-\pi)}_\ell(z)}\:\frac{\mathrm{d}\Psi^{(\varphi\,=\,-\pi)}_\ell(z)}{\mathrm{d}z}\Bigg\} = \pm \,\frac{\sqrt{\lvert c_n\rvert}}{m} \,\lvert z\rvert^{\frac{n}{2}}\qquad \text{for } z\to \pm \infty\, , 
\end{align}
being positive (right-moving) for $z\to\infty$ and negative (left-moving) for $z\to -\infty$. Note that, throughout this work, complex conjugation is denoted by an overhead bar. The above realization retrospectively explains why figure~\ref{fig:AnalyticContinuationQuarticPotential} shows a continuation to $\varphi=-\pi$ instead of $\varphi=\pi$, as the opposite rotation of the Stokes wedges would result in purely incoming Gamow--Siegert boundary conditions, describing anti-resonant states. This way, depending on whether the solutions decay in the Stokes sector just above or just below the real axis, we obtain solutions possessing either purely incoming or outgoing boundary conditions, being precisely the two possibilities that arise when imposing Gamow--Siegert boundary conditions. \\

\begin{figure}[H]
    \centering
    \includegraphics[width=\textwidth]{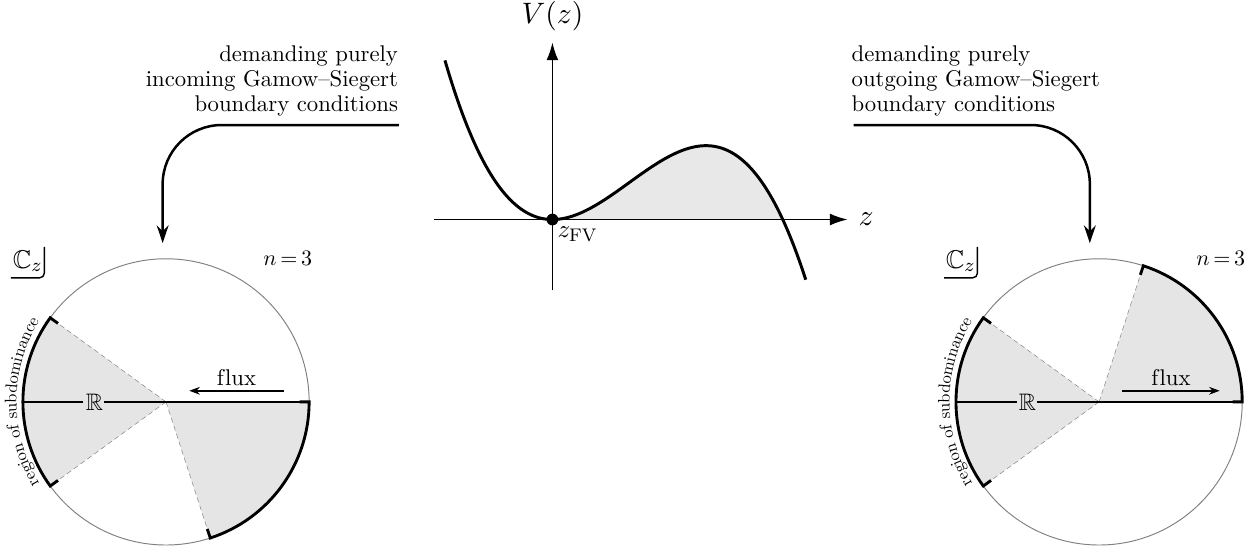}
    \caption{Illustration of the behavior of an odd-degree polynomial, with the example shown being a cubic potential with real coefficients. Once more, in the spatial direction for which barrier penetration effects are relevant, the real axis resides at the boundary of two Stokes sectors. Depending on whether to instate incoming or outgoing Gamow--Siegert boundary conditions, the eigenvalue problem is defined in the sector below or above the real axis.}
    \label{fig:OddPoweredPolynomial}
\end{figure}
\noindent 
A similar situation arises when the potential is unbounded from below in only one spatial direction, allowing tunneling exclusively in that direction, as depicted in figure~\ref{fig:OddPoweredPolynomial}. While this situation cannot be accessed by analytically continuing a globally stable potential, the identical conclusions to the previous case can be drawn. The real axis lies precisely on the boundary between two Stokes wedges, for which the eigenfunctions will asymptotically have a purely oscillatory behavior when viewed as functions on $\mathbb{R}$. Demanding the eigenfunctions to be subdominant either in the Stokes sectors above or below the real axis consequently amounts to imposing either a purely incoming or outgoing flux. Thus, the eigenfunctions arising from solving the boundary value problems illustrated on the left and the right of figure~\ref{fig:OddPoweredPolynomial} amount to anti-resonant or resonant states, respectively.

\section{Quantum mechanics on a generalized contour}
\label{sec:3_QM_on_ComplexContour}

As we have clarified in the previous section, the resonant ground-state energy---determining the sought-after decay rate---can be attained by solving a generalized eigenvalue problem in the complex plane, for which the eigenfunctions are demanded to be subdominant in two non-adjacent Stokes sectors. In order to efficiently capture the desired spectral information, it proves useful to introduce the notion of a propagator associated with the eigenvalue problem at hand, which one can ultimately express in terms of a functional integral. This feat can be achieved utilizing mostly standard tools from quantum mechanics with remarkably little adjustments. While the subsequently shown techniques could potentially be extended to more general Sturm--Liouville problems, let us for now stay in the realm of quantum mechanics and suppose our boundary value problem assumes the usual form of a time-independent Schrödinger equation
\begin{align}
	\scalebox{1.1}{\bigg\{}\!\!\!\!\!\!\!\!\!\underbrace{-\,\frac{\hbar^2}{2m}\frac{\mathrm{d}^2}{\mathrm{d}z^2} + V(z)}_{\displaystyle{\text{original Hamiltonian }\widehat{H}}}\!\!\!\!\!\!\!\!-\, E_\ell\scalebox{1.1}{\bigg\}} \,\Psi_\ell(z) = 0\, , \qquad\quad \Psi_\ell(\eta z)\xrightarrow{\,\eta\to\infty\,} 0 \;\;\;\; \mathrm{for} \;z\in S_\pm\, .
	\label{eq:Schrödinger_EV_Problem}
\end{align}
Keeping the discussion in line with our previous considerations, let us demand the potential $V(z)$ to be a polynomial with arbitrary complex coefficients. As before, we denote the two non-adjacent Stokes sectors in which the eigenfunctions $\Psi_\ell(z)$ are (asymptotically) demanded to exponentially decay by $S_\pm$. Let us emphasize that the following derivation is entirely general and holds for any arbitrary choice of (non-adjacent) Stokes sectors $S_\pm$ used to define the eigenvalue problem. While the final result clearly depends on this choice, both the structure and validity of the derivation itself remain unchanged.\\ 

\begin{figure}[H]
	\centering
	\includegraphics[width=0.4\textwidth]{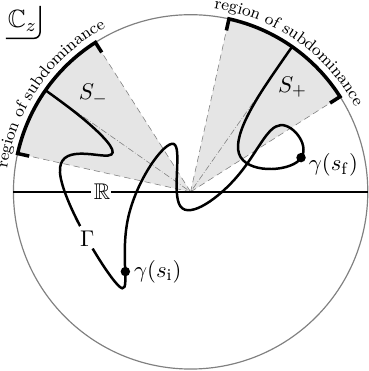}
	\caption{Exemplary choice of the complex contour $\Gamma$, parametrized through the smooth surjection ${\gamma(\sbullet):\mathbb{R}\to\Gamma}$, asymptotically terminating in the two non-adjacent Stokes wedges $S_\pm$ chosen for the definition of the generic Schrödinger-type eigenvalue problem~\eqref{eq:Schrödinger_EV_Problem}. Any such contour allows us to define the propagator $\smash{K_\gamma\big(s_\mathrm{i},\mathrm{s}_f\:\! ;T\big)}$, representing the probability of a quantum-mechanical point particle initially located at $\gamma(s_\mathrm{i})$ to be found at $\gamma(s_\mathrm{f})$ after time $T$ when evolving under the influence of the original Hamiltonian $\widehat{H}$, assuming the particles motion is fully constrained to the contour $\Gamma$.}
	\label{fig:CustomContour_Gamma}
\end{figure}

\noindent 
The first step in attributing a (quantum-mechanical) propagator to the generalized eigenvalue problem~\eqref{eq:Schrödinger_EV_Problem} is to restrict our view to a real, one-dimensional subspace of the complex plane, serving as the confining space for the propagation of our quantum particle. To analyze the problem along a single dimension, we choose an arbitrary smooth contour $\Gamma\subset \mathbb{C}$ asymptotically starting and ending inside the Stokes wedges $S_\pm$ as illustrated in figure~\ref{fig:CustomContour_Gamma}. The contour is the image of the smooth function $\gamma(s)$, with positions on $\Gamma$ parametrized by the real coordinate $s$.\footnote{Note that $\gamma(\sbullet)$ is not necessarily defined over the entire complex plane, it suffices for the function to solely be defined on the contour $\Gamma$.} A proper parametrization furthermore demands $\gamma'(s)\neq 0$, whereas the requirements of $\Gamma$ ending asymptotically in the particular Stokes sectors $S_\pm$ can be represented as $\exists s_+\in \mathbb{R}^+ \text{ s.t. } \forall s>s_+: \gamma(s)\in S_+$ and $\exists s_-\in \mathbb{R}^- \text{ s.t. } \forall s<s_-: \gamma(s)\in S_-$. Even though we will later constrain the admissible contours $\Gamma$ further, let us for now proceed under the premise that any such choice is equally valid. Note that the presented idea of mapping non-Hermitian eigenvalue problems to the real line by choosing a complex contour is not novel and has been applied by Mostafazadeh, see e.g.~\cite{MostafazadehComplexContour} or his review on pseudo-Hermitian quantum mechanics~\cite{MostafazadehReview}.

\noindent 
The introduction of a contour $\Gamma$ allows us to transform the initially complex eigenvalue problem~\eqref{eq:Schrödinger_EV_Problem} to one defined on the real axis, as the ODE can be cast into the form 
\begin{align}
	\Bigg\{\!\underbrace{-\:\!\frac{\hbar^2}{2m}\frac{1}{\gamma'(s)}\, \frac{\mathrm d}{\mathrm{d}s}\bigg[\frac{1}{\gamma'(s)}\, \frac{\mathrm d}{\mathrm{d}s}\bigg] + V\!\!\:\big[\gamma(s)\big]}_{\displaystyle{\text{effective Hamiltonian } \widehat{H}_\gamma} \text{ on $\Gamma$}}-\,E_\ell\Bigg\}\, \underbrace{\Psi_\ell\big[\gamma(s)\big]\vphantom{\Big[}}_{\displaystyle{\eqqcolon \psi_\ell(s)}} = 0\, .
	\label{eq:Hamiltonian}
\end{align}
Here, we introduced the effective Hamiltonian $\widehat{H}_\gamma$ as well as the ``contour-constrained'' wave function $\psi_\ell(s)$. By construction of $\Gamma$, we have $\smash{\Psi_\ell\big[\gamma(s)\big]\,\raisebox{-0.5pt}{$\xrightarrow{s\to\pm\infty}$}\,0}$, thus enforcing $\smash{\psi_\ell(s) \,\raisebox{-0.5pt}{$\xrightarrow{s\to\pm\infty}$}\, 0}$, i.e. we know the functions $\psi_\ell(s)$ to decay at $s$-spatial infinity. We have thus recast the eigenvalue problem into the conventional form familiar from standard quantum mechanics---defined on the real axis, with normalizable eigenfunctions $\psi_\ell(s)$ vanishing at the boundaries. At this point, we therefore may temporarily forget about the existence of the contour $\Gamma$ and treat the spectral problem as if it would be formulated on $\mathbb{R}$, with the crucial information being encapsulated in the transformed Hamiltonian $\widehat{H}_\gamma$, which, for $\Gamma\neq \mathbb{R}$, generally ceases to be Hermitian independently of the chosen scalar product for the underlying function space $L^2(\mathbb{R})$.\footnote{While in the previous discussion of resonant states, Hermiticity of $\widehat{H}$ was violated due to the Gamow--Siegert boundary conditions yielding additional boundary terms when integrating by parts, in the present case Hermiticity would be explicitly violated by $\widehat{H}_\gamma$ itself due to $\gamma(s)$ being manifestly complex. In case $\gamma(s)$ is real but non-linear, Hermiticity would be restored by choosing the measure of the $L^2\big(\mathbb{R},\mu\big)$-norm accordingly as $\mu(s)=\gamma'(s)$.} To proceed, we select a convenient scalar product, for which we will deliberately pick the conventional inner product defined through
\begin{align}
	\braket[\rbig]{\phi\!\!\;|\psi} \coloneqq \!\!\:\mathlarger{\int}_{-\infty}^{\infty}\, \closure{\phi(s)}\,\psi(s)\,\mathrm{d}s\, .
	\label{eq:StandardScalarProduct}
\end{align}
Let us clearly state that the foregoing scalar product once more constitutes a choice. While we could in principle introduce an additional real weight function $\mu(s)$, the subsequent results are seen to be practically independent of this arbitrary choice, which is briefly discussed in appendix~\ref{sec:B_ScalarProduct_Discussion}. It proves to be most convenient to work with the standard scalar product~\eqref{eq:StandardScalarProduct}, thereby preserving Hermiticity of the momentum operator $\hat{p}_\gamma\coloneqq -i\hbar \,\mathrm{d}/\mathrm{d}s$, being canonically conjugate to the position operator $\hat{x}_\gamma$ on the contour. As in traditional quantum mechanics, we can define a complete set of position and momentum eigenstates denoted by $\ket{s}$ and $\ket{p}$ respectively. Given the standard scalar product~\eqref{eq:StandardScalarProduct} on $\Gamma$, these states satisfy the usual identities $\hat{x}_\gamma\ket{s}=s\ket{s}$, $\hat{p}_\gamma\ket{p}=p\ket{p}$, $\braket{s\!\!\;|s'}=\delta(s-s')$, $\braket{p\!\!\;|p'}=2\pi\hbar\,\delta(p-p')$, $\braket{s\!\!\;|p}=e^{ips/\hbar}$ as well as the two completeness relations 
\begin{align}
	\mathds{1}=\mathlarger{\int}_{-\infty}^{\infty} \ket{s}\bra{s}\,\mathrm{d}s = \frac{1}{2\pi\hbar}\mathlarger{\int}_{-\infty}^{\infty} \ket{p}\bra{p}\,\mathrm{d}p\, .
	\label{eq:CompletenessRelations_PosMom}
\end{align}
Note that the given relations have to be altered accordingly in case a weight function $\mu(s)$ is introduced into the definition of the scalar product, see appendix~\ref{sec:B_ScalarProduct_Discussion}. Having established the foregoing, one can introduce a slightly modified definition of the quantum-mechanical propagator as
\begin{align}
	K_{\gamma,\theta}\big(s_\mathrm{i},s_\mathrm{f}\:\!;T\big) &\coloneqq \frac{1}{\gamma'(s_\mathrm{i})}\,\braket[\rbigg]{s_\mathrm{f} | \,\exp\!\bigg(\!\!-\!\frac{i e^{-i\theta} \widehat{H}_\gamma T}{\hbar}\bigg)\;\!\! |\:\! s_\mathrm{i} }\, ,
	\label{eq:Definition_Propagator}
\end{align}
characterizing the transition amplitude of a point particle propagating from $s_\mathrm{i}$ to $s_\mathrm{f}$ in a given time interval $T$. With the potential $V(z)$ assumed to be time-independent, the propagator only depends on the total time difference $T$. The additional factor $\gamma'(s_\mathrm{i})^{-1}$ is introduced solely for convenience, as will become clear in due course. For generality, we introduced the Wick-rotation angle $\theta\in (0,\pi/2]$, interpolating between the usual real-time propagator, attained in the limit $\theta\to 0^+$, and the Euclidean one, obeying $\theta\in \pi/2$.\footnote{In general, $\theta$ is not constrained to the interval $(0,\pi/2]$ but can be chosen in the full range $(-\pi,\pi]$, as long as convergence of all expressions is controlled. In certain scenarios, this additional freedom is even required.} The broader applicability of the resulting formulas outweighs the minor notational overhead suffered by introducing this additional angular parameter. Similar to conventional quantum mechanics, we will find that the so-defined propagator~\eqref{eq:Definition_Propagator} can be expressed via a spectral representation as well as a generalized path integral. Equating both expressions allows us to infer vital information about the spectrum of the theory using purely functional techniques. Going forward, let us first deal with the easily attainable spectral representation in section~\ref{sec:3_1_Spectral_Representation} before delving into the path integral derivation portrayed in section~\ref{sec:3_2_PathIntegral_Representation}.

\subsection{Spectral representation}
\label{sec:3_1_Spectral_Representation}

Recall that the spectral representation in ordinary quantum mechanics is trivially attained by inserting a resolution of unity in terms of the complete set of eigenstates $\ket{\psi_\ell}$ to the given Hamiltonian $\widehat{H}_\gamma$. However, the usual construction crucially relies on the fact that the Hamiltonian is Hermitian, which $\widehat{H}_\gamma$ generally does not satisfy. Instead, simple integration by parts yields the relation
\begin{align}
	\braket[\cbig]{\phi\!\!\:|\widehat{H}_\gamma\psi}&=-\frac{\hbar^2}{2m} \mathlarger{\int}_{-\infty}^{\infty} \closure{\phi(s)} \: \frac{1}{\gamma'(s)}\, \frac{\mathrm d}{\mathrm{d}s}\bigg[\frac{1}{\gamma'(s)}\, \frac{\mathrm d}{\mathrm{d}s}\bigg] \psi(s) \,\mathrm{d}s + \mathlarger{\int}_{-\infty}^{\infty} \closure{\phi(s)} \,  V\!\!\:\big[\gamma(s)\big]\:\! \psi(s) \,\mathrm{d}s \nonumber \\[0.1cm]
	&= -\frac{\hbar^2}{2m} \mathlarger{\int}_{-\infty}^{\infty} \psi(s) \:\! \frac{\mathrm d}{\mathrm{d}s}\cbigg\{\frac{1}{\gamma'(s)}\, \frac{\mathrm d}{\mathrm{d}s}\bigg[\frac{1}{\gamma'(s)}\,\closure{\phi(s)}\bigg]\!\!\:\cbigg\}   \,\mathrm{d}s + \mathlarger{\int}_{-\infty}^{\infty}\psi(s)\, \closure{\closure{V\!\!\:\big[\gamma(s)\big]}\,\phi(s)} \,\mathrm{d}s \nonumber \\[0.1cm] 
	&= \braket[\cbig]{\widehat{H}_\gamma^\dagger\phi\!\!\:|\psi} = \closure{\braket[\cbig]{\psi\!\!\:|\widehat{H}_\gamma^\dagger\phi}}  \, ,
    \label{eq:Comp_HermitianConjugate}
\end{align}
with all boundary terms vanishing under the assumption of the wave functions $\phi(s)$ and $\psi(s)$ decaying at $s$-spatial infinity, allowing us to read off the Hermitian conjugate as 
\begin{align}
	\widehat{H}_\gamma^\dagger &= -\frac{\hbar^2}{2m}\frac{\mathrm d}{\mathrm{d}s}\cbigg[\frac{1}{\closure{\gamma'(s)}} \frac{\mathrm d}{\mathrm{d}s}\frac{1}{\closure{\gamma'(s)}}\cbigg] + \,\closure{V\!\!\:\big[\gamma(s)\big]} \, . 
	\label{eq:ConjugateHamiltonian}
\end{align}
The vital difference to the standard Hermitian case is that the eigenfunctions $\psi_\ell(s)$ of $\widehat{H}_\gamma$ are not orthogonal with respect to the prescribed scalar product~\eqref{eq:StandardScalarProduct}. An appropriate resolution of the identity thus requires the use of a so-called \emph{bi-orthogonal basis}~\cite{BirkhoffBiorthogonalBasis,PellBiorthogonalBasis,BariBiorthogonalBasis,SternheimBiorthogonalBasis}; a modern review on the topic has been provided by Brody~\cite{BrodyBiorthogonalBasis}. The eigenfunctions $\psi_\ell(s)$ of $\widehat{H}_\gamma$ are thereby supplemented by a second set of eigenfunctions $\phi_\ell(s)$ of the Hermitian adjoint $\widehat{H}_\gamma^\dagger$, satisfying 
\begin{align}
	\widehat{H}_\gamma \:\!\psi_\ell(s)&=E_\ell \, , & \widehat{H}_\gamma^\dagger \,\phi_\ell(s)&=\closure{E_\ell} \, .
	\label{eq:EigenvalueEquationDualBasis}
\end{align}
Assuming the spectrum possesses no degeneracies, it can be shown that the two bases $\big\{\psi_\ell(\sbullet)\big\}_{\scalebox{0.75}{$\ell\!\!\:\in\!\!\:\mathbb{N}^0$}}$ and $\big\{\phi_\ell(\sbullet)\big\}_{\scalebox{0.75}{$\ell\!\!\:\in\!\!\:\mathbb{N}^0$}}$ are mutually orthogonal, satisfying $\braket{\phi_\ell\!\!\:|\psi_m}\propto \delta_{\ell m}$. This in turn allows us to establish the desired resolution of the identity as\footnote{Beware that these considerations only rigorously apply to finite-dimensional Hilbert spaces~\cite{BrodyBiorthogonalBasis}. Expecting no mathematical atrocities to hide in the present case, we will proceed without the proper foundation backing up this otherwise straightforward generalization.}
\begin{align}
	\mathds{1}=\mathlarger{\sum}_{\ell=0}^\infty\: \frac{\ket{\psi_\ell}\bra{\phi_\ell}}{\braket{\phi_\ell\:\!\!|\psi_\ell}}\, .
	\label{eq:BiorthogonalResolutionOfUnity}
\end{align}
The important realization in the present case is the fact that the two sets of eigenfunctions are related, with the dual eigenfunctions $\phi_\ell(s)$ satisfying
\begin{align}
	\phi_\ell(s)=\closure{\gamma'(s)\:\! \psi_\ell(s)}\, ,
	\label{eq:RelationDualEigenfunctions}
\end{align}
which can be verified by plugging the ansatz~\eqref{eq:RelationDualEigenfunctions} into the eigenvalue relation~\eqref{eq:EigenvalueEquationDualBasis}, explicitly using the form of $\smash{\widehat{H}_\gamma}$ and $\smash{\widehat{H}_\gamma^\dagger}$ given in equations~\eqref{eq:Hamiltonian} and~\eqref{eq:ConjugateHamiltonian} respectively. Inserting the resolution of unity~\eqref{eq:BiorthogonalResolutionOfUnity} into the definition~\eqref{eq:Definition_Propagator} of the propagator $K_{\gamma,\theta}\big(s_\mathrm{i},s_\mathrm{f}\:\!;T\big)$ on the contour $\Gamma$, we arrive at the simple result
\begin{align}
	K_{\gamma,\theta}\big(s_\mathrm{i},s_\mathrm{f}\:\!;T\big) &=\frac{1}{\gamma'(s_\mathrm{i})}\:\mathlarger{\mathlarger{\sum}}_{\ell=0}^\infty \; \braket[\rbigg]{s_\mathrm{f} | \,\exp\!\bigg(\!\!-\!\frac{i e^{-i\theta} \widehat{H}_\gamma T}{\hbar}\bigg)\;\!\! |\, \psi_\ell} \, \frac{\braket[\rbig]{\phi_\ell\!\!\:|s_\mathrm{i}}\;}{\braket[\rbig]{\phi_\ell\!\!\:|\psi_\ell}} \nonumber \\
	&= \frac{1}{\gamma'(s_\mathrm{i})}\: \mathlarger{\mathlarger{\sum}}_{\ell=0}^\infty \; \frac{\closure{\phi_\ell(s_\mathrm{i})}\,\psi_\ell(s_\mathrm{f})}{\braket[\rbig]{\phi_\ell\!\!\:|\psi_\ell}} \: \exp\!\bigg(\!\!-\!\frac{i e^{-i\theta} E_\ell T}{\hbar}\bigg) \nonumber\\
	&= \mathlarger{\mathlarger{\sum}}_{\ell=0}^\infty \:\, \Psi_\ell\big[\gamma(s_\mathrm{i})\big]\,\Psi_\ell\big[\gamma(s_\mathrm{f})\big] \:\Bigg\{\mathlarger{\int}_{\Gamma}\:\! \,\Psi_\ell(z)^2\,\mathrm{d}z\Bigg\}^{\!-1} \: \exp\!\bigg(\!\!-\!\frac{i e^{-i\theta} E_\ell T}{\hbar}\bigg)\, .
	\label{eq:Propagator_SpectralRepresentation}
\end{align}
In the last line, we expressed the ``contour-constrained'' eigenfunctions $\psi_\ell(s)$ in terms of the holomorphic solutions $\Psi_\ell(z)$ to the initial eigenvalue problem~\eqref{eq:Schrödinger_EV_Problem} in the complex plane, while using relation~\eqref{eq:RelationDualEigenfunctions} to express $\phi_\ell(s)$ in terms of $\psi_\ell(s)$. Additionally, we utilized the important fact that the normalization $\braket[\rbig]{\phi_\ell\!\!\:|\psi_\ell}$ can be represented as the complex contour integral 
\begin{align}
	\!\!\!\:\scalebox{0.97}{$\displaystyle{\braket[\rbig]{\phi_\ell\!\!\:|\psi_\ell}=\!\!\:\mathlarger{\int}_{-\infty}^{\infty}\, \closure{\phi_\ell(s)}\,\psi_\ell(s)\,\mathrm{d}s=\!\!\:\mathlarger{\int}_{-\infty}^{\infty}\!\: \gamma'(s)\!\: \psi_\ell(s)^2\!\:\mathrm{d}s=\!\!\:\mathlarger{\int}_{-\infty}^{\infty} \!\:\gamma'(s)\!\: \Psi_\ell\big[\gamma(s)\big]^2\!\:\mathrm{d}s =\!\!\:\mathlarger{\int}_{\Gamma}\, \Psi_\ell(z)^2\!\:\mathrm{d}z\, ,}$}
	\label{eq:Normalization_ContourIntegral}
\end{align}
again making explicit use of the relation between the bi-orthogonal basis functions as expressed in equation~\eqref{eq:RelationDualEigenfunctions}.\footnote{Note that the emerging contour integral is non-vanishing everywhere aside from exceptional points, since zeros of the given wave function normalization~\eqref{eq:Normalization_ContourIntegral} in the complex coupling plane correspond to Bender--Wu branch points where level crossings occur, see e.g. the seminal work by the eponymous authors~\cite{BenderAnharmonicOscillator}.} We find that the propagator~\eqref{eq:Propagator_SpectralRepresentation} depends solely on the endpoints $\gamma(s_\mathrm{i,f})$ in the complex plane, retrospectively showing why it was convenient to introduce an additional factor $\gamma'(s_\mathrm{i})^{-1}$ in the initial definition~\eqref{eq:Definition_Propagator} of the propagator, neutralizing the otherwise leftover factor $\gamma'(s_\mathrm{i})$ entering through the conversion of $\closure{\phi_\ell(s_\mathrm{i})}$ into $\psi_\ell(s_\mathrm{i})$. All remaining $\gamma$-dependence has been absorbed into the normalization factor~\eqref{eq:Normalization_ContourIntegral}, for which the contour $\Gamma$ can be freely deformed without altering the value of the complex contour integral, provided $\Gamma$ ends in the desired regions of subdominance of the holomorphic eigenfunctions $\Psi_\ell(z)$. This concludes the derivation of the spectral representation, which was found to be virtually independent of the chosen contour $\Gamma$ and its parameterization. Even though the absence of any complex conjugation in the final representation~\eqref{eq:Normalization_ContourIntegral} of the propagator seems to contradict the usual quantum-mechanical case, note that this is indeed not the case. In case one deals with a real, confining, Hermitian Hamiltonian $\widehat{H}$, allowing for the definition of the eigenvalue problem on the real axis, i.e. enabling the choice $\Gamma=\mathbb{R}$, the eigenfunctions can be chosen to be entirely real, thus the emerging complex conjugation in the usual formula can be dropped. All other representations of the eigenfunctions can only differ by an overall (global) phase, which drops out of the propagator once the normalization is performed accordingly.\footnote{This also clarifies that the quantum-mechanical propagator in case of time-independent potentials $V(z)$ is indeed a manifestly analytic quantity in all its parameters, which is not immediately evident given the usual discussion still involving complex conjugation.}\\

\noindent 
As a brief final note, one should keep in mind that the given spectral representation~\eqref{eq:Propagator_SpectralRepresentation} is only well-defined as long as the sum converges, effectively restricting the admissible values of $\theta$. Given the asymptotic growth of the polynomial potential $V(z)\sim c_n z^n$ for $\lvert z\rvert \to \infty$, the argument of the eigenvalues asymptotically approaches
\begin{align}
    \arg(E_\ell)\xrightarrow{\ell\to\infty} \vartheta_\infty \coloneqq \frac{2\pi}{n+2}\bigg[\frac{n+2}{2}-\big(k_++k_-\big)\bigg]+\frac{2\arg(c_n) }{n+2} \quad \text{mod } 2\pi\, .
\end{align}
In this context, we indexed the two Stokes sectors in which we demand the eigenfunctions to be subdominant by $k_\pm$, i.e. we identified $S_\pm\ \eqqcolon S_{k_\pm}$ utilizing the defining relation~\eqref{eq:DefinitionStokesSectors}.\footnote{Note that in the ordinary self-adjoint case, for which all $c_\ell$ are real and the eigenvalue problem is defined on the real axis (thus $n$ being even), one arrives at $k_-=\pm (n+2)/2$ and $k_+=0$, thus indeed recovering the correct result of the eigenvalues being (asymptotically) real.} The discussion on how to arrive at this result, together with the full leading-order behavior~\eqref{eq:EigenvalueAsymptoticsE} of the eigenvalues $E_\ell$ for $\ell\to\infty$, is provided in appendix~\ref{sec:D_LargeOrderBehavior_Eigenvalues}, hinging on earlier works by Sibuya~\cite{SibuyaEigenvalueProblems} and Shin~\cite{ShinPTEigenvalues,ShinLargeEigenvalues}. Demanding convergence of the sum~\eqref{eq:Propagator_SpectralRepresentation} subsequently requires
\begin{align}
    \arg\!\big(i e^{-i\theta} E_\ell\big)\xrightarrow{\ell\to\infty} \frac{\pi}{2}-\theta+\vartheta_\infty \in \bigg(\!\!-\!\frac{\pi}{2},\frac{\pi}{2}\bigg) \quad\! \Longrightarrow \!\quad \theta\in \Big(\vartheta_\infty,\pi+\vartheta_\infty\Big)\, ,
\end{align}
in turn constraining the admissible choices of $\theta$. In the desired case of describing resonant states encoding false vacuum decay, one obtains $k_++k_-=(n+3)/2$ (odd $n$) or $k_++k_-=(n+4)/2$ (even $n$) together with $c_n\in \mathbb{R}$, such that the argument $\vartheta_\infty$ is negative. Note that this is consistent with expectations, as the resonant eigenenergies possess a negative imaginary part. Thus, the choice of $\theta\in \big(0,\pi/2\big]$ is fully unconstrained, which would not be the case if we were interested in anti-resonant states.

\subsection{Path integral representation}
\label{sec:3_2_PathIntegral_Representation}

With the spectral representation~\eqref{eq:Propagator_SpectralRepresentation} of the propagator $K_{\gamma,\theta}\big(s_\mathrm{i},s_\mathrm{f}\:\!;T\big)$ under sufficient control, let us turn our focus toward its associated path integral counterpart. As usual, the path integral derivation hinges on inserting a sequence of identity resolutions, expressed in terms of position and momentum operators~\eqref{eq:CompletenessRelations_PosMom}, into the time evolution operator, after which an appropriate continuum limit is taken. Representing the effective Hamiltonian~\eqref{eq:Hamiltonian} in terms of the canonical operators $\hat{x}_\gamma$ and $\hat{p}_\gamma=-i\hbar \,\mathrm{d}/\mathrm{d}s$ yields
\begin{align}
	\widehat{H}_\gamma=H_\gamma\big(\hat{x}_\gamma,\hat{p}_\gamma\big) 
	&=\frac{1}{2m}\frac{1}{\gamma'(\hat{x}_\gamma)^2}\,\hat{p}_\gamma^2+\frac{i\hbar}{2m}\frac{\gamma''(\hat{x}_\gamma)}{\gamma'(\hat{x}_\gamma)^3}\, \hat{p}_\gamma + V\!\!\:\big[\gamma(\hat{x}_\gamma)\big]\, .
	\label{eq:NonStandard_Hamiltonian}
\end{align}
Throughout the work, we adopt the convention that pure operators are indicated with overhead wedges, while functions of operators are left unadorned. We furthermore made the deliberate choice to \emph{normal order} all operators, meaning momentum operators are commuted to the right; a notion also referred to as \emph{standard ordering}~\cite{LangoucheFunctionalIntegration}. The arising discretized path integral thus employs the so-called \emph{post-point discretization scheme}.\footnote{It is crucial to point out that this ordering choice does not influence the final result, as the initial Hamiltonian $\widehat{H}_\gamma$ is fully specified. While a different ordering prescription leads to altered intermediate expressions, utilizing the appropriate substitution rules discussed in appendix~\ref{sec:SubstitutionRules} restores uniqueness in the continuum limit $N\to\infty$, see e.g.~\cite{LangoucheFunctionalIntegration} and references therein.} Splitting the analytically continued time evolution operator entailed in the propagator~\eqref{eq:Definition_Propagator} into $N$ equal parts and inserting full sets of position eigenstates, one arrives at the intermediate discretized path integral representation
\begin{align}
	K_{\gamma,\theta}\big(s_\mathrm{i},s_\mathrm{f}\:\!;T\big) &= \frac{1}{\gamma'(s_\mathrm{i})}\lim_{N \to \infty} \mathlarger{\mathlarger{\int}}_{-\infty}^{\infty} \, \mathlarger{\prod}_{k=1}^{N-1} \:\!\mathrm{d}s_{k} \; \Bigg\{\prod_{\ell=1}^N \braket[\cbigg]{s_\ell|\: \exp\!\!\!\;\bigg[\!\!\:-\!\!\:\frac{i e^{-i\theta} T}{N\hbar}\, \widehat{H}_\gamma\bigg]\!\!\;|\,s_{\ell-1}\!}\Bigg\}_{\!\substack{\,\scalebox{0.8}{$s_0\:\!\!=\:\!\!s_\mathrm{i}$}\;\;\;\, \\ \scalebox{0.8}{$s_{\!\!\:N}\:\!\!=\:\!\!s_\mathrm{f}$}\;\;\;\,}}\nonumber \\ 
    &= \frac{1}{\gamma'(s_\mathrm{i})}\lim_{N \to \infty} \mathlarger{\mathlarger{\int}}_{-\infty}^{\infty} \, \mathlarger{\prod}_{k=1}^{N-1} \:\!\mathrm{d}s_{k} \; \Bigg\{\prod_{\ell=1}^N \Big[\gamma'(s_{\ell-1})\:\!K_{\gamma,\theta}\big(s_{\ell-1},s_{\ell}\:\! ; T/N\big)\Big]\Bigg\}_{\!\substack{\,\scalebox{0.8}{$s_0\:\!\!=\:\!\!s_\mathrm{i}$}\;\;\;\, \\ \scalebox{0.8}{$s_{\!\!\:N}\:\!\!=\:\!\!s_\mathrm{f}$}\;\;\;\,}} \nonumber \\ 
    &= \lim_{N \to \infty} \mathlarger{\mathlarger{\int}}_{-\infty}^{\infty} \, \mathlarger{\prod}_{k=1}^{N-1} \:\!\cbig[\gamma'(s_{k})\,\mathrm{d}s_{k}\cbig] \,\Bigg\{\prod_{\ell=1}^N K_{\gamma,\theta}\big(s_{\ell-1},s_{\ell}\:\! ; T/N\big)\Bigg\}_{\!\substack{\,\scalebox{0.8}{$s_0\:\!\!=\:\!\!s_\mathrm{i}$}\;\;\;\, \\ \scalebox{0.8}{$s_{\!\!\:N}\:\!\!=\:\!\!s_\mathrm{f}$}\;\;\;\,}} \!\!.
	\label{eq:DiscretizedPI_Start} 
\end{align}
Accounting for the additional term $\gamma'(s_\mathrm{i})^{-1}$ instated in the original definition~\eqref{eq:Definition_Propagator} of $K_{\gamma,\theta}$, the intermediate propagators are similarly dressed with inverse weight factors. In the last line, we simply redistributed these additional weight factors in a convenient way. The upcoming crucial step is the study of the short-time propagator $K_{\gamma,\theta}\big(s_{\ell-1},s_{\ell}\:\! ; T/N\big)$, requiring some subtle trickery to arrive at the correct result. As it is rather instructive, we choose to provide two different derivations of the desired result, hinging on techniques reviewed in appendix~\ref{sec:A_PathIntegralTechniques}. In the present section, we will simply apply these methods, referring the inclined reader to the associated appendix chapter for a more in-depth discussion.

\subsubsection{$1^{\text{st}}$ option: Na\"{\i}ve propagator evaluation \& use of substitution rules}

Following conventional path integral derivations, one inserts a set of momentum eigenstates~\eqref{eq:CompletenessRelations_PosMom} into the short-time propagator $K_{\gamma,\theta}\big(s_{\ell-1},s_{\ell}\:\! ; T/N\big)$ and na\"{\i}vely evaluates the emerging expression as if the infinitesimal time evolution operator would already be normal ordered. This procedure is usually justified in either of two ways: 
\begin{itemize}
    \item For Hamiltonians with split position- and momentum-dependence, one utilizes a standard Suzuki-Trotter decomposition~\cite{TrotterProduct,SuzukiProduct} to insert the position and momentum-resolutions between the split exponential pieces, leading to the fact that all arising matrix elements can be evaluated trivially. The overall effect of this trick is then precisely that of using~\eqref{eq:DiscretizedPI_Start} with the quirk that the infinitesimal time evolution operators entailed in the short-time propagators readily possess a fixed ordering. However, in the present case, with the presence of position-dependent kinetic terms, this strategy does not suffice.
    \item A straightforward but mathematically misguided way of obtaining the leading-order behavior of the propagator for short times is to na\"{\i}vely expand the exponential time evolution operator to order $T/N$, after which the matrix element can be evaluated easily. As a final step, one re-exponentiates the attained result, leading to conventional expressions for the propagator. A more sophisticated study would however reveal that the dropped terms yield non-negligible contributions, thus at first glance invalidating this approach. However, when using the initially erroneous result for the short-time propagator in the full discretized path integral~\eqref{eq:DiscretizedPI_Start}, the correct macroscopic result miraculously emerges~\cite{LangoucheFunctionalIntegration}. The rather subtle discussion required to arrive at this conclusion is presented in appendix~\ref{sec:A_PathIntegralTechniques}, utilizing the operator-ordering approach conceptualized by Graham~\cite{GrahamPathIntegralMethods1,GrahamPathIntegralMethods2}.\footnote{Beware that the results presented in appendix~\ref{sec:A_PathIntegralTechniques} constitute well-established knowledge, with some of the necessary techniques having been developed by DeWitt as early as the 1950s~\cite{deWittPointTransformations,deWittCurvedSpaces}; other notable early works include~\cite{ItoIntegral,StratonovichIntegral,McLaughlinSchulmanCurvedSpaces,GervaisPointCanonicalTrafos,HirshfeldCovariantPI,WeissGravPathIntegral,LangouchePathIntegralTechniques,LangoucheFunctionalIntegration}. Despite these results being well-known in communities dealing with stochastic calculus or path integrals on curved manifolds, we include the procedure to arrive at these findings to keep the work fully self-contained.}
\end{itemize}
We conclude that the second option allows us to treat the current case as if the infinitesimal time evolution operators are indeed normal ordered, leading to correct results once the continuum limit has been taken. Symbolizing equivalence up to contributions that vanish when taking the continuum limit $N\to\infty$ with $\doteq$, a notion introduced by DeWitt~\cite{deWittCurvedSpaces}, one arrives at the relation  
\begin{align}
	K_{\gamma,\theta}\big(s_{\ell-1},s_{\ell}\:\! ; T/N\big)&\doteq\frac{1}{\gamma'(s_{\ell-1})}\mathlarger{\mathlarger{\int}}_{-\infty}^{\infty}\frac{\mathrm{d}p_\ell}{2\pi\hbar} \: \braket[\rBigg]{s_\ell|\: \normordbig\exp\!\!\:\cbigg[-\frac{i e^{-i\theta} T}{N\hbar}\, H_\gamma\big(\hat{x}_\gamma,\hat{p}_\gamma\big)\cbigg]\normordbig|\,p_\ell\!} \braket[\rbig]{\!\:p_\ell\!\!\:|s_{\ell-1}}\nonumber \\[0.05cm]
	&= \frac{1}{\gamma'(s_{\ell-1})}\mathlarger{\mathlarger{\int}}_{-\infty}^{\infty}\frac{\mathrm{d}p_\ell}{2\pi\hbar} \: \exp\!\!\:\cbigg[-\:\!\frac{i e^{-i\theta} T}{N\hbar}\, H_\gamma\big(s_\ell,p_\ell\big)\cbigg] \braket[\rbig]{s_\ell\:\!\!|p_\ell}\:\! \braket[\rbig]{\!\:p_\ell\!\!\:|s_{\ell-1}} \nonumber \\[0.05cm]
	&=\frac{1}{\gamma'(s_{\ell-1})} \exp\cbigg\{-\:\!\frac{i e^{-i\theta} T}{N\hbar}\,V\!\!\:\big[\gamma(s_\ell)\big]\!\!\:\cbigg\}\nonumber \\[0.1cm]
    &\quad \times \!\mathlarger{\mathlarger{\int}}_{-\infty}^{\infty}\frac{\mathrm{d}p_\ell}{2\pi\hbar} \:\exp\!\Bigg\{\frac{ip_\ell \big(s_\ell-s_{\ell-1}\big)}{\hbar}-\frac{i e^{-i\theta} T}{N\hbar}\cbigg[\frac{1}{2m}\frac{p_\ell^2}{\gamma'(s_\ell)^2}+\frac{i\hbar p_\ell}{2m}\frac{\gamma''(s_\ell)}{\gamma'(s_\ell)^3}\cbigg]\!\!\:\Bigg\} \nonumber \\
	&= \frac{1}{\gamma'(s_{\ell-1})} \cbigg[\frac{mN\gamma'(s_\ell)^2}{2\pi\hbar ie^{-i\theta}T}\cbigg]^{\!\!\:\frac{1}{2}}\: \label{eq:ShortTimePropagator}\\ 
    &\quad\times \exp\!\!\:\Bigg\{\frac{ie^{i\theta}mN\gamma'(s_\ell)^2}{2\hbar T}\cbigg[s_{\ell}-s_{\ell-1}-\frac{ie^{-i\theta}T\hbar\gamma''(s_\ell)}{2mN\gamma'(s_\ell)^3}\cbigg]^{\!\!\:2}-\,\frac{i e^{-i\theta} T}{N\hbar}\,V\!\!\:\big[\gamma(s_\ell)\big]\!\!\:\Bigg\} \, . 
    \nonumber
\end{align} 
In the first line, we denoted normal ordering of the operator with colons. Note that convergence of the Gaussian integral in question requires
\begin{align}
	\mathrm{Re}\bigg[\frac{ie^{-i\theta}}{\gamma'(s_\ell)^2}\bigg]> 0 \qquad \Longrightarrow \qquad \mathrm{arg}\cbig[\gamma'(s_\ell)^2\cbig]\:\!\in \!\!\:\cbig(-\!\:\theta,\pi-\theta\cbig)\, ,
    \label{eq:Restriction_Contour}
\end{align} 
which retrospectively constrains the admissible choices of the contour $\Gamma$ by restricting its slope at each point. The illustrated example contour depicted in figure~\ref{fig:CustomContour_Gamma} would certainly violate this condition due to the entailed loop. We will comment on the implications of this restriction in appendix~\ref{sec:C_Restrictions_Contour}, arguing that in the relevant cases one can indeed always find a representative $\Gamma$ to enforce the necessary condition~\eqref{eq:Restriction_Contour} to be met. For the time being, let us assume $\Gamma$ has been chosen accordingly for all integrals to properly converge. The provided short-time propagator~\eqref{eq:ShortTimePropagator} in its current form is ill-suited for taking the continuum limit $N\to\infty$, as a na\"{\i}ve limiting procedure would arrive at erroneous results. Thus, let us first massage the given expression~\eqref{eq:ShortTimePropagator} by utilizing two techniques that were also heavily employed in appendix~\ref{sec:A_PathIntegralTechniques} to justify the na\"{\i}ve evaluation of the matrix elements entailed in the discretized propagator~\eqref{eq:DiscretizedPI_Start}: 
\begin{itemize}
    \item For all short-time propagators~\eqref{eq:ShortTimePropagator} entering expression~\eqref{eq:DiscretizedPI_Start}, each appearance of $\big(\Delta s_\ell\big)^2$ scales like a contribution of order $\Delta t$, where we introduced the abbreviations $\Delta s_\ell\coloneqq s_\ell-s_{\ell-1}$ and $\Delta t\coloneqq T/N$. This allows us to conveniently expand the exponential, neglecting terms that would yield overall contributions of order $\Delta t^{3/2}$ and higher, as those would vanish when subsequently taking the continuum limit $N\to\infty$ of the discretized path integral. 
    \item So-called \emph{substitution rules} allow us to replace certain expressions in each infinitesimal propagator by alternative terms while leaving the full path integral expression~\eqref{eq:DiscretizedPI_Start} invariant, as the arising differences only enter higher-order terms that vanish when taking the continuum limit~\cite{LangoucheFunctionalIntegration,dePireyPathIntegralMethods}. While these substitution rules nontrivially modify the short-time propagators, the integrations over the intermediate positions $s_k$ effectively cancel the arising differences to the desired order. In the present case, the required substitution rules take the form 
    \begin{subequations}
        \begin{align}
        \big(\Delta s_\ell\big)^{\!\!\: 2} &\doteq 2ie^{-i\theta}\hbar \cbig[2m\gamma'(s_\ell)^2\cbig]^{-1} \:\! \Delta t\, , \label{eq:SubstitutionRules_ContourCase1} \\
        \Delta t^{-1}\big(\Delta s_\ell\big)^{\!\!\: 3} &\doteq 6ie^{-i\theta} \hbar\cbig[2m\gamma'(s_\ell)^2\cbig]^{-1} \:\! \Delta s_\ell \, , \label{eq:SubstitutionRules_ContourCase2} \\
        \Delta t^{-2}\big(\Delta s_\ell\big)^{\!\!\: 4} &\doteq -12e^{-2i\theta}\hbar^2 \cbig[2m\gamma'(s_\ell)^2\cbig]^{-2} \:\! \Delta t\, , \label{eq:SubstitutionRules_ContourCase3} \\
        \Delta t^{-3}\big(\Delta s_\ell\big)^{\!\!\: 6}&\doteq -120ie^{-3i\theta} \hbar^3 \cbig[2m\gamma'(s_\ell)^2\cbig]^{-3} \:\! \Delta t\, , \label{eq:SubstitutionRules_ContourCase4}
    \end{align}
    \label{eq:SubstitutionRules_ContourCase}%
    \end{subequations}
    where we reinstated the appropriate powers of $\hbar$ when compared to expression~\eqref{eq:SubstitutionRules}. Their derivation for the most general Hamiltonian~\eqref{eq:GeneralQuadraticHamiltonian} quadratic in the momentum $\hat{p}$ can be found in appendix~\ref{sec:SubstitutionRules}. As before, $\doteq$ indicates equivalence in the path integral sense, stating that both sides of the equality yield identical results when appropriately applied to the infinite-dimensional integral~\eqref{eq:DiscretizedPI_Start}.\footnote{Note that these substitution rules need to be instated solely ``outside'' of exponentials, as otherwise the odd-powered substitution rule~\eqref{eq:SubstitutionRules_ContourCase2} inherits additional contributions.}
\end{itemize}
Strategically expanding some of the undesirable terms, the infinitesimal propagator~\eqref{eq:ShortTimePropagator} can be expressed as
\begin{align}
    \scalebox{0.97}{$\displaystyle{\!\!\!K_{\gamma,\theta}\big(s_{\ell-1},s_{\ell}\:\!;\!\!\;\Delta t\big) \!\!\:\doteq\!\!\: \frac{\gamma'(s_\ell)}{\gamma'(s_{\ell-1})} \bigg(\frac{2\pi\hbar ie^{-i\theta}\!\!\;\Delta t}{m}\bigg)^{\!\!-\frac{1}{2}} \exp\!\!\:\Bigg\{\!\!\;\frac{ie^{i\theta}m}{2\hbar \Delta t}\:\!\cbig[\gamma'(s_\ell) \Delta s_\ell\cbig]^2-\:\!\frac{ie^{-i\theta}\!\!\;\Delta t}{\hbar}\:\!V\!\!\:\big[\gamma(s_\ell)\big]\!\!\;\Bigg\} \phantom{\, ,}}$}\nonumber \\ 
    \scalebox{0.97}{$\displaystyle{\times \Bigg\{1+\frac{\gamma''(s_\ell) \Delta s_\ell}{2\gamma'(s_\ell)}+\frac{\gamma''(s_\ell)^2 \big(\Delta s_\ell\big)^2}{8\gamma'(s_\ell)^2}-\frac{ie^{-i\theta}\hbar\gamma''(s_\ell)^2 \Delta t}{8m\gamma'(s_\ell)^4}\Bigg\}\:\! ,}$} 
    \label{eq:ShortTimeProp1}
\end{align}
where, as introduced previously, we set $\Delta s_\ell=s_\ell-s_{\ell-1}$ and $\Delta t=T/N$ to make contact with the notation employed in appendix~\ref{sec:SubstitutionRules}. At last, we na\"{\i}vely factored the complex square root prefactor  while setting $\smash{\big[\gamma'(s_\ell)^2\big]^{1/2}=\gamma'(s_\ell)}$.\footnote{Let us ignore additional factors of $-1$ that emerge in case the contour runs ``backwards'', i.e. $\mathrm{Re}\big[\gamma'(s_\ell)\big]<0$. While for rather exotic eigenvalue problems, for which both relevant Stokes sectors $S_\pm$ are entirely contained in the same half-plane defined by the imaginary axis, such issues cannot be avoided, in the sought-after cases such pathological situations do not occur. In the special cases for which these considerations become relevant, one could argue that the additional factors can be absorbed in the path integral measure, eventually dropping out when computing ratios.} Using the substitution rule~\eqref{eq:SubstitutionRules_ContourCase1}, one finds that the two $\gamma''(s_\ell)^2$ contributions entering equation~\eqref{eq:ShortTimeProp1} fully cancel, thus they are dropped in all subsequent expressions. Expanding the fraction $\gamma'(s_\ell)/\gamma'(s_{\ell-1})$ to order $\smash{\big(\Delta s_\ell\big)^{\!\!\:2}}$ then grants the intermediate result 
\begin{align}
    K_{\gamma,\theta}\big(s_{\ell-1},s_{\ell}\:\!;\Delta t\big) \doteq\bigg(\frac{2\pi\hbar ie^{-i\theta}\Delta t}{m}\bigg)^{\!\!-\frac{1}{2}} \:\! \exp\!\!\:\Bigg\{\!\!\;\frac{ie^{i\theta}m}{2\hbar \Delta t}\:\!\cbig[\gamma'(s_\ell) \Delta s_\ell\cbig]^{\!\!\;2}-\:\!\frac{ie^{-i\theta}\Delta t}{\hbar}\:\!V\!\!\:\big[\gamma(s_\ell)\big]\!\!\;\Bigg\} \phantom{\, ,}\nonumber \\ 
    \times \Bigg\{1+\frac{3\gamma''(s_\ell)}{2\gamma'(s_\ell)}\:\! \Delta s_\ell+\frac{3\gamma''(s_\ell)^2-\gamma'(s_\ell)\gamma'''(s_\ell)}{2\gamma'(s_\ell)^2}\:\!\big(\Delta s_\ell\big)^{\!\!\:2}\Bigg\}\, . 
    \label{eq:ShortTimeProp2}
\end{align}
At this stage, it would not be too obvious how to manipulate the expression to reduce it to its eventual canonical form. Let us therefore present the final result directly, which ultimately takes the form
\begin{align}
    \scalebox{0.99}{$\displaystyle{\!\!\!K_{\gamma,\theta}\big(s_{\ell-1},s_{\ell}\:\!;\Delta t\big) \doteq\bigg(\frac{2\pi\hbar ie^{-i\theta}\!\!\;\Delta t}{m}\bigg)^{\!\!-\frac{1}{2}} \:\! \exp\!\!\:\Bigg\{\!\!\;\frac{ie^{i\theta}m}{2\hbar \Delta t}\:\!\cbig[\gamma(s_\ell)-\gamma(s_{\ell-1})\cbig]^{\!\!\;2}-\:\!\frac{ie^{-i\theta}\!\!\;\Delta t}{\hbar}\:\!V\!\!\:\big[\gamma(s_\ell)\big]\!\!\;\Bigg\} \, .}$} 
    \label{eq:ShortTimeProp_Final}
\end{align}
We will now show retrospectively that the right-hand side of equation~\eqref{eq:ShortTimeProp_Final} indeed matches the previous result~\eqref{eq:ShortTimeProp2} after instating the appropriate substitution rules 
\eqref{eq:SubstitutionRules_ContourCase}. At first, one utilizes the expansion
\begin{align}
    \cbig[\gamma(s_\ell)-\gamma(s_{\ell-1})\cbig]^{\!\!\;2}=\cbig[\gamma'(s_\ell)\Delta s_\ell\cbig]^{\!\!\;2}&-\,\gamma'(s_\ell)\gamma''(s_\ell)\big(\Delta s_\ell\big)^3 \nonumber \\[0.1cm] 
    &+\frac{3\gamma''(s_\ell)^2+4\gamma'(s_\ell)\gamma'''(s_\ell)}{12}\,\big(\Delta s_\ell\big)^4+\mathcal{O}\cbig[\big(\Delta s_\ell\big)^4\cbig]\, ,
    \label{eq:ExpansionGammaPrimeSquared}
\end{align}
with which the required equality (in the path integral sense) takes the form 
\begin{align}
    &\!\!\!\,\exp\!\!\:\Bigg\{\!\!\;\frac{ie^{i\theta}m}{2\hbar \Delta t}\:\!\cbig[\gamma'(s_\ell)-\gamma'(s_{\ell-1})\cbig]^{\!\!\;2}\Bigg\} \nonumber \\[0.05cm]
    &\qquad\doteq \Bigg\{1-\frac{ie^{i\theta}m}{2\hbar \Delta t}\,\gamma'(s_\ell)\gamma''(s_\ell)\big(\Delta s_\ell\big)^3-\frac{e^{2i\theta}m^2}{8\hbar^2 \Delta t^2}\,\gamma'(s_\ell)^2\gamma''(s_\ell)^2\big(\Delta s_\ell\big)^6 \nonumber \\[-0.1cm]
    &\qquad\phantom{\doteq \Bigg\{1\;} +\frac{ie^{i\theta}m}{2\hbar \Delta t}\frac{3\gamma''(s_\ell)^2+4\gamma'(s_\ell)\gamma'''(s_\ell)}{12}\,\big(\Delta s_\ell\big)^4\Bigg\}\exp\!\!\:\Bigg\{\!\!\;\frac{ie^{i\theta}m}{2\hbar \Delta t}\:\!\cbig[\gamma'(s_\ell) \Delta s_\ell\cbig]^{\!\!\;2}\Bigg\} \nonumber \\[0.15cm]
    &\qquad\overset{?}{\doteq}\Bigg\{1+\frac{3\gamma''(s_\ell)}{2\gamma'(s_\ell)}\:\! \Delta s_\ell +\frac{3\gamma''(s_\ell)^{2}-\gamma'(s_\ell)\gamma'''(s_\ell)}{2\gamma'(s_\ell)^2}\:\!\big(\Delta s_\ell\big)^{\!\!\:2}\Bigg\} \exp\!\!\:\Bigg\{\!\!\;\frac{ie^{i\theta}m}{2\hbar \Delta t}\:\!\cbig[\gamma'(s_\ell) \Delta s_\ell\cbig]^{\!\!\;2}\Bigg\} \, . 
    \label{eq:ExponentialComparison}
\end{align}
At this point the substitution rules~\eqref{eq:SubstitutionRules_ContourCase} enter the game again. Combining~\eqref{eq:SubstitutionRules_ContourCase3} and~\eqref{eq:SubstitutionRules_ContourCase4} with~\eqref{eq:SubstitutionRules_ContourCase1}, we can infer the replacements
\begin{align}
    \Delta t^{-2}\big(\Delta s_\ell\big)^{\!\!\: 4} &\doteq \frac{3ie^{-i\theta}\hbar}{m\gamma'(s_\ell)^2} \:\! \big(\Delta s_\ell\big)^{\!\!\: 2}\, , &
    \Delta t^{-3}\big(\Delta s_\ell\big)^{\!\!\: 6}&\doteq -\frac{15e^{-2i\theta}\hbar^2}{m^2\gamma'(s_\ell)^4}\:\!\big(\Delta s_\ell\big)^{\!\!\: 2} \, .
    \label{eq:SubstitutionRules_ContourCase_Combined}
\end{align}
Together with the default substitution rule~\eqref{eq:SubstitutionRules_ContourCase2}, one indeed finds that the last equality of expression~\eqref{eq:ExponentialComparison} holds, concluding our (first) derivation of the short-time propagator~\eqref{eq:ShortTimeProp_Final}.\footnote{Expansion~\eqref{eq:ExpansionGammaPrimeSquared} has to be instated with great care, as the substitution rules~\eqref{eq:SubstitutionRules_ContourCase} only apply in case the exponent of the exponential piece multiplying the terms one wants to substitute away is precisely given by $i e^{i\theta}m \gamma'(s)^2\Delta s^2/(2\hbar \Delta t)$. In case one would argue the other way round, expanding $\gamma'(s)\Delta s$ around $\gamma(s)-\gamma(s-\Delta s)$, the new exponent $\smash{i e^{i\theta}m \big[\gamma(s)-\gamma(s-\Delta s)\big]^2/(2\hbar \Delta t)}$ would lead to altered substitution rules. A na\"{\i}ve application of the former ones~\eqref{eq:SubstitutionRules_ContourCase} then fails to recover the correct result.}

\subsubsection{$2^{\text{nd}}$ option: Starting directly with the ``correct'' short-time propagator}

A slightly simpler way of attaining the identical result~\eqref{eq:ShortTimeProp_Final} emerges when we utilize the complete expression~\eqref{eq:SchematicForm_InfPropagator} for the infinitesimal propagator obtained in appendix~\ref{sec:RigorousEstimate}, allowing us to bypass the use of substitution rules altogether. Inserting the coefficient functions entering the non-standard Hamiltonian~\eqref{eq:NonStandard_Hamiltonian} into equations~\eqref{eq:SchematicForm_InfPropagator},~\eqref{eq:CoefficientFunctionsComplete1} and~\eqref{eq:CoefficientFunctionsComplete2}, the infinitesimal propagator takes the form 
\begin{align}
    \scalebox{0.94}{$\displaystyle{K_{\gamma,\theta}\big(s_{\ell-1},s_\ell\:\!;\!\!\;\Delta t\big)}$}\,&\scalebox{0.94}{$\displaystyle{\doteq\!\!\; \frac{1}{\gamma'(s_{\ell-1})}\cbigg[\frac{m\gamma'(s_\ell)^2}{2\pi\hbar ie^{-i\theta} \Delta t}\cbigg]^{\!\!\:\frac{1}{2}} \, \exp\!\!\:\Bigg\{\!\!\;\frac{ie^{i\theta}m}{2\hbar \Delta t}\:\!\cbig[\gamma'(s_\ell) \Delta s_\ell\cbig]^2-\:\!\frac{ie^{-i\theta}\Delta t}{\hbar}\:\!V\!\!\:\big[\gamma(s_\ell)\big]\!\!\;\Bigg\}}$} \nonumber \\ 
    &\scalebox{0.94}{$\displaystyle{\quad\; \times \Bigg\{1-\frac{\gamma''(s_\ell)}{\gamma'(s_\ell)}\,\Delta s_\ell+\frac{\gamma'''(s_\ell)}{2\gamma'(s_\ell)}\,\big(\Delta s_\ell\big)^{\!\!\:2}-\frac{ie^{i\theta}m}{2\hbar \Delta t}\,\gamma'(s_\ell)\gamma''(s_\ell)\big(\Delta s_\ell\big)^{\!\!\:3} }$}\label{eq:ShortTimePropContour} \\[-0.05cm] 
    &\scalebox{0.94}{$\displaystyle{\quad\; \phantom{\times\Bigg\{1} + \frac{ie^{i\theta}m}{2\hbar \Delta t}\frac{15\gamma''(s_\ell)^2+4\gamma'(s_\ell)\gamma'''(s_\ell)}{12}\,\big(\Delta s_\ell\big)^{\!\!\:4}-\frac{e^{2i\theta}m^2}{8\hbar^2\Delta t^2} \,\gamma'(s_\ell)^2\gamma''(s_\ell)^2\big(\Delta s_\ell\big)^{\!\!\:6}\Bigg\}\:\! ,}$} \nonumber
\end{align}
where for convenience we already exponentiated the $F_{1,0}$ term, solely entailing the potential contribution; all other terms entering $F_{1,0}$ fully cancel. Once more, mind the additional factor $\gamma(s_{\ell-1})^{-1}$ in definition~\eqref{eq:Definition_Propagator} of the propagator on the contour $\Gamma$, which is absent in the derivation provided in appendix~\ref{sec:RigorousEstimate}. Notice that we can factor the $\smash{\big(\Delta s_\ell\big)^j}$ terms into $j=1,2$ and $j=3,4,6$ contributions (beware of the cross-term), yielding 
\begin{align}
    K_{\gamma,\theta}\big(s_{\ell-1},s_\ell\:\!;\!\!\;\Delta t\big)&\doteq \frac{\gamma'(s_{\ell})}{\gamma'(s_{\ell-1})}\bigg(\frac{2\pi\hbar ie^{-i\theta}\!\!\;\Delta t}{m}\bigg)^{\!\!-\frac{1}{2}} \, \exp\!\!\:\Bigg\{\!\!\;\frac{ie^{i\theta}m}{2\hbar \Delta t}\:\!\cbig[\gamma'(s_\ell) \Delta s_\ell\cbig]^{\!\!\;2}-\:\!\frac{ie^{-i\theta}\Delta t}{\hbar}\:\!V\!\!\:\big[\gamma(s_\ell)\big]\!\!\;\Bigg\} \nonumber \\ 
    &\quad\; \times \Bigg\{1-\frac{\gamma''(s_\ell)}{\gamma'(s_\ell)}\,\Delta s_\ell+\frac{\gamma'''(s_\ell)}{2\gamma'(s_\ell)}\,\big(\Delta s_\ell\big)^{\!\!\:2}\Bigg\} \Bigg\{1-\frac{ie^{i\theta}m}{2\hbar \Delta t}\,\gamma'(s_\ell)\gamma''(s_\ell)\big(\Delta s_\ell\big)^3 \nonumber \\[-0.1cm]
    &\quad\; -\frac{e^{2i\theta}m^2}{8\hbar^2 \Delta t^2}\,\gamma'(s_\ell)^2\gamma''(s_\ell)^2\big(\Delta s_\ell\big)^6 +\frac{ie^{i\theta}m}{2\hbar \Delta t}\frac{3\gamma''(s_\ell)^2+4\gamma'(s_\ell)\gamma'''(s_\ell)}{12}\,\big(\Delta s_\ell\big)^4\Bigg\}\:\! ,
    \label{eq:ShortTimePropContourFull}
\end{align}
where we once more na\"{\i}vely applied $\smash{\big[\gamma'(s_\ell)^2\big]^{1/2}=\gamma'(s_\ell)}$. One can now utilize the relation 
\begin{align}
    \frac{\gamma'(s_{\ell})}{\gamma'(s_{\ell-1})}\:\!\Bigg\{1-\frac{\gamma''(s_\ell)}{\gamma'(s_\ell)}\,\Delta s_\ell+\frac{\gamma'''(s_\ell)}{2\gamma'(s_\ell)}\,\big(\Delta s_\ell\big)^{\!\!\:2}\Bigg\} = 1+\mathcal{O}\cbig[\big(\Delta s_\ell\big)^{\!\!\;3}\:\!\cbig]\, ,
\end{align}
together with the previous expansion~\eqref{eq:ExpansionGammaPrimeSquared}, which in its exponential form was provided in equation~\eqref{eq:ExponentialComparison}. Together, one once more arrives at relation~\eqref{eq:ShortTimeProp_Final}, constituting the reduced final result.

\subsubsection{Continuum limit \& further obstructions to the functional integral}

With the short-time propagator given in its canonical form~\eqref{eq:ShortTimeProp_Final}, we can proceed by taking the continuum limit $N\to\infty$. Inserting the result into the previous relation~\eqref{eq:DiscretizedPI_Start} yields the discretized path integral 
\begin{align}
	\!\!K_{\gamma,\theta}\big(s_\mathrm{i},s_\mathrm{f}\:\!;T\big) &= \lim_{N \to \infty}  \bigg(\frac{2\pi\hbar ie^{-i\theta}\!\!\;\Delta t}{m}\bigg)^{\!\!-\frac{N}{2}} \!\mathlarger{\mathlarger{\int}}_{-\infty}^{\infty} \, \mathlarger{\prod}_{k=1}^{N-1} \:\!\cbig[\gamma'(s_{k})\:\!\mathrm{d}s_{k}\cbig] \nonumber \\ 
    &\qquad\qquad \times \exp\cBigg(\:\!\mathlarger{\sum}_{\ell=1}^N\,\cbigg\{\!\!\;\frac{ie^{i\theta}m}{2\hbar \Delta t}\:\!\cbig[\gamma(s_\ell)-\gamma(s_{\ell-1})\cbig]^{\!\!\;2}-\:\!\frac{ie^{-i\theta}\!\!\;\Delta t}{\hbar}\:\!V\!\!\:\big[\gamma(s_\ell)\big]\!\!\;\cbigg\}\!\!\:\cBigg)_{\!\substack{\,\scalebox{0.78}{$\Delta t\:\!\!=\:\!\! T/N$}\\[0.025cm] \,\scalebox{0.8}{$s_0\:\!\!=\:\!\!s_\mathrm{i}$}\;\;\;\, \\[0.025cm] \scalebox{0.8}{$s_{\!\!\:N}\:\!\!=\:\!\!s_\mathrm{f}$}\;\;\;\,}} \nonumber \\[0.05cm] 
    &= \lim_{N \to \infty}  \bigg(\frac{2\pi\hbar ie^{-i\theta}\!\!\;\Delta t}{m}\bigg)^{\!\!-\frac{N}{2}} \!\mathlarger{\mathlarger{\int}}_{\Gamma} \; \mathlarger{\prod}_{k=1}^{N-1} \,\mathrm{d}z_k \nonumber \\
    &\qquad\qquad \times \exp\cBigg(\frac{ie^{i\theta}}{\hbar}\,\mathlarger{\sum}_{\ell=1}^N \;\Delta t\, \cbigg\{\!\!\;\frac{m}{2}\:\!\bigg(\frac{z_\ell-z_{\ell-1}}{\Delta t}\bigg)^{\!\!2}-e^{-2i\theta}\:\!V(z_\ell)\!\!\;\cbigg\}\!\!\:\cBigg)_{\substack{\!\!\!\!\!\!\!\!\!\:\scalebox{0.78}{$\Delta t\:\!\!=\:\!\! T/N$}\\[0.025cm] \,\scalebox{0.8}{$z_0\:\!\!=\:\!\!\gamma(s_\mathrm{i})$}\;\;\;\, \\ \scalebox{0.8}{$z_{\!\!\:N}\:\!\!=\:\!\!\gamma(s_\mathrm{f})$}\;\;\;\,}}\!\!\!\! .
	\label{eq:DiscretizedPI_End} 
\end{align}
The crucial step is to notice that all intermediate $s_k$-integrals could be expressed as complex contour integrals, ranging over $\Gamma$. We especially recover the feature that the expression for the propagator is independent of the chosen contour $\Gamma$ and its parametrization $\gamma$, depending solely on the endpoints $\gamma(s_{\mathrm{i,f}})$ in the complex plane. Recall that the identical observation was previously made for the spectral representation~\eqref{eq:Propagator_SpectralRepresentation}, constituting an important consistency check. Remarkably, the discretized expression~\eqref{eq:DiscretizedPI_End} coincides exactly with the ordinary result, with the sole exception of the intermediate position integrals ranging over $\Gamma$ instead of $\mathbb{R}$. This, in particular, enables us to take the continuum limit $N\to\infty$ straightforwardly, using the standard procedure of inferring the replacements 
\begin{align}
	\begin{split}
		z_\ell\:\!&\xrightarrow{\:\!N\to\infty\:\!}\:\! z(t) \, , \hspace{-1.5cm} \qquad\qquad\qquad\qquad\qquad\qquad\qquad\qquad \,\!\!\: \sum_{\ell=1}^{N}\,\Delta t\,\xrightarrow{\:\!N\to\infty\:\!} \int_{0}^{T} \text{d}t \, ,\\[0.15cm]
		\frac{z_\ell-z_{\ell-1}}{\Delta t} \:\!&\xrightarrow{\:\!N\to\infty\:\!} \:\! \dot{z}(t) \, , \hspace{-1.5cm}\qquad\qquad\qquad\qquad\!\!\!\: \bigg(\frac{2\pi\hbar ie^{-i\theta}\!\!\;\Delta t}{m}\bigg)^{\!\!-\frac{N}{2}}\, \prod_{k=1}^{N-1}\mathrm{d}z_k \:\!\xrightarrow{\:\!N\to\infty\:\!}\:\! \mathcal{D}_\theta\llbracket z\rrbracket \, .\qquad\;\;
	\end{split}
\end{align}
The formally $\theta$-dependent path integral measure is subsequently denoted by $\mathcal{D}_\theta\llbracket z\rrbracket$, where we chose to highlight functional relations by double square brackets. Gathering the results, we finally arrive at the desired path integral representation of the analytically continued propagator, given by 
\begin{align}
    K_{\gamma,\theta}\big(s_\mathrm{i},s_\mathrm{f}\:\!;T\big)=\mathlarger{\mathlarger{\int}}_{\scalebox{0.72}{$\mathcal{C}\big([0,T],\Gamma\big)$}}^{\substack{z(0)\:\!=\:\!\gamma(s_\mathrm{i})\\[0.05cm] z(T)\:\!=\:\!\gamma(s_\mathrm{f})\,}} \, \mathcal{D}_\theta\llbracket z\rrbracket\; \exp\Bigg\{\frac{ie^{i\theta}}{\hbar}\mathlarger{\int}_0^T\bigg[\frac{m}{2}\:\!\dot{z}(t)^2-e^{-2i\theta}\:\!V\!\!\;\scalebox{1.1}{\big(}z(t)\!\!\;\scalebox{1.1}{\big)}\bigg]\mathrm{d}t\Bigg\}\, .
    \label{eq:Propagator_PathIntegralRepresentation}
\end{align}
The functional integral ranges over all continuous functions mapping from the time interval $[0,T]$ into the target space $\Gamma$, for which we denote the underlying function space by $\mathcal{C}\big([0,T],\Gamma\big)$, subject to the standard Dirichlet boundary conditions at both endpoints. It is rather striking that, apart from the functional integration contour, there are virtually no changes to the path integral representation of the (partially) Wick-rotated propagator when comparing to a scenario for which the original eigenvalue problem~\eqref{eq:Schrödinger_EV_Problem} is defined on the real line. The information which Stokes wedges $S_\pm$ are chosen to define the underlying eigenvalue problem is entirely contained in the integration contour, with the action of the theory being independent of $\Gamma$.\\

\noindent 
In order for the path integral derivation to go through unobstructed, we especially require the individual $z_k$-integrals to be convergent, bestowing yet another constraint on the contour $\Gamma$ upon us. Once more assuming $V(z)$ to be a polynomial of order $n \geq 3$, the contour $\Gamma$ is required to asymptotically end in regions for which one has
\begin{align}
    \mathrm{Re}\cbig[ie^{-i\theta}\hbar^{-1} V(z)\cbig] \,\xrightarrow{\lvert z\rvert \to \infty}\, \mathrm{Re}\cbig[ie^{-i\theta} \hbar^{-1}c_n z^n\cbig] > 0\, .
\end{align}
This condition again provides us with wedge-shaped regions in the complex plane, given by 
\begin{align}
	\mathcal{S}_k\coloneqq\Bigg\{z\in\mathbb{C}: \cbigg\lvert\, \mathrm{arg}(z)-\frac{2\pi k}{n}+\frac{\mathrm{arg}\big(\hbar^{-1}c_n\big)}{n}+\frac{\pi-2\theta}{2n}\cbigg\lvert <\frac{\pi}{2n}\Bigg\}\, , \qquad \text{with }k\in \mathbb{Z}_{n}\, ,
    \label{eq:DefinitionStokesSectors_PI}
\end{align}
in which $\Gamma$ is required to terminate asymptotically. While there were $n+2$ Stokes sectors $S_k$, out of which one chooses $S_\pm$ defining the boundary value problem at hand, there are an additional $n$ narrower sectors $\mathcal{S}_k$ that further constrain the contour. Note that the sectors $\mathcal{S}_k$ are sufficiently abundant for there to always exist an overlap with both $S_+$ and $S_-$, guaranteeing $S_\pm \cap \big(\mathcal{S}_0 \cup \ldots \cup \mathcal{S}_n\big)$ to not be empty, enabling the path integral procedure for arbitrary polynomial potentials. Other implications on the contour $\Gamma$ are briefly discussed in appendix~\ref{sec:C_Restrictions_Contour}.

\subsection{Combining the different representations: Master formulas}
\label{sec:3_3_Equating_Representations}

We are now fit to collect our results. Given the Schrödinger-type eigenvalue problem~\eqref{eq:Schrödinger_EV_Problem} with a polynomial potential $V(z)$, with the holomorphic eigenfunctions $\Psi_\ell(z)$ demanded to decay exponentially in two no-adjacent Stokes sectors $S_\pm$, one arrives at the \myuline{exact} equality 
\begin{equation}
	\!\!\scalebox{0.98}{$\displaystyle{\mathlarger{\mathlarger{\int}}_{\scalebox{0.72}{$\mathcal{C}\big([0,T],\Gamma\big)$}}^{\substack{z(0)\:\!=\:\!z_\mathrm{i}\\[0.05cm] z(T)\:\!=\:\! z_\mathrm{f}\,}} \, \mathcal{D}_\theta\llbracket z\rrbracket\,\exp\!\!\:\bigg(\frac{iS_\theta\llbracket z\rrbracket}{\hbar}\bigg)\:\!\!=\mathlarger{\mathlarger{\sum}}_{\ell=0}^\infty \: \exp\!\bigg(\!\!-\!\!\:\frac{ie^{-i\theta}E_\ell T}{\hbar}\bigg) \:\! \Psi_\ell(z_\mathrm{i}) \, \Psi_\ell(z_\mathrm{f})\,\scalebox{1.1}{\bigg\{}\!\int_\Gamma \Psi_\ell(z)^2\,\mathrm{d}z\scalebox{1.1}{\bigg\}}^{\!\!\!\:-1}\! \, ,}$}
    \label{eq:Final_Relation_Propagator}
\end{equation}
holding true as long as both sides of the equation are well-defined. In the above relation, $\Gamma$ is an arbitrary complex contour encompassing $z_{\mathrm{i,f}}$, asymptotically ending in the two angular sectors $S_\pm \cap \big(\mathcal{S}_0 \cup \ldots \cup \mathcal{S}_n\big)$ of the complex plane. The portrayed expression~\eqref{eq:Final_Relation_Propagator} directly emerges from equating~\eqref{eq:Propagator_SpectralRepresentation} and~\eqref{eq:Propagator_PathIntegralRepresentation}, simplifying the appearances of $\gamma(s_{\mathrm{i,f}})$ by $z_{\mathrm{i,f}}$. We furthermore defined the analytically continued action functional 
\begin{align}
    S_\theta\llbracket z\rrbracket \coloneqq e^{i\theta}\mathlarger{\int}_0^T\bigg[\frac{m}{2}\:\!\dot{z}(t)^2-e^{-2i\theta}\:\!V\!\!\;\scalebox{1.1}{\big(}z(t)\!\!\;\scalebox{1.1}{\big)}\bigg]\mathrm{d}t\, .
\end{align}
In similar fashion, we can obtain the trace of the analytically continued time evolution operator $\widehat{U}_{\gamma,\theta}$ on $\Gamma$ by formally setting $z_\mathrm{i}=z_\mathrm{f}$ in the above relation~\eqref{eq:Final_Relation_Propagator} and subsequently integrating over $\Gamma$. This can be seen by inferring
\begin{align}
    Z_\theta(T) &\coloneqq  \mathrm{tr} \cbigg[\exp\!\bigg(\!\!-\!\frac{i e^{-i\theta} \widehat{H}_\gamma T}{\hbar}\bigg)\cbigg] \nonumber \\[0.1cm]
    &\phantom{:}= \mathlarger{\mathlarger{\int}}_{-\infty}^\infty \braket[\rbigg]{s | \,\exp\!\bigg(\!\!-\!\frac{i e^{-i\theta} \widehat{H}_\gamma T}{\hbar}\bigg)\;\!\! |\, s } \: \mathrm{d}s = \mathlarger{\mathlarger{\int}}_{-\infty}^\infty K_{\gamma,\theta}\big(s,s\:\!;T\big) \, \gamma'(s)\,\mathrm{d}s\, ,
\end{align}
then utilizing the fact that the propagator $K_{\gamma,\theta}\big(s,s;T\big)$ solely depends on $\gamma(s)$, thus the remaining endpoint integral can once more be represented as a contour integral. This leads to another significant relation of the form
\begin{equation}
	Z_\theta(T)=\mathlarger{\mathlarger{\int}}_{\scalebox{0.72}{$\mathcal{C}\big([0,T],\Gamma\big)$}}^{z(0)\:\!=\:\!z(T)} \, \mathcal{D}_\theta\llbracket z\rrbracket\,\exp\!\!\:\bigg(\frac{iS_\theta\llbracket z\rrbracket}{\hbar}\bigg)=\mathlarger{\mathlarger{\sum}}_{\ell=0}^\infty \: \exp\!\bigg(\!\!-\!\!\:\frac{ie^{-i\theta}E_\ell T}{\hbar}\bigg) \, .
    \label{eq:Final_Relation_Trace}
\end{equation}
This equality naturally generalizes the usual notion of the partition function, given by a path integral over periodic field configurations. As the propagator constitutes the primary building block for all other quantities commonly encountered in standard quantum mechanics, such as the resolvent, these can be derived in a similar manner utilizing the result~\eqref{eq:Final_Relation_Propagator}. Note that the previously encountered freedom of choosing the contour $\Gamma$ now resides in the ability to smoothly deform the half-dimensional integration contour $\mathcal{C}\big([0,T],\Gamma\big)$ within the infinite-dimensional, complexified function space $\mathcal{C}\big([0,T],\mathbb{C}\big)\eqqcolon \mathcal{C}^\mathbb{C}\big([0,T]\big)$, provided the deformation respects the prescribed regions of convergence at the contour boundaries. 

\section{Applying the ideas to quantum tunneling}
\label{sec:4_Quantum_Tunneling_Revisited}

With the key results of this work, encapsulated in equations~\eqref{eq:Final_Relation_Propagator} and~\eqref{eq:Final_Relation_Trace}, now firmly established, we return to the original motivation of investigating quantum tunneling. Our previous endeavors will directly culminate in the sought-after relation between the real-time picture of quantum-mechanical decay and the ordinary instanton method. To this end, let us briefly recall the usual heuristic argument brought forward when justifying the instanton method, which subsequently will be brought onto firm ground utilizing the previous results.

\subsection{Recap of the traditional argument: Potential-deformation method}
\label{sec:4_1_PotentialDeformationMethod}

\begin{figure}[H]
    \centering
    \includegraphics[width=0.9\textwidth]{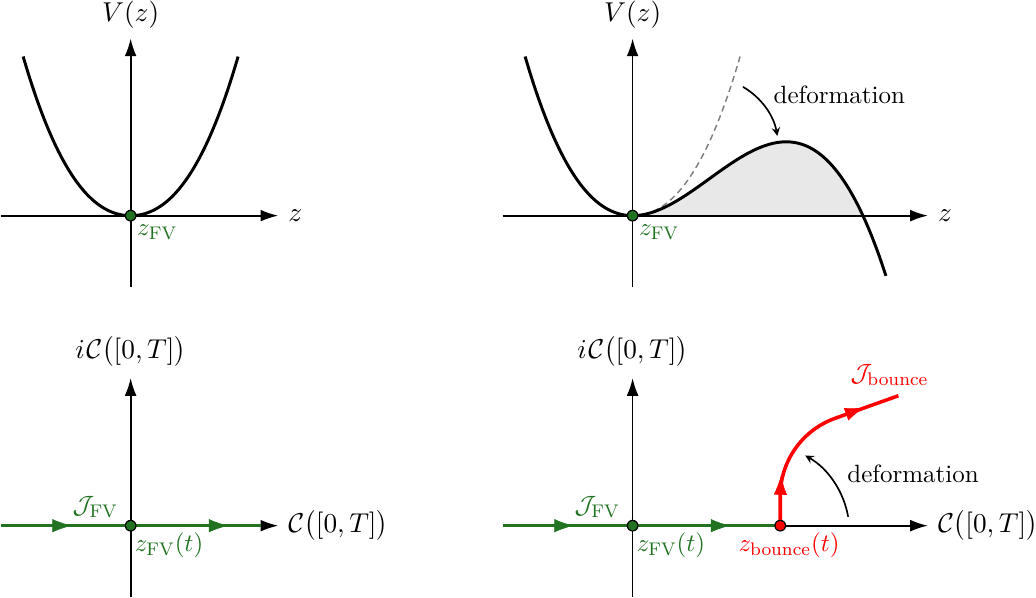}
    \caption{Schematic illustration of the potential-deformation approach by Callan \& Coleman~\cite{CallanColemanFateOfFalseVac2}. In case the false vacuum $z_\mathrm{FV}$ is stabilized, as shown in the left panel, the sole critical trajectory starting and ending at the (then global) minimum $z_\mathrm{FV}$ for $T\to\infty$ is the trivial FV trajectory $z_\mathrm{FV}(t)=z_\mathrm{FV}$. The associated steepest-descent thimble $\mathcal{J}_\mathrm{FV}$ in the complexified function space $\mathcal{C}^\mathbb{C}\big([0,T]\big)$ then simply coincides with the real function space $\mathcal{C}\big([0,T]\big)$. Once the potential is deformed to be unbounded, allowing for quantum tunneling to occur, one finds an additional saddle point in guise of the classical bounce trajectory $z_\mathrm{bounce}(t)$, satisfying both the equation of motion as well as the demanded Dirichlet boundary conditions. By heuristic arguments, Callan \& Coleman argue that the functional integration contour of the emerging path integral has to be distorted into the complex domain in order to render the resulting negative mode integration arising from the bounce saddle convergent, capturing the crucial factor of $1/2$ multiplying the imaginary bounce contribution in the process. This is schematically portrayed in the lower right panel, restricting our view to the single representative dimension along which the FV trajectory and the bounce are connected by the gradient flow.}
    \label{fig:ColemanDeformationArgument}
\end{figure}

\begin{figure}[H]
    \centering
    \includegraphics[width=\textwidth]{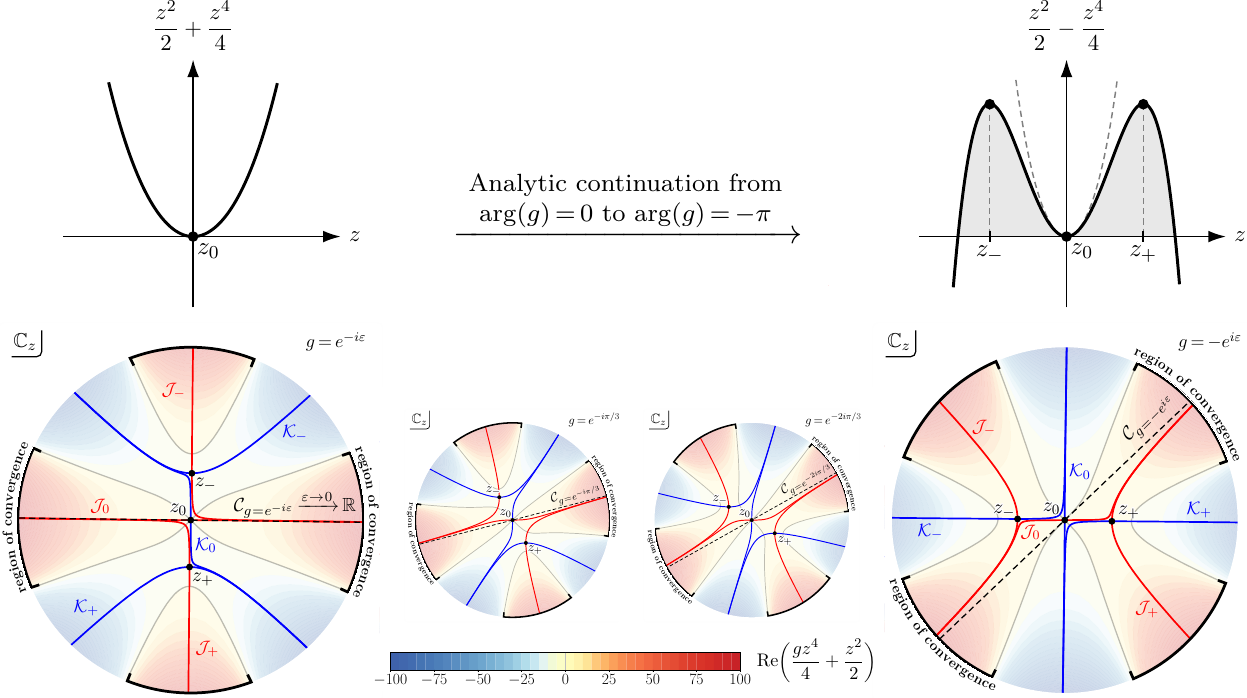}
    \caption{Illustration of the analytic continuation of the single-dimensional integral $\mathcal{Z}_\mathrm{E}(g)$ from $g=1$ to $g=-1$ along $\mathrm{arg}(g)<0$. The saddle points $z_{0,\pm}$ as well as their associated steepest ascent (blue, $\mathcal{K}_{0,\pm}$) and descent (red, $\mathcal{J}_{0,\pm}$) contours are shown for four different values of $\mathrm{arg}(g)$, together with the tilted integration contour $\mathcal{C}_g=e^{-i\,\mathrm{arg}(g)/4}\mathbb{R}$. We lift the degeneracy of the steepest ascent/descent contours in the case $g\in\mathbb{R}$ by instating an infinitesimal imaginary part to $g$. While for $\mathrm{arg}(g)=0$ the integration contour coincides with the real line, in the limit $\mathrm{arg}(g)\to -\pi$ the integral receives an imaginary contribution from the saddle points $z_\pm$, constituting the equivalent to bounces in the infinite-dimensional setting. Following Callan \& Coleman, this identification comprises the explanation of the contour deformation shown in figure~\ref{fig:ColemanDeformationArgument}.}
    \label{fig:AnalyticCont_ToyModel}
\end{figure}

As originally introduced by Callan \& Coleman in their seminal work on the instanton method~\cite{CallanColemanFateOfFalseVac2}, the potential-deformation argument has been the staple explanation of the functional treatment until the introduction of the direct method by Andreassen et al.~\cite{SchwartzDirectMethod}, which will be addressed in section~\ref{sec:5_1_Flaws_DirectMethod}. Following the traditional argument, the key quantity under consideration is the Euclidean FV-to-FV transition amplitude for late times, from which Callan \& Coleman formally extract the ground-state decay rate by projecting onto the ground-state energy via
\begin{equation}
	\Gamma_0 = 2 \lim_{T \rightarrow \infty} \mathrm{Im}\bigg\{T^{-1} \log\!\Big[K_\mathrm{E}\scalebox{1.15}{$\big($}z_\mathrm{FV},z_\mathrm{FV}\:\!;T\scalebox{1.15}{$\big)$}\Big] \!\bigg\} \, .
	\label{eq:Ground_State_DecayRate_Coleman}
\end{equation}
In all subsequent chapters, we adopt the standard convention of denoting the Euclidean propagator by $K_\mathrm{E}\eqqcolon K_{\pi/2}$, while the real-time propagator is expressed without an index, i.e. $K\eqqcolon K_0$. Working with an unbounded potential as shown in the right panel of figure~\ref{fig:ColemanDeformationArgument}, Callan \& Coleman argue that the deformation of a globally stable potential into an unstable one results in a distortion of the functional integration contour past the emerging nontrivial bounce saddle point $z_\mathrm{bounce}(t)$, shown schematically in the lower right panel of figure~\ref{fig:ColemanDeformationArgument}. In this manner, the propagator acquires an imaginary part due to traversing half of the steepest-descent thimble $J_\mathrm{bounce}$, emanating from $z_\mathrm{bounce}(t)$, resulting in the desired decay rate when invoking the appropriate weight factor of $1/2$ multiplying the bounce contribution.
\noindent
The illustrated deformation argument is regularly accompanied by a single-dimensional toy example exhibiting the conjectured behavior of the path integral analog, usually in the form of the simple quartic integral
\begin{align}
    \mathcal{Z}_\mathrm{E}(g)\coloneqq \mathlarger{\mathlarger{\int}}_{\scalebox{0.85}{$\mathcal{C}_g$}} \exp\cbigg\{-\:\!\frac{1}{\hbar}\,\bigg(\frac{gz^4}{4}+\frac{z^2}{2}\bigg)\!\!\;\cbigg\}\, \mathrm{d}z \, , \qquad \text{with } \;  \mathcal{C}_{\scalebox{0.75}{$g\!\!\:>\!\!\:0$}}=\mathbb{R}\, ,
\end{align}
see e.g. the lucid treatments~\cite{SchulmanPathIntegrals,MarinoAdvancedQM,SchwartzPrecisionDecayRate}. Similarly to our previous discussion of the analytic continuation of eigenvalue problems as introduced in section~\ref{sec:2_2_AnalyticContinuation_EigenvalueProblems}, the analytic continuation of the ``partition function'' $\mathcal{Z}_\mathrm{E}(g)$ from positive to negative values of $g$ is accompanied by a rotation of the integration contour $\mathcal{C}_g$ in the complexified $z$-plane, see figure~\ref{fig:AnalyticCont_ToyModel}. Performing a steepest-descent decomposition of the integration contour for positive and negative values of the coupling constant $g$ subsequently reveals an identical distortion to that conjectured by Callan \& Coleman for the path integral case. Whereas for positive values of the coupling $g$, the steepest-descent contour $\mathcal{J}_0$ from the (global) minimum $z_0$ coincides with the real axis, for negative $g$ the contour $\mathcal{C}_g\equiv\mathcal{J}_0$ picks up contributions from all three saddles $z_{0}$ and $z_\pm$, as the system is ``on a Stokes line''.\footnote{Formally meaning that the imaginary part of the exponent evaluated at multiple saddle points coincides, resulting in a degeneracy/merging of several steepest descent/ascent curves. In other words, the gradient flow connects at least one pair of critical points of the exponent.} The two additional saddle points $z_\pm$, constituting local minima of the full exponent when viewing the situation on the real line, mimic the role of the bounce in the previous distortion argument illustrated in figure~\ref{fig:ColemanDeformationArgument}.
\noindent 
Even though the prescribed procedure is successful in reproducing the correct result for the ground-state decay rate $\Gamma_0$ given arbitrary potentials, the heuristic contour-deformation argument possesses several unresolved issues which require further understanding to close the immanent gaps.

\subsection{Pending questions regarding the conventional explanation}
\label{sec:4_2_OpenQuestions_ColemanApproach}

While we will ultimately find the previously introduced computation scheme to be entirely correct, the portrayed argument is yet incomplete, leaving several conceptual questions unaddressed:
\begin{enumerate}[label= \roman*)]
    \item Why exactly does the functional integration contour have to be deformed as shown in figure~\ref{fig:ColemanDeformationArgument}? While the single-dimensional toy example mimics the desired behavior, it is yet to be shown that the infinite-dimensional functional case follows suit. Callan \& Coleman moreover acknowledge that there is the ambiguity to deform the potential in the conjugate manner, leading to a deformation of the integration contour into the lower complex half-plane. Consequently, the imaginary part of the propagator acquires the opposite sign, which in turn affects the sign of the projected ground-state energy. How is the prescription of the sign, encapsulating the physical scenario of the probability current either leaving or entering the FV region, imprinted in the deformation process?
    \item How is the illustrated procedure and the emerging complex ground-state energy related to the notion of resonant states utilized for practically all other decay rate calculations? Despite matching results, a direct connection between e.g. WKB computations of the decay rate and the functional instanton method has remained obscure. Does the equivalence between results based on solutions to the Schrödinger equation and the instanton method hold to all orders, even nonperturbatively, or do these approaches only agree up to some fixed order?
    \item How does one treat the case for which the tunneling potential takes the form as illustrated in the left panel of figure~\ref{fig:GenericMetastablePot+TimeEvolution}, i.e. when the potential is not globally unstable, but rather possesses a true vacuum state. In that case, one necessarily encounters an additional saddle point dubbed the ``shot''~\cite{SchwartzPrecisionDecayRate, PatrascioiuComplexTime}, see figure~\ref{fig:DifferentClassicalTrajectories}, resulting in the projection onto the global ground state instead of the metastable one when studying the late time behavior of the Euclidean propagator. Only once the steepest-descent thimble associated to the shot saddle is dismissed does one end up with the correct result. How does this prescription of dropping such saddle points arise from first principles?
    \item What makes the false vacuum point $z_\mathrm{FV}$ special when it serves as both the initial and final point of the Euclidean propagator? How does the contour deformation argument change in case one chooses different endpoints for the propagator when projecting onto the ground state? 
\end{enumerate}
All of these questions can be effectively addressed by invoking the previous developments into path integrals associated with generalized eigenvalue problems in the complex plane, along with the fact that the desired resonant energies are directly linked to such nontrivial boundary value problems as seen in section~\ref{sec:2_2_AnalyticContinuation_EigenvalueProblems}.

\subsection{Shedding light on Callan \& Coleman's functional contour prescription}
\label{sec:4_3_FormalizingColemansIdeas}

\begin{figure}[H]
    \centering
    \includegraphics[width=0.98\textwidth]{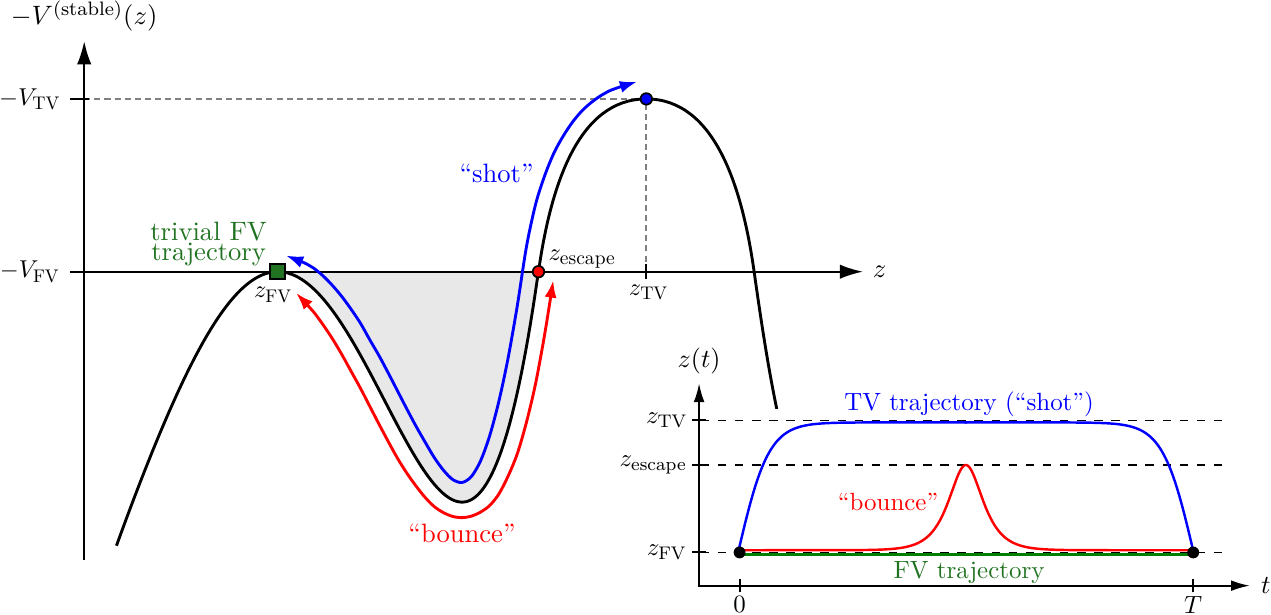}
    \caption{Depiction of the relevant (real) classical trajectories emerging when investigating the Euclidean propagator $\smash{K_{\mathrm{E}}^{(\mathrm{global})}\big(z_\mathrm{FV},z_\mathrm{FV}\:\!;T\big)}$ for large $T$, given a potential possessing both a metastable FV as well as a TV region. Whereas the ``shot'' directly probes the properties of the global minimum of the potential $V^{(\mathrm{stable})}(z)$, the constant FV trajectory and the ``bounce'' are associated with properties of the FV region.}
    \label{fig:DifferentClassicalTrajectories}
\end{figure}

\begin{figure}[H]
    \centering
    \includegraphics[width=\textwidth]{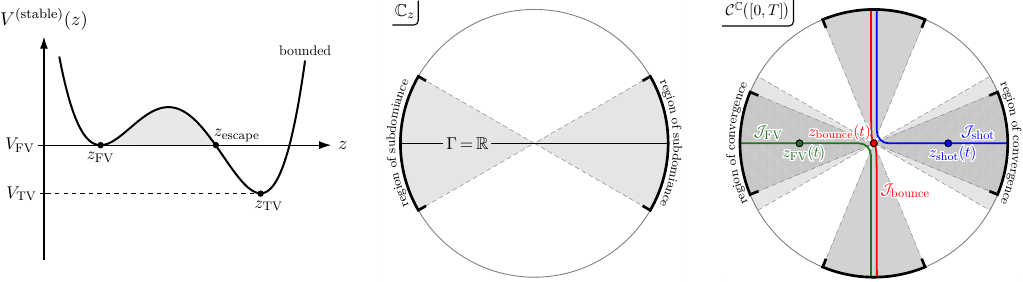}
    \caption{Schematic structure of the relevant steepest-descent thimbles in the complexified function space $\mathcal{C}^\mathbb{C}\big([0,T]\big)$, arising in the study of the propagator $\smash{K_{\mathrm{E}}^{(\mathrm{global})}\big(z_\mathrm{FV},z_\mathrm{FV}\:\!;T\big)}$ for the (globally) stable potential $V^{(\mathrm{stable})}(z)$ shown on the left. The degeneracy of the thimbles, due to the three saddle points $z_\mathrm{FV}(t)$, $z_\mathrm{bounce}(t)$ and $z_\mathrm{shot}(t)$ being connected by the gradient flow, has been lifted for better comprehension. With the eigenvalue problem defined on the real line, the integration contour of the associated path integral is simply the underlying real function space $\mathcal{C}\big([0,T]\big)$, which can be decomposed as a sum of the three contours $\mathcal{J}_\mathrm{FV}$, $\mathcal{J}_\mathrm{bounce}$ and $\mathcal{J}_\mathrm{shot}$. The shaded light gray regions illustrate the Stokes wedges $S_\pm$ defining the eigenvalue problem, whereas the dark gray wedges depict the regions of convergence of the functional integral, practically imprinting the angular sectors $\mathcal{S}_0\cup \mathcal{S}_1\cup \ldots \cup \mathcal{S}_n$.}
    \label{fig:Tunneling_Thimble_Structure_stable}
\end{figure}

With all the key pieces of the puzzle, provided in sections~\ref{sec:2_Resonant_State_Introduction} and~\ref{sec:3_3_Equating_Representations}, in place, we are now fit to assemble them into the complete picture. To this end, let us investigate the rather generic situation depicted in figure~\ref{fig:GenericMetastablePot+TimeEvolution} (or similarly in the left panel of figure~\ref{fig:Tunneling_Thimble_Structure_stable}), for which the globally stable potential $V^{(\mathrm{stable})}(z)$ admits both a metastable FV region together with a global minimum $z_\mathrm{TV}$. Prior to detailing the simple procedure that ultimately gives rise to the desired contour deformation prescription proposed by Callan \& Coleman~\cite{CallanColemanFateOfFalseVac2}, we offer a few additional remarks:
\begin{itemize}
    \item For clarity of exposition, let us assume that quantum tunneling is only viable in a single spatial direction; all subsequent arguments directly carry over to the case for which a trapped particle is allowed to tunnel into both directions. 
    \item The frequently studied scenario for which the potential is unbounded from below at the outset proves simpler than the more intricate present case involving an additional TV region. Due to the absence of a global minimum, the first step in the ensuing analysis can essentially be bypassed.
    \item Despite our previous endeavors being mostly confined to polynomial potentials, it entails no loss of generality to apply the provided results to non-polynomial potentials, as we can effectively approximate the relevant potential region to arbitrary accuracy using high-order polynomials.\footnote{This subliminally assumes the barrier region to have a finite extent, i.e. for the potential to not be asymptotically flat with $\lim_{z\to\infty} V(z)=V(z_\mathrm{FV})$, which potentially spells trouble~\cite{PatrascioiuComplexTime}.} 
    \item For all subsequent developments, we will initially ignore the role of multi-instantons, instead focusing on the schematic structure of the functional contour deformation within the single-instanton approximation. A more sophisticated analysis of the full thimble structure, including multi-bounces, is left to future work.
\end{itemize}
\noindent 
Paralleling the discussion contained in our previous work~\cite{WagnerExcitedStateTunneling}, let us formally distinguish between the global energy eigenvalues $\Englob$ and the sought-after resonant energies $\Enres$, formerly dubbed $\Enloc$. Utilizing relation~\eqref{eq:Propagator_SpectralRepresentation} for the ordinary case of defining the eigenvalue problem on the real axis given a globally stable potential, the Euclidean FV-to-FV propagator in the full potential $V^{(\mathrm{stable})}(z)$ attains the form
\begin{align}
		\!\! \scalebox{0.985}{$\displaystyle{K_{\mathrm{E}}^{(\mathrm{global})}\big(z_\mathrm{FV},z_\mathrm{FV}\:\!;T\big)=\mathlarger{\mathlarger{\sum}}_{\ell=0}^\infty \: \exp\cbigg\{-\,\frac{E^{(\mathrm{global})}_\ell T}{\hbar}\cbigg\} \, \Psi^{(\mathrm{global})}_\ell(z_\mathrm{FV})^2 \,\scalebox{1.1}{\bigg\{}\!\int_{-\infty}^{\infty} \Psi^{(\mathrm{global})}_\ell(z)^2\,\mathrm{d}z\scalebox{1.1}{\bigg\}}^{\!\!\!\:-1} ,}$}
        \label{eq:SpectralRepresentationPropagatorGlobal}
\end{align}
allowing us to project out the (global) ground state energy using the late time-limit
\begin{align}
    E_0^{(\mathrm{global})}=-\hbar \lim_{T\to\infty} \bigg\{T^{-1} \log\!\Big[K_\mathrm{E}^{(\mathrm{global})}\scalebox{1.15}{$\big($}z_\mathrm{FV},z_\mathrm{FV}\:\!;T\scalebox{1.15}{$\big)$}\Big]\!\!\;\bigg\} \, .
    \label{eq:Ground_State_Energy_Global}
\end{align}
As realized by Patrascioiu~\cite{PatrascioiuComplexTime} and rediscovered by Andreassen et al.~\cite{SchwartzPrecisionDecayRate}, when studying the relevant critical trajectories of the arising path integral, one generally finds the three important motions depicted in figure~\ref{fig:DifferentClassicalTrajectories}---the trivial FV motion, ``bounce'' and ``shot''.\footnote{In case the potential is asymptotically flat, the ``shot''-like solution will look slightly different. Note that this case was originally studied by Patrascioiu~\cite{PatrascioiuComplexTime}, dubbing that particular third motion the ``slide''.} As illustrated in the right panel of figure~\ref{fig:Tunneling_Thimble_Structure_stable}, the (greatly simplified) thimble decomposition of the original integration contour thereby attains the schematic form
\begin{align}
    \mathcal{C}\big([0,T]\big)\equiv \mathcal{J}_\mathrm{FV}+\mathcal{J}_\mathrm{bounce}+\mathcal{J}_\mathrm{shot}\, .
\end{align}
With the imaginary contribution arising from the bounce saddle being fully offset by the overlap of the relevant steepest-descent thimbles $\mathcal{J}_\mathrm{FV}$, $\mathcal{J}_\mathrm{bounce}$ and $\mathcal{J}_\mathrm{shot}$,\footnote{Note that this is perfectly in line with the expected behavior known from (real) single-dimensional Laplace-type integrals, for which the only contributions arise from the minima of the exponent in the integration domain. Contributions from maxima of the exponent, despite constituting critical points, can be safely neglected due to the overlap of the steepest-descent contours in question.} the emerging result for the global ground state energy $\Eglob$ is indeed fully real, reading
\begin{align}
    E_0^{(\mathrm{global})}=-\hbar \lim_{T\to\infty} \bigg\{T^{-1} \log\!\Big[K_\mathrm{E}^{(\mathrm{FV})}\scalebox{1.15}{$\big($}z_\mathrm{FV},z_\mathrm{FV}\:\!;T\scalebox{1.15}{$\big)$}+K_\mathrm{E}^{(\mathrm{shot})}\scalebox{1.15}{$\big($}z_\mathrm{FV},z_\mathrm{FV}\:\!;T\scalebox{1.15}{$\big)$}\Big]\!\!\;\bigg\} \, .
    \label{eq:Ground_State_Energy_Global_Semiclassics}
\end{align}
The superscript on the propagators in this case highlights the classical solution about which the functional integral is expanded when performing a semiclassical evaluation. For large times $T$, the shot dominates relation~\eqref{eq:Ground_State_Energy_Global_Semiclassics}, effectively probing the TV region of the potential. Indeed, regardless of the endpoints $z_\mathrm{i,f}$, one will always find a shot-like solution, which, in the limit $T\to\infty$, converges pointwise to the constant TV trajectory, thereby ensuring the accuracy of equality~\eqref{eq:Ground_State_Energy_Global} for arbitrary choices of $z_\mathrm{i,f}$.

\begin{figure}[H]
\includegraphics[width=\textwidth]{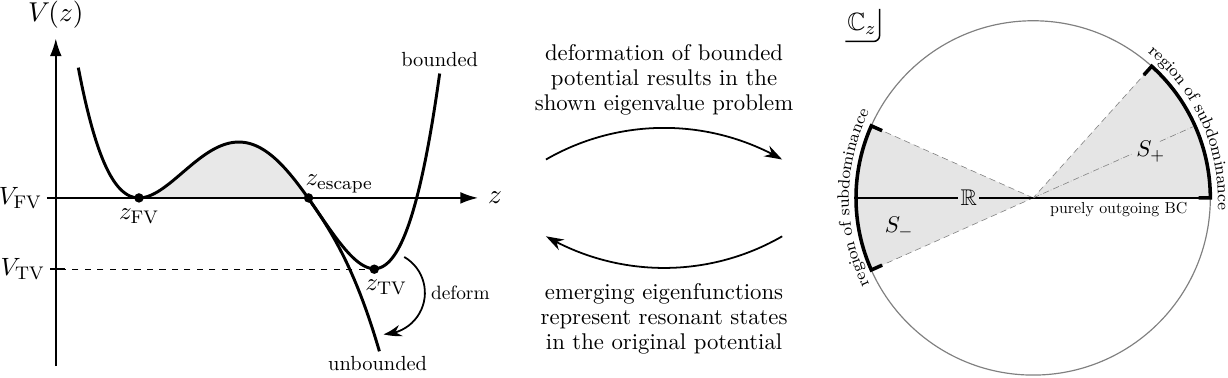}
\caption{Schematic procedure on how to compute the resonant spectrum in the metastable FV region of an otherwise stabilized potential $V^{(\mathrm{stable})}(z)$. One initially deforms the original potential past the classical escape point $z_\mathrm{escape}$ to be unbounded from below, rendering the potential unstable in the process. The emerging eigenvalue problem associated to the unstable potential $V^{(\mathrm{unstable})}(z)$, defined in a Stokes sector that only touches the real axis in the half-plane where the deformation was instated, yields the sought-after Gamow--Siegert states that describe the desired long-lived quasi-bound states in the original potential $V^{(\mathrm{stable})}(z)$.}
\label{fig:SchematicCycleDepiction}
\end{figure}

\begin{figure}[H]
    \centering
    \includegraphics[width=\textwidth]{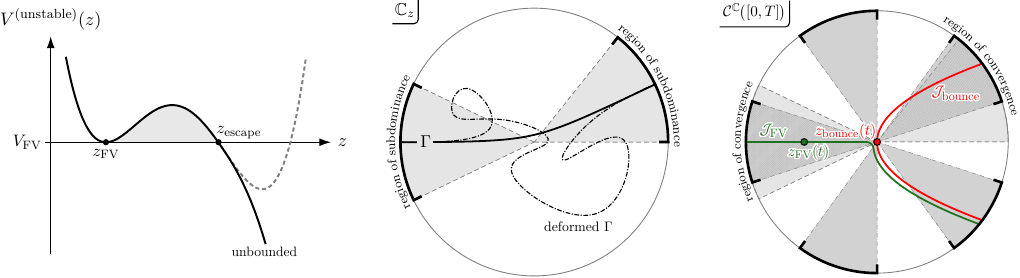}
    \caption{Expected structure of the steepest-descent thimbles for the deformed, unstable potential $V^{(\mathrm{unstable})}(z)$. The emerging eigenvalue problem is defined in tilted Stokes wedges bordering the real axis in all asymptotic regions into which the particle can tunnel. Consequently, the functional integration contour is similarly distorted into the upper complex half-plane, arising from the fact that the hatched overlap between the lightly shaded regions of subdominance $S_\pm$ defining the eigenvalue problem and the dark gray regions of convergence of the discretized functional integral exclude the real function space $\mathcal{C}\big([0,T]\big)$ to be an admissible integration cycle. Accordingly, the correct complexified integration contour $\mathcal{C}\big([0,T],\Gamma\big)$ only receives contributions from $\mathcal{J}_\mathrm{FV}$ and $\mathcal{J}_\mathrm{bounce}$ due to the shot solution being absent, which is schematically depicted on the right. Beware that the shown thimble decomposition is an over-simplified representation, intended to convey the basic idea while omitting the true complexity of the infinite-dimensional setting.}
    \label{fig:Tunneling_Thimble_Structure_unstable}
\end{figure}

\noindent 
To access the sought-after resonant energies $\Enres$, we desire to adjust the initial eigenvalue problem granting $\Englob$ by imposing outgoing Gamow--Siegert boundary conditions toward the TV region. Building upon the insights arising from our previous discussion in section~\ref{sec:2_2_AnalyticContinuation_EigenvalueProblems}, we find that a deformation of the stable potential $V^{(\mathrm{stable})}(z)$ into an unbounded one, as shown in figure~\ref{fig:SchematicCycleDepiction}, will yield an eigenvalue problem posed in a Stokes wedge bordering the positive real axis, thereby naturally enforcing purely outgoing boundary conditions. By instating this potential deformation only past the classical escape point $z_\mathrm{escape}$, thus leaving the barrier region fully invariant, the prescribed eigenvalue problem involving the new, unbounded potential $V^{(\mathrm{unstable})}(z)$ precisely encodes the desired intermediate quasi-bound states in the original potential $V^{(\mathrm{stable})}(z)$.
\noindent
The procedure for extracting the desired resonant ground state energy $\Eres$ then mirrors that used previously for the global ground state $\Eglob$. Once more utilizing the spectral representation~\eqref{eq:SpectralRepresentationPropagatorGlobal}, albeit with the replacements $(\mathrm{global})\mapsto (\mathrm{resonant})$ and $\int_{-\infty}^{\infty}\mapsto \int_\Gamma$, the sought-after eigenvalue $\Eres$ is projected out via
\begin{align}
    E_0^{(\mathrm{resonant})}=-\hbar \lim_{T\to\infty} \bigg\{T^{-1} \log\!\Big[K_\mathrm{E}^{(\mathrm{resonant})}\scalebox{1.15}{$\big($}z_\mathrm{FV},z_\mathrm{FV}\:\!;T\scalebox{1.15}{$\big)$}\Big]\!\!\;\bigg\} \, .
    \label{eq:Ground_State_Energy_Resonant}
\end{align}
As for the global eigenenergies, this step essentially hinges on the fact that, among the tower of resonant energies $\Enres$, the resonant ground state $\Eres$ possesses the smallest real part. Relative to the previous discussion surrounding equation~\eqref{eq:Ground_State_Energy_Global}, two main differences emerge: 
\begin{itemize}
    \item The potential entering the Euclidean path integral entailed in equation~\eqref{eq:Ground_State_Energy_Resonant} is now the deformed potential $V^{(\mathrm{unstable})}(z)$. Crucially, due to the lack of a global minimum, the shot solution is naturally absent from the discussion, constituting no critical trajectory anymore. The relevant remaining saddle points are thus the trivial FV motion as well as the bounce. It is worth noting that, with the barrier region unchanged, the bounce solution remains unchanged as well.
    \item The arising functional integral is defined on the complexified contour $\mathcal{C}\big([0,T],\Gamma\big)$ instead of $\mathcal{C}\big([0,T]\big)$. Hereby, $\Gamma$ once more constitutes a contour that asymptotically terminates in the angular regions $S_\pm \cap \big(\mathcal{S}_0 \cup \ldots \cup \mathcal{S}_n\big)$, as illustrated in the central panel of figure~\ref{fig:Tunneling_Thimble_Structure_unstable}.
\end{itemize}

\noindent 
The absence of the shot goes hand in hand with the fact that the real function space $\mathcal{C}\big([0,T]\big)$ ceases to be an admissible integration cycle for the (resonant) Euclidean FV-to-FV propagator, which is schematically depicted in the right panel of figure~\ref{fig:Tunneling_Thimble_Structure_unstable}. Instead, we are led to believe that the proper integration contour $\mathcal{C}\big([0,T],\Gamma\big)$ asymptotically terminates in the identical region of convergence as the bounce thimble $\mathcal{J}_\mathrm{bounce}$.\footnote{We emphasize that this merely constitutes a well-motivated conjecture, as a full thimble decomposition in the infinite-dimensional, complexified function space $\mathcal{C}^\mathbb{C}\big([0,T]\big)$ is yet far out of reach. Determining the relevant intersection numbers between the original integration contour $\mathcal{C}\big([0,T],\Gamma\big)$ and the infinite number of steepest-ascent thimbles $\mathcal{K}_\mathrm{crit}$ remains the subject of active research and has yet to be fully realized.} This way, the (na\"{\i}ve) thimble decomposition takes the form 
\begin{align}
    \mathcal{C}\big([0,T],\Gamma\big)\equiv \mathcal{J}_\mathrm{FV}+\mathcal{J}_\mathrm{bounce}\, ,
\end{align}
consequently resulting in the well-known relation
\begin{align}
    E_0^{(\mathrm{resonant})}=-\hbar \lim_{T\to\infty} \Bigg\{T^{-1} \log\!\bigg[K_\mathrm{E}^{(\mathrm{FV})}\scalebox{1.15}{$\big($}z_\mathrm{FV},z_\mathrm{FV}\:\!;T\scalebox{1.15}{$\big)$}+\frac{1}{2}\,K_\mathrm{E}^{(\mathrm{bounce})}\scalebox{1.15}{$\big($}z_\mathrm{FV},z_\mathrm{FV}\:\!;T\scalebox{1.15}{$\big)$}\bigg]\!\!\;\Bigg\} \, .
    \label{eq:Ground_State_Energy_Resonant_Semiclassics}
\end{align}
As illustrated in the right panel of figure~\ref{fig:Tunneling_Thimble_Structure_unstable}, the factor $1/2$ arises from the remaining overlap between both $\mathcal{J}_\mathrm{FV}$ and $\mathcal{J}_\mathrm{bounce}$ thimbles, which are still degenerate. It is immanent that this simple argument formalizes the prescription put forward by Callan \& Coleman, previously resting on heuristic arguments. We especially see that, in case the original potential under investigation is unbounded to start with, the instated potential deformation from $V^{(\mathrm{stable})}(z)$ to $V^{(\mathrm{unstable})}(z)$ can be waived, while in the predominant case the resonant energy can be attained by formally dropping the steepest-descent thimble $\mathcal{J}_\mathrm{shot}$ from the evaluation of the propagator $\smash{K_\mathrm{E}^{(\mathrm{global})}}$. It is worth noting that similar reasoning applies to alternative computational approaches, for instance those that extract the decay rate from the Euclidean partition function~\eqref{eq:Final_Relation_Trace}. Additionally, the given explanation can now be employed to address the previously posed questions stated in section~\ref{sec:4_2_OpenQuestions_ColemanApproach}.
\begin{enumerate}
    \item[i--iii)] The first three questions have effectively been addressed in the preceding paragraphs. As shown, the (Euclidean) instanton method explicitly extracts the resonant ground state energy, elucidating why the usual decay rate computations based on semiclassical solutions to the Schrödinger equation precisely match those of the instanton procedure. Meanwhile, the functional contour prescription by Callan \& Coleman precisely encodes the desired outgoing Gamow--Siegert boundary conditions, whereas a distortion of the integration contour into the negative complex half-plane, resulting in the opposite imaginary part for the energy eigenvalues, is related to incoming boundary conditions, i.e. anti-resonant states. Starting from a globally stable potential $V^{(\mathrm{stable})}(z)$, shot-like solutions turn out to be naturally absent from the discussion surrounding decay rates, as the extraction of \mbox{(anti-)resonant} energies explicitly demands an unbounded potential in order to encapsulate the radiating boundary conditions. After instating a formal potential deformation $V^{(\mathrm{stable})}(z)\mapsto V^{(\mathrm{unstable})}(z)$ required to access the sought-after spectral information, the previously dominating shot-like solutions are no longer present. Due to the requirement of leaving the relevant meta-stable FV region as well as the barrier invariant, the aforementioned deformation can for all practical purposes be waived, as simply computing the bounce-solution in the original, globally stable potential $V^{(\mathrm{stable})}(z)$ suffices.\footnote{Utilizing a semiclassical evaluation scheme for the functional integrals at play, in the formal limit $\hbar\to 0^+$, all relevant contributions necessarily arise from an infinitesimal neighborhood around the relevant saddle points. Therefore, the computed decay rate is, to all orders in a perturbative expansion in $\hbar$, independent of how the potential deformation is instated precisely. However, there will arise nonperturbative corrections depending on how exactly the potential is deformed. Hence, there exists a natural ambiguity on how the desired decay rate is defined, which would similarly emerge for other semiclassical computation schemes, e.g. for WKB procedures.} 
    \item[iv)] To examine the distinguishing features of the choices $z_\mathrm{i}=z_\mathrm{f}=z_\mathrm{FV}$ and $\theta=\pi/2$, let us formulate the most general version of the previously stated relations~\eqref{eq:Ground_State_Energy_Global} and~\eqref{eq:Ground_State_Energy_Resonant}. Fully spelled out, one arrives at the two equalities\\[-1.2cm]
    \begin{adjustwidth}{-1.75cm}{-0.75cm} 
    \begin{subequations}
    \begin{align}
    E_0^{(\mathrm{global})}\!\!\:&=ie^{i\theta}\hbar \lim_{T\to\infty}\!\!\: \left\{\frac{1}{T} \rule{0pt}{1.1cm}\log\!\left(\!\!\:\rule{0pt}{0.95cm}\smash{\mathlarger{\mathlarger{\int}}_{\scalebox{0.72}{$\mathcal{C}\big([0,T]\big)$}}^{\substack{z(0)\:\!=\:\!z_\mathrm{i}\\[0.05cm] z(T)\:\!=\:\!z_\mathrm{f}\,}}} \,\!\!\: \mathcal{D}_\theta\llbracket z\rrbracket\, \exp\!\!\:\Bigg\{\frac{ie^{i\theta}}{\hbar}\mathlarger{\int}_0^T\bigg[\frac{m}{2}\:\!\dot{z}(t)^2-e^{-2i\theta}\:\!V^{(\mathrm{stable})}\scalebox{1.1}{\big(}z(t)\!\!\;\scalebox{1.1}{\big)}\bigg]\mathrm{d}t\Bigg\}\!\!\:\right)\!\!\:\right\} ,
    \label{eq:Ground_State_Energy_Global_FullySpelledOut} \\
    E_0^{(\mathrm{resonant})}\!\!\:&=ie^{i\theta}\hbar \lim_{T\to\infty}\!\!\: \left\{\frac{1}{T} \rule{0pt}{1.1cm}\log\!\left(\!\!\:\rule{0pt}{0.95cm}\smash{\mathlarger{\mathlarger{\int}}_{\scalebox{0.72}{$\mathcal{C}\big([0,T],\Gamma\big)$}}^{\substack{z(0)\:\!=\:\!z_\mathrm{i}\\[0.05cm] z(T)\:\!=\:\!z_\mathrm{f}\,}}} \,\!\!\: \mathcal{D}_\theta\llbracket z\rrbracket\, \exp\!\!\:\Bigg\{\frac{ie^{i\theta}}{\hbar}\mathlarger{\int}_0^T\bigg[\frac{m}{2}\:\!\dot{z}(t)^2-e^{-2i\theta}\:\!V^{(\mathrm{unstable})}\scalebox{1.1}{\big(}z(t)\!\!\;\scalebox{1.1}{\big)}\bigg]\mathrm{d}t\Bigg\}\!\!\:\right)\!\!\:\right\} . \label{eq:Ground_State_Energy_Resonant_FullySpelledOut} \\[-1.5cm] 
    \nonumber
    \end{align}
    \label{eq:Ground_State_Energy_Both_FullySpelledOut}%
    \end{subequations}
    \end{adjustwidth}
    While the above relations~\eqref{eq:Ground_State_Energy_Global_FullySpelledOut} and~\eqref{eq:Ground_State_Energy_Resonant_FullySpelledOut} are universally valid for arbitrary $z_\mathrm{i,f}\in \mathbb{C}$ and $\theta\in (0,\pi/2]$, a convenient evaluation is feasible only when imposing specific values for these free parameters. Compared to other picks, the choice $\theta=\pi/2$ and $z_\mathrm{i}=z_\mathrm{f}=z_\mathrm{FV}$ exhibits the unique feature that all relevant critical trajectories are real-valued. Even though differing real choices of the endpoints $z_\mathrm{i,f}$ inside the FV region also result in real critical trajectories, the arising motions would possess less symmetry, unnecessarily complicating a proper evaluation.  

    While it is certainly convenient to work with real critical trajectories, the traditional procedure offers a far more significant advantage: We never have to make the potential deformation $V^{(\mathrm{stable})}(z)\mapsto V^{(\mathrm{unstable})}(z)$ explicit. Due to the real bounce motion probing the potential solely up to the classical escape point $z_\mathrm{escape}$, the distinction between the original bounded potential and the deformed, unstable one is irrelevant. This is vital, as a semiclassical evaluation generally mandates the incorporation of complex critical trajectories, which latently requires the potential to be a holomorphic function. However, if one demands the potential deformation to leave the potential fully invariant over an extended region of the real axis, then the two potentials $V^{(\mathrm{stable})}(z)$ and $V^{(\mathrm{unstable})}(z)$ cannot be simultaneously analytic, as their difference vanishes on a strip along the real line. Therefore, in most situations, a proper semiclassical analysis is only viable for one of the two functional integrals. By working in Euclidean signature and fixing the endpoints $z_\mathrm{i,f}$ conveniently, one bypasses this issue without the requirement to explicitly address it. Lastly, it seems indispensable for the relevant thimbles (in the predominant Euclidean case $\mathcal{J}_\mathrm{FV}$ and $\mathcal{J}_\mathrm{bounce}$) to be degenerate, i.e. for the system to be located on a Stokes line. This way, the overlap of the steepest-descent thimbles ensures that both the real and imaginary part of the desired quantity $\Eres$ respectively are fully encoded by two separate thimbles. Otherwise, significant imaginary contributions may be obscured, hidden in the exponentially suppressed tails of thimbles whose dominant real parts overshadow the nonperturbatively small imaginary components. Again, this vital feature does not occur generically and requires finely tuned parameters to emerge.
    
    In summary, the traditional instanton approach involves a considerable degree of subtlety, as only very specific choices for $\theta$ and $z_\mathrm{i,f}$ lead to a broadly applicable result. It is remarkable that Callan \& Coleman were able to arrive at this prescription with the tools available at the time.
\end{enumerate}
\noindent 
Beware that, while the preceding discussion clarifies the connection between resonant states and the instanton method, there remains the open question on how to rigorously establish the full thimble decomposition of the generalized contour $\mathcal{C}\big([0,T],\Gamma\big)$ in the complexified function space $\mathcal{C}^\mathbb{C}\big([0,T]\big)$. Such a feat however requires major advances in Picard--Lefschetz theory, that are, to the best of our knowledge, not yet within reach. As a brief final remark, we note that the results presented here also shed light on why our previous work~\cite{WagnerExcitedStateTunneling} correctly recovered the expressions for excited-state decay rates. In this context, we emphasize that what was actually computed in that work is not the previously stated relation~(3.3) found within~\cite{WagnerExcitedStateTunneling}, but rather the slightly altered, exact projection
\begin{align}
	E_\ell^{(\mathrm{resonant})} &= \frac{i e^{i\theta}\hbar}{T}  \;\log\!\!\:\Bigg\{\mathlarger{\int}_\Gamma\mathrm{d}z_\mathrm{i} \mathlarger{\int}_\Gamma \mathrm{d}z_\mathrm{f}\; \Psi_\ell^{(\mathrm{resonant})}(z_\mathrm{i}) \:\Psi_\ell^{(\mathrm{resonant})}(z_\mathrm{f})\, K_\theta^{(\mathrm{resonant})}\mathlarger{\big(}z_\mathrm{i},z_\mathrm{f}\:\!;T\mathlarger{\big)}\Bigg\} \label{eq:Resonant_Energy_Conjecture} \\ 
    &= \frac{i e^{i\theta}\hbar}{T} \;\log\!\!\:\scalebox{1.3}{\Bigg\{}\!\!\:\mathlarger{\mathlarger{\int}}_{\vphantom{\xi}\smash{\scalebox{0.75}{$\mathcal{C}\big([0,T],\Gamma\big)$}}} \mathcal{D}_\theta\llbracket z\rrbracket \; \Psi_\ell^{(\mathrm{resonant})}\big[z(0)\big] \,\Psi_\ell^{(\mathrm{resonant})}\big[z(T)\big]\:\! \exp\!\!\:\bigg(\frac{iS_\theta \llbracket z\rrbracket}{\hbar}\bigg)\!\scalebox{1.3}{\Bigg\}} \, , \nonumber 
\end{align}
assuming the wave functions to be properly normalized to unity on $\Gamma$. Since the resonant wave functions $\smash{\Psi_\ell^{(\mathrm{resonant})}(z)}$, to leading-order in $\hbar$, have their dominant support near $z_\mathrm{FV}$, the formerly instated normalized uniform WKB solution precisely captured the required leading-order behavior to faithfully compute~\eqref{eq:Resonant_Energy_Conjecture}.\footnote{Note that the reality of the uniform WKB approximation in the relevant domain ensured the complex conjugation in our former relation (3.3), provided in \cite{WagnerExcitedStateTunneling}, to drop out, which was vital to arrive at the correct result, since the proper formulation~\eqref{eq:Resonant_Energy_Conjecture} prominently does not entail any complex conjugation.} The natural absence of all shot-like contributions implicitly took care of both the potential deformation and the functional contour distortion, which was not yet explicitly accounted for in our previous treatment.

\section{Assessment of earlier attempts to unravel the instanton method}
\label{sec:5_Assessment_Earlier_Attempts}

Recent years have seen a surge of interest centered around studying tunneling from a real-time perspective, with the goal to relate the arising results to the traditional instanton method. The most promising result to date is the ``direct method'' proposed by Andreassen et al.~\cite{SchwartzDirectMethod}, claiming to provide the long-sought connection between real-time dynamics and instanton calculus. In section~\ref{sec:5_1_Flaws_DirectMethod}, we will argue that their argument involves a notable oversight and several gaps that merit closer examination, effectively stripping the procedure of its expressive power. The following subsection~\ref{sec:5_2_Flaws_Steadyons} critically examines the work by Steingasser \& Kaiser~\cite{SteingasserRealTimeInstantons} regarding their ``steadyon'' picture of real-time tunneling. At last, we compile some shorter comments on other recent results, that, in our view, also fall short of providing a clear one-to-one correspondence between real-time dynamics and instanton results in section~\ref{sec:5_3_Other_Approaches}.

\subsection{Limitations of the ``direct method'' by Andreassen et al.}
\label{sec:5_1_Flaws_DirectMethod}

We devote this subsection to the ``direct method'' of obtaining the ground-state decay rate introduced by Andreassen et al.~\cite{SchwartzDirectMethod}, as it similarly arose as an attempt to address some of the questions posed by the traditional instanton method discussed in subsection~\ref{sec:4_2_OpenQuestions_ColemanApproach}. Instead of relying on the computation of the imaginary part to the resonant ground-state energy, the procedure attempts to directly capture the survival probability $P_\text{FV}(T)$ of the particle to be located inside the FV region after some time $T$ has elapsed. This is achieved by studying the real-time evolution of a $\delta$-peak initially located at $z_\mathrm{FV}$ and subsequently performing a formal analytic continuation to Euclidean time, intricately manipulating the arising objects. We will find that their method unfortunately falls short of achieving the desired task, as several technical concerns in their derivation spoil the overall argument. To this end, let us briefly review the procedure, adapting their notation to our conventions, thereby allowing us to address some intermediate steps that appear to warrant further explanation as they arise. When referencing equations, we will refer to the extended write-up of their method provided in reference~\cite{SchwartzPrecisionDecayRate}, which includes clarifying steps which were omitted in their original exposition~\cite{SchwartzDirectMethod}.\\ 

\noindent
As we briefly discussed in section~\ref{sec:2_Resonant_State_Introduction}, the decay rate is only well-defined for intermediate times $T_\text{sloshing} \ll T \ll T_\text{back-reaction}$, for which the FV probability 
\begin{align}
    P_\mathrm{FV}(T)=\mathlarger{\int}_{\mathrm{FV}} \,\cbig\lvert \Psi\big(z,T\big)\cbig\rvert^2 \, \text{d}z 
\end{align}
faithfully exhibits the simple exponential behavior $P_\text{FV}(T)\propto e^{-\Gamma_0 T}$, with $\Gamma_0$ constituting the desired ground-state decay rate. The two timescales $\smash{T_\text{sloshing}\approx \sqrt{m/V''(z_\mathrm{FV})}}$ and $T_\text{back-reaction}$ respectively capture the approximate duration of wave function oscillations inside the FV region as well as the scale at which back-reaction and other non-linear effects become non-negligible. Assuming the initial wave function $\Psi_{T=0}(z)$, solely supported inside the FV region, to be generic, the condition $T_\text{sloshing} \ll T$ guarantees that the FV ground state dominates the behavior of the survival probability $P_\text{FV}(T)$ in the given time window.\footnote{Again, note that when preparing very specific initial wave functions that closely resemble a purely excited state in the FV region (if it were stabilized), the exponential decay will similarly be primarily governed by that particular excited-state decay rate instead of the ground-state decay rate.} Instead of the depletion of probability in the FV region, one can similarly investigate the probability growth in the true vacuum region, letting Andreassen et al. to arrive at the simple relation
\begin{equation}
	\Gamma_0 = \lim_{\substack{\scalebox{0.75}{$T/T_\text{back-reaction}\!\!\:\!\!\:\rightarrow\!\!\: 0$} \\ \scalebox{0.75}{$\!\!\!\!\!\!\!\! T/T_\text{sloshing}\!\!\:\rightarrow\!\!\:\infty$} }} \left[\frac{1}{P_\text{FV}(T)} \frac{\text{d}P_\text{TV}(T)}{\text{d}T}\right] \:\! .
\end{equation}
While they generalize to the case of potentially having multiple regions $R$ for the particle to tunnel into, breaking up the spatial integral contained in $P_\text{TV}(T)$ further and thus defining the partial decay widths $\Gamma_\mathrm{R}$, let us concentrate on the case of investigating the total width. Choosing the initial wave function to be $\Psi_{T=0}(z)=\delta\big(z-z_\text{FV}\big)$, thereby removing the additional convolution integrals when expressing the time evolution using the real-time propagator $K\!\!\:\big(z_\mathrm{i},z_\mathrm{f}\:\!;T\big)$, one arrives at the two relations\footnote{Beware that choosing the initial wave-function in the present form requires a regulator, as the false vacuum probability at time 0 would be divergent due to integrating the squared Dirac distribution, yielding $P_\mathrm{FV}(T\!=\!0)=\delta(0)$. We will ignore this fact and assume the divergence has been properly cured, as only ratios will eventually enter the expression for the decay rate, for which any initial normalization will drop out.} 
\begin{align}
    P_\text{FV}(T)=\mathlarger{\int}_{\text{FV}}\: \Big\lvert K\!\!\:\big(z_\text{FV},z\:\!;T\big) \Big\rvert^2\, \text{d}z\, , \qquad\qquad P_\text{TV}(T)=\mathlarger{\int}_{\text{TV}}\: \Big\lvert K\!\!\:\big(z_\text{FV},z\:\!;T\big) \Big\rvert^2\, \text{d}z\, .
    \label{eq:Probability_PropagatorRep}
\end{align}
Andreassen et al. now massage the expression for the escape probability $P_\text{TV}(T)$ by utilizing the fact that the arising propagators can be represented as a sum over trajectories that start at $z_\mathrm{FV}$ and eventually are required to leave the FV region, as the endpoint-integral over $z$ only includes the TV region. Due to continuity and the fact that we deal with purely real paths here, any trajectory $z(t)$ contributing to the propagator is required to traverse the classical escape point $z_\text{escape}$ that splits the FV from the TV region at some intermediate time $0< t^\star \leq T$. Since there could potentially be multiple crossings, one takes $t^\star$ to be the time of the first traversal, thereby defining the functional $t_\mathrm{escape}\llbracket z\rrbracket$ that assigns the time $t^\star$ of the first traversal of $z_\mathrm{escape}$ to any path $z(t)$. This allows Andreassen et al. to split the full propagator appearing in equation~\eqref{eq:Probability_PropagatorRep} using 
\begin{align}
	K\!\!\:\big(z_\text{FV},z\:\!;T\big) &= \mathlarger{\mathlarger{\int}}_{0}^{T} \scalebox{1.15}{\Bigg\{}\mathlarger{\mathlarger{\int}}_{\scalebox{0.75}{$z(0)\!=\!z_\text{FV}$}}^{\scalebox{0.75}{$z(T)\!=\!z$}}\mathcal{D}\llbracket z\rrbracket \,
	\exp\!\!\:\bigg(\frac{i\:\! S\llbracket z\rrbracket}{\hbar}\bigg) \,  \delta\big(t_\mathrm{escape}\llbracket z\rrbracket-t^\star\big)\scalebox{1.15}{\Bigg\}}\,\text{d}t^\star \nonumber \\ 
    &= \mathlarger{\int}_{0}^{T}\, \widetilde{K}\!\!\:\big(z_\text{FV},z_\text{escape}\:\!;t^\star\big)\:\! K\!\!\:\big(z_\text{escape},z\:\!;T-t^\star\big) \, \text{d}t^\star \, ,
	\label{eq:Split_Propagator}
\end{align}
defining the modified propagator $\widetilde{K}$ as the formal path integral expression
\begin{equation}
	\widetilde{K}\!\!\:\big(z_\text{FV},z_\text{escape}\:\!;t^\star\big) = \mathlarger{\mathlarger{\int}}_{\scalebox{0.75}{$z(0)\!=\!z_\text{FV}$}}^{\scalebox{0.75}{$z(t^\star)\!=\!z_\text{escape}$}}\mathcal{D}\llbracket z\rrbracket \,
	\exp\!\!\:\bigg(\frac{i\:\! S\llbracket z\rrbracket}{\hbar}\bigg) \,  \delta\big(t_\mathrm{escape}\llbracket z\rrbracket-t^\star\big) \, .
\end{equation}
We see that the modified propagator only ranges over paths hitting $z_\text{escape}$ precisely at $t^\star$ for the first time, such that the split~\eqref{eq:Split_Propagator} is unambiguous. In section~\ref{sec:5_2_Flaws_Steadyons}, this restriction on the paths contributing to $\widetilde{K}$ will be referred to as the ``crossing condition''. Apart from the fact that one should regulate the $\delta$-distribution entailed in $\widetilde{K}$ to make the expression well-defined, this intermediate result is still exact. One thus arrives at the equality 
\begin{align}
    P_\text{TV}(T)&=\mathlarger{\int}_{0}^{T}\mathrm{d}t^\star_1\mathlarger{\int}_{0}^{T}\mathrm{d}t^\star_2 \;\:\! \widetilde{K}\!\!\:\big(z_\text{FV},z_\text{escape}\:\!;t^\star_1\big)\:\!\closure{\widetilde{K}\!\!\:\big(z_\text{FV},z_\text{escape}\:\!;t^\star_2\big)} \nonumber \\ 
    &\qquad\qquad\qquad\; \times \Bigg\{\mathlarger{\int}_{\text{TV}}\text{d}z\: K\!\!\:\big(z_\text{escape},z\:\!;T-t^\star_1\big) \:\! \closure{K\!\!\:\big(z_\text{escape},z\:\!;T-t^\star_2\big)} \Bigg\} \, ,
\end{align}
again representing complex conjugation with an overhead bar. Andreassen et al. now state that, in the limit $T/T_\text{back-reaction}\to 0$, one can employ the simplification 
\begin{align}
    &\,\mathlarger{\int}_{\text{TV}}\text{d}z\; \closure{K\!\!\:\big(z_\text{escape},z\:\!;T-t^\star_2\big)} \, K\!\!\:\big(z_\text{escape},z\:\!;T-t^\star_1\big)  \nonumber \\[0.05cm]
    = &\, \mathlarger{\int}_{\text{TV}} \text{d}z \; \braket[\cbig]{z_\text{escape}\!\!\:|U^\dagger\big(T-t^\star_2\big)\!\!\:| z} \braket[\cbig]{z\!\!\:|U\big(T-t^\star_1\big)\!\!\:| z_\text{escape}} \nonumber \\[0.05cm] 
    =&\: \cbigg\langle z_\text{escape} \,\cbigg\vert\, U\big(t^\star_2-T\big)  \:\! \bigg\{\int_{\text{TV}} \text{d}z \,\ket{z}\bra{z}\bigg\} \, U\big(T-t^\star_1\big)\!\!\:\, \cbigg\vert  \, z_\text{escape} \cbigg\rangle \nonumber \\[0.05cm] 
    \overset{?}{\approx} &\: \braket[\cbig]{z_\text{escape}\!\!\:|U\big(t^\star_2-t^\star_1\big)\!\!\:| z_\text{escape}} = K\!\!\:\big(z_\text{escape},z_\text{escape}\:\!;t^\star_2-t^\star_1\big) \, .
\end{align}
They argue that replacing the projector $\smash{\displaystyle{\scalebox{1.3}{$\int$}}_{\!\!\mathrm{FV}} \:\!\text{d}z \,\ket{z}\bra{z}}$ by $\smash{\displaystyle{\scalebox{1.3}{$\int$}}_{\!\!-\infty}^{\infty} \text{d}z \,\ket{z}\bra{z}}=\mathds{1}$ only modifies the expression by exponentially small terms of order $\Gamma_0$ itself, which they deem justified due to contributions for which $z$ lies inside the FV region being associated with back-tunneling events. To us, it is by no means clear that this approximation goes through unobstructed, as we deal with oscillatory real-time propagators instead of Euclidean ones, for which the presumed exponential suppression cannot be argued easily.\footnote{Also note that in the usual instanton method, one has to acknowledge contributions of back-and-forth tunneling events in the form of multi-instantons due to zero-mode enhancements, thus it is unclear if such configurations can be neglected in this real-time approach.} Additionally, there could arise non-negligible contributions from the region close to $z_\mathrm{escape}$, for which the suppression factor (even in the Euclidean case) would presumably be small. Since it seems hard to provide a precise error estimate for the above simplification, we see the necessity to revisit this step, as this result is also utilized by works building upon the direct method, see e.g.~\cite{SteingasserFiniteTemp, SteingasserRealTimeInstantons}. As the above simplification will however not be our primary concern, let us assume the approximation to be well-justified, proceeding with their derivation, having arrived at the relation 
\begin{align}
    P_\text{TV}(T)&\approx \mathlarger{\int}_{0}^{T}\mathrm{d}t^\star_1\mathlarger{\int}_{0}^{T}\mathrm{d}t^\star_2 \; \widetilde{K}\!\!\:\big(z_\text{FV},z_\text{escape}\:\!;t^\star_1\big)\,\closure{\widetilde{K}\!\!\:\big(z_\text{FV},z_\text{escape}\:\!;t^\star_2\big)} \, K\!\!\:\big(z_\text{escape},z_\text{escape}\:\!;t^\star_2-t^\star_1\big) \, .
\end{align}
One can now cleverly split the integration region into the two parts $t_1^\star-t_2^\star > 0$ and $t_1^\star-t_2^\star \leq 0$, after which one can employ the two relations 
\begin{equation}
	\int_0^{T} \text{d}t_2^\star \int_{t_2^\star}^T \text{d}t_1^\star = \int_0^T \text{d}t_1^\star \int_{0}^{t_1^\star} \text{d}t_2^\star  \, ,\qquad \mathrm{and}\qquad  K\!\!\:\big(z_\text{f},z_\text{i}\:\!;t\big)=\closure{K\!\!\:\big(z_\text{i},z_\text{f}\:\!;-t\big)} \, .
\end{equation}
This lets us arrive at their intermediate result 
\begin{align}
    \!\!\!P_\text{TV}(T)&\approx \mathlarger{\int}_{0}^{T}\mathrm{d}t^\star_1\mathlarger{\int}_{0}^{t_1^\star}\mathrm{d}t^\star_2 \; \widetilde{K}\!\!\:\big(z_\text{FV},z_\text{escape}\:\!;t^\star_1\big)\,\closure{\widetilde{K}\!\!\:\big(z_\text{FV},z_\text{escape}\:\!;t^\star_2\big) \:\! K\!\!\:\big(z_\text{escape},z_\text{escape}\:\!;t^\star_1-t^\star_2\big)} \nonumber \\ 
    &+ \mathlarger{\int}_{0}^{T}\mathrm{d}t^\star_2\mathlarger{\int}_{0}^{t_2^\star}\mathrm{d}t^\star_1 \; \widetilde{K}\!\!\:\big(z_\text{FV},z_\text{escape}\:\!;t^\star_1\big)\,\closure{\widetilde{K}\!\!\:\big(z_\text{FV},z_\text{escape}\:\!;t^\star_2\big)} \, K\!\!\:\big(z_\text{escape},z_\text{escape}\:\!;t^\star_2-t^\star_1\big)  \\  
    &= \mathlarger{\int}_{0}^{T}\mathrm{d}t^\star \; \widetilde{K}\!\!\:\big(z_\text{FV},z_\text{escape}\:\!;t^\star\big)\,\closure{K\!\!\:\big(z_\text{FV},z_\text{escape}\:\!;t^\star\big)} + \closure{\widetilde{K}\!\!\:\big(z_\text{FV},z_\text{escape}\:\!;t^\star\big)}\,K\!\!\:\big(z_\text{FV},z_\text{escape}\:\!;t^\star\big) \, , \nonumber
\end{align}
where one utilizes the previous definition~\eqref{eq:Split_Propagator} of the modified propagator to get rid of the inner $t^\star$-integration. Differentiating the given expression, one arrives at the desired approximation
\begin{align}
    \frac{\text{d}P_\text{TV}(T)}{\text{d}T}\approx 2\, \mathrm{Re}\scalebox{1.4}{\Bigg\{} \!\!\left[ \mathlarger{\mathlarger{\int}}_{\scalebox{0.75}{$z(0)\!=\!z_\text{FV}$}}^{\scalebox{0.75}{$z(T)\!=\!z_\text{escape}$}} \mathcal{D}\llbracket z\rrbracket \,
		\exp\!\!\:\bigg(\frac{i\:\! S\llbracket z\rrbracket}{\hbar}\bigg) \,  \delta\big(t_\mathrm{escape} \llbracket z\rrbracket-T\big)\right] \qquad\qquad\qquad \nonumber \\
		\times \closure{\left[ \mathlarger{\mathlarger{\int}}_{\scalebox{0.75}{$z(0)\!=\!z_\text{FV}$}}^{\scalebox{0.75}{$z(T)\!=\!z_\text{escape}$}} \mathcal{D}\llbracket z\rrbracket \,
			\exp\!\!\:\bigg(\frac{i\:\! S\llbracket z\rrbracket}{\hbar}\bigg)\right]}\scalebox{1.4}{\Bigg\}}\, .
    \label{eq:TVProbability_Derivative}
\end{align}
To make contact to conventional instanton calculus, Andreassen et al. seek to re-express the real-time path integrals via Euclidean pendants, requiring an analytic continuation. Assuming the propagator $K\!\!\:\big(z_\text{i},z_\text{f}\:\!;T\big)$ to be entire in $T$, they utilize the two formal relations 
\begin{subequations}
    \begin{align}
    K\!\!\:\big(z_\text{i},z_\text{f}\:\!;T\big) &= \Big[K\!\!\:\big(z_\text{i},z_\text{f}\:\!;-i\mathcal{T}\big)\Big]_{\substack{\scalebox{0.75}{$\mathcal{T}\!\!\:>\!\!\:0$}\;\;\\[0.05cm] \scalebox{0.75}{$\mathcal{T}\!\!\:=\!\!\:iT$}}}= \Big[K_\mathrm{E}\big(z_\text{i},z_\text{f}\:\!;\mathcal{T}\big)\Big]_{\substack{\scalebox{0.75}{$\mathcal{T}\!\!\:>\!\!\:0$}\;\;\\[0.05cm] \scalebox{0.75}{$\mathcal{T}\!\!\:=\!\!\:iT$}}}\, , \\[0.1cm] 
    \closure{K\!\!\:\big(z_\text{i},z_\text{f}\:\!;T\big)} &= K\!\!\:\big(z_\text{f},z_\text{i}\:\!;-T\big) = \Big[K\!\!\:\big(z_\text{f},z_\text{i}\:\!;-i\mathcal{T}'\big)\Big]_{\substack{\scalebox{0.75}{$\mathcal{T}'\!\!\:>\!\!\:0$}\;\;\;\;\,\\[0.05cm] \scalebox{0.75}{$\mathcal{T}'\!\!\:=\!\!\:-iT$}}}= \Big[K_\mathrm{E}\big(z_\text{f},z_\text{i}\:\!;\mathcal{T}'\big)\Big]_{\substack{\scalebox{0.75}{$\mathcal{T}'\!\!\:>\!\!\:0$}\;\;\;\;\,\\[0.05cm] \scalebox{0.75}{$\mathcal{T}'\!\!\:=\!\!\:-iT$}}}\, ,
\end{align}
\label{eq:AnalyticCont_Propagators}%
\end{subequations}
which, up to an additional phase factor $\eta_\pm$ arising from the $\delta$-distribution, also hold for the modified propagator $\widetilde{K}$. The above relations~\eqref{eq:AnalyticCont_Propagators} should be understood as follows: One desires to evaluate the real-time propagator, given a positive time interval $T\in\mathbb{R}^+$. To achieve this, one formally sets $T=-i\mathcal{T}$, thereby expressing the real-time propagator in terms of its Euclidean pendant involving the (initially imaginary) time variable $\mathcal{T}=iT$. Since the functional integral associated to the Euclidean propagator is only amenable to calculation given real, positive values of $\mathcal{T}$, one carries out the evaluation under that condition. Assuming analyticity of the resulting function in $\mathcal{T}$ then justifies extending the attained relation to complex $\mathcal{T}$. Therefore, only after the Euclidean propagator has been computed for positive real $\mathcal{T}$, does one substitute back $\mathcal{T}=iT$ to recover the desired real-time result.\footnote{Beware that this step requires knowledge of the full analytic form of the function $K_\mathrm{E}\big(z_\text{i},z_\text{f}\:\!;\mathcal{T}\big)$ for $\mathcal{T}>0$, demanding access to the exact result of the functional integral. This presents a notable limitation of the provided procedure, as will be discussed toward the end of this section.} Hereby, one formally Wick-rotates the propagator and its conjugated counterpart in opposite directions, which subsequently enables us to combine the product of propagators without the requirement to introduce a proper Keldysh-contour. Note that we will meticulously keep track of $\mathcal{T}$ and $\mathcal{T}'$, as they are related to $T$ in conjugate ways---a lack of this distinction is ultimately the principal gap in the derivation provided by Andreassen et al. Utilizing time translation invariance in the first path integral while employing relations~\eqref{eq:AnalyticCont_Propagators} for both $K$ and $\widetilde{K}$, we arrive at the intermediate result
\begin{align}
    \frac{\text{d}P_\text{TV}(T)}{\text{d}T}&\approx 2\, \mathrm{Re}\scalebox{1.4}{\Bigg\{} \!\!\left[ \mathlarger{\mathlarger{\int}}_{\scalebox{0.75}{$z(-T)\!=\!z_\text{FV}$}}^{\scalebox{0.75}{$z(0)\!=\!z_\text{escape}$}} \mathcal{D}\llbracket z\rrbracket \,
		\exp\!\!\:\bigg(\frac{i\:\! S\llbracket z\rrbracket}{\hbar}\bigg) \,  \delta\big(t_\mathrm{escape} \llbracket z\rrbracket\big)\right]  \nonumber \\
		&\qquad\qquad\qquad\qquad\qquad\qquad\qquad\quad\times \closure{\left[ \mathlarger{\mathlarger{\int}}_{\scalebox{0.75}{$z(0)\!=\!z_\text{FV}$}}^{\scalebox{0.75}{$z(T)\!=\!z_\text{escape}$}} \mathcal{D}\llbracket z\rrbracket \,
			\exp\!\!\:\bigg(\frac{i\:\! S\llbracket z\rrbracket}{\hbar}\bigg)\right]}\scalebox{1.4}{\Bigg\}} \nonumber \\ 
        &=2\, \mathrm{Re}\scalebox{1.4}{\Bigg\{} \!\!\left[ \mathlarger{\mathlarger{\int}}_{\scalebox{0.75}{$z(-\mathcal{T})\!=\!z_\text{FV}$}}^{\scalebox{0.75}{$z(0)\!=\!z_\text{escape}$}} \mathcal{D}_\mathrm{E}\llbracket z\rrbracket \,
		\exp\!\!\:\bigg(\!\!\!\:-\!\!\:\frac{S_\mathrm{E}\llbracket z\rrbracket}{\hbar}\bigg) \,  \eta_+ \delta\big(\tau_\mathrm{escape}\llbracket z\rrbracket\big)\right] \nonumber \\
		&\qquad\qquad\qquad\qquad\qquad\qquad\qquad\quad\times \left[ \mathlarger{\mathlarger{\int}}_{\scalebox{0.75}{$z(0)\!=\!z_\text{escape}$}}^{\scalebox{0.75}{$z(\mathcal{T}')\!=\!z_\text{FV}$}} \mathcal{D}_\mathrm{E}\llbracket z\rrbracket \,
			\exp\!\!\:\bigg(\!\!\!\:-\!\!\:\frac{S_\mathrm{E}\llbracket z\rrbracket}{\hbar}\bigg)\right]\!\!\scalebox{1.4}{\Bigg\}} _{\substack{\scalebox{0.75}{$\mathcal{T},\mathcal{T}'\!\!\:>\!\!\:0$}\\[0.05cm] \scalebox{0.75}{$\mathcal{T}\:\!\!\:=\!\!\:+ iT$}\\[0.05cm] \scalebox{0.75}{$\mathcal{T}'\!\!\:=\!\!\:-iT$}}} \nonumber \\
        &=2\, \mathrm{Re}\scalebox{1.4}{\Bigg\{} \eta_+\mathlarger{\mathlarger{\int}}_{\scalebox{0.75}{$z(-\mathcal{T})\!=\!z_\text{FV}$}}^{\scalebox{0.75}{$z(\mathcal{T}')\!=\!z_\text{FV}$}} \mathcal{D}_\mathrm{E}\llbracket z\rrbracket \,
		\exp\!\!\:\bigg(\!\!\!\:-\!\!\:\frac{S_\mathrm{E}\llbracket z\rrbracket}{\hbar}\bigg) \,  \delta\big(\tau_\mathrm{escape} \llbracket z\rrbracket\big) \scalebox{1.4}{\Bigg\}} _{\substack{\scalebox{0.75}{$\mathcal{T},\mathcal{T}'\!\!\:>\!\!\:0$}\\[0.05cm] \scalebox{0.75}{$\mathcal{T}\:\!\!\:=\!\!\:+ iT$}\\[0.05cm] \scalebox{0.75}{$\mathcal{T}'\!\!\:=\!\!\:-iT$}}} \, .
        \label{eq:IntermediateResult_Andreassen}
\end{align}
Note that we omitted the factor of $i$ inside the argument of the $\delta$-distribution when passing from $t_\mathrm{escape}\llbracket z\rrbracket$ to $ i\tau_\mathrm{escape}\llbracket z\rrbracket$, as it has been taken care of by the overall phase factor $\eta_+$. In the last line, we also conveniently combined both functional integrals.\footnote{This merging only works due to the explicit demand of all paths to traverse precisely the point $z_\mathrm{escape}$ at $\tau=0$, otherwise one would require an additional integral over the starting-/end-point position to combine the two path integrals into a single one (hinging on an insertion of unity in terms of position eigenstates).} The additional phase $\eta_+$ entering the expression can be obtained from investigating a single-dimensional toy integral mimicking the structure of the path integral. Given $\alpha\in \mathbb{R}^*$ and real-valued functions $S(z)$ and $t(z)$, one arrives at the formal equality
\begin{align}
	\mathlarger{\int} \Big[A\:\! e^{i\alpha S(z)}+\closure{A}\:\! e^{-i\alpha S(z)}\Big] \, \delta\big[\alpha \:\! t(z)\big] \, \text{d}z & = \frac{A\:\! e^{i\alpha S(z^\star)}+\closure{A}\:\! e^{-i\alpha S(z^\star)}}{\big\lvert \alpha \:\! t'(z^\star)\big\rvert} \nonumber\\
    &\!\!\!\!\!\!\overset{!}{=} \frac{1}{\big\lvert \alpha \:\! t'(z^\star)\big\rvert} \!\:\cbigg\{\frac{\eta_-\:\!A}{-i\;\! \text{sign}(\alpha)}\:  e^{i\alpha S(z^\star)} + \frac{\eta_+\:\!\closure{A}}{i\;\! \text{sign}(\alpha)}\: e^{-i\alpha S(z^\star)} \cbigg\} \nonumber \\[0.05cm]
    &\!\!\!\!\!\!= \frac{1}{\big\lvert t'(z^\star)\big\rvert} \!\:\cbigg\{\frac{\eta_-\:\!A}{r}\: e^{-r S(z^\star)} + \frac{\eta_+\:\!\closure{A}}{s}\: e^{-s S(z^\star)}  \cbigg\}_{\substack{\scalebox{0.75}{$r\!\!\:=\!\!\:-i\alpha$}\\[0.03cm] \scalebox{0.75}{$s\!\!\:=\!\!\:+i\alpha$}}} \\[0.05cm]
	&\!\!\!\!\!\!= \Bigg\{\mathlarger{\int} \eta_- A\:\! e^{-r S(z)} \;\! \delta\big[r \:\! t(z)\big] + \eta_+\closure{A}\:\! e^{-s S(z)} \;\! \delta\big[s \:\! t(z)\big] \, \text{d}z \Bigg\}_{\substack{r,s\:>\:0 \:\\[0.03cm] \scalebox{0.75}{$r\!\!\:=\!\!\:-i\alpha$}\\[0.03cm] \scalebox{0.75}{$s\!\!\:=\!\!\:+i\alpha$} }} \:\! , \nonumber
\end{align}
where $z^\star$ is assumed to be the sole point in the domain for which $t(z^\star)=0$ holds. As before, we utilized the prescription put forward when discussing equation \eqref{eq:AnalyticCont_Propagators}. The original integral, involving real $\alpha$, is re-expressed using the new parameters $\pm i\alpha$. The arising integrations are then performed as if these combinations $\pm i\alpha$ would be real and positive, after which one analytically continues the result to their actual values. Consequently, we see that $\eta_\pm=\pm i \,\mathrm{sign}(\alpha)$ yields the desired equality between the expressions, which is then na\"{\i}vely extended to the path integral case. Note that for the single integral at hand, we would not require the introduction of two different variables $r$ and $s$, as one could get away with a single one, yielding relation (4.15) provided in reference \cite{SchwartzPrecisionDecayRate}, taking the form
\begin{align}
	\mathrm{Re}\:\!\Bigg\{\mathlarger{\int} A\:\! e^{i\alpha S(z)}\, \delta\big[\alpha \:\! t(z)\big] \, \text{d}z\Bigg\} &= -\, \mathrm{sign}(\alpha)\,\mathrm{Im}\Bigg\{\mathlarger{\int} A\:\! e^{-r S(z)} \;\! \delta\big[r \:\! t(z)\big]\, \text{d}z \Bigg\}_{\substack{\scalebox{0.75}{$r\!\!\:>\!\!\:0 \;\;\;\;$}\\[0.03cm] \scalebox{0.75}{$r\!\!\:=\!\!\:-i\alpha$} }} \, .
\end{align}
Inserting $\eta_+=i$ into our previous intermediate result~\eqref{eq:IntermediateResult_Andreassen} grants us the representation\footnote{Note that Andreassen et al. take the absolute value of the provided expression, arguing there to be a sign ambiguity arising from a branch cut, which they state has to be fixed by physical considerations. While this step is not completely clear to us, it does not interfere with any arguments of the remaining part of the section. However, a first-principles derivation should leave no sign ambiguity, as has been seen in section~\ref{sec:4_3_FormalizingColemansIdeas}, where the demand of outgoing Gamow--Siegert boundary conditions provides the desired sign prescription.} 
\begin{align}
    \frac{\text{d}P_\text{TV}(T)}{\text{d}T}&\approx -2\, \mathrm{Im}\scalebox{1.3}{\Bigg\{} \mathlarger{\mathlarger{\int}}_{\scalebox{0.75}{$z(-\mathcal{T})\!=\!z_\text{FV}$}}^{\scalebox{0.75}{$z(\mathcal{T}')\!=\!z_\text{FV}$}} \mathcal{D}_\mathrm{E}\llbracket z\rrbracket \, \exp\!\!\:\bigg(\!\!\!\:-\!\!\:\frac{S_\mathrm{E}\llbracket z\rrbracket}{\hbar}\bigg) \,  \delta\big(\tau^\star \llbracket z\rrbracket\big) \scalebox{1.3}{\Bigg\}}_{\substack{\scalebox{0.75}{$\mathcal{T},\mathcal{T}'\!\!\:>\!\!\:0$}\\[0.05cm] \scalebox{0.75}{$\mathcal{T}\:\!\!\:=\!\!\:+\!\:iT$}\\[0.05cm] \scalebox{0.75}{$\mathcal{T}'\!\!\:=\!\!\:-\!\:iT$}}} \, .
        \label{eq:FinalResult_Andreassen}
\end{align}
At this point, our result deviates considerably from the one portrayed by Andreassen et al., as there are still two different variables $\mathcal{T}$ and $\mathcal{T}'$ at play, both of which cannot be trivially equated. Comparing to their expression (4.16), they appear to have set $\mathcal{T}=\mathcal{T}'$, continuing both variables back to real time via $\mathcal{T}=iT$, which is however not permitted. The fact that for a single integral one can get away with one variable, as illustrated in their equation (4.15), does not transfer to the relevant case of a product of two path integrals which respectively require a Wick-rotation in opposite direction. Owing to this sign discrepancy, the procedure loses its expressive power, as one would have to solve a Euclidean path integral for two arbitrary positive time scales $\mathcal{T},\mathcal{T}'$, after which one would have to analytically continue the individual parameters in conjugate ways. Due to time translation invariance explicitly broken by the entailed $\delta$-contribution, this will likely be a sophisticated endeavor, as will become clear shortly. Before giving some concluding remarks on their evaluation of the encountered expressions, let us mention some further concerns about their remaining steps, which we also feel need to be addressed: 
\begin{itemize}
    \item While they claim that the numerator $P_\mathrm{FV}(T)$ can be manipulated in the identical manner, this is not too obvious due to the missing time derivative. A na\"{\i}ve attempt to recover their result would amount to rewriting
    \begin{align}
    P_\text{FV}(T)&=\mathlarger{\int}_{\text{FV}}\: \Big\lvert K\!\!\:\big(z_\text{FV},z\:\!;T\big) \Big\rvert^2\, \text{d}z
     \nonumber \\ 
     &=\mathlarger{\int}_{\text{FV}}\: \braket[\cbig]{z_\text{FV}\!\!\:|U^\dagger(T)\!\!\:| z} \braket[\cbig]{z\!\!\:|U(T)\!\!\:| z_\text{FV}} \, \text{d}z
     \nonumber \\  
     &=\cbigg\langle z_\text{FV} \,\cbigg\vert\, U(-T)  \:\! \bigg\{\int_{\text{FV}} \text{d}z \,\ket{z}\bra{z}\bigg\} \, U(T)\!\!\:\, \cbigg\vert  \, z_\text{FV} \cbigg\rangle \nonumber \\[0.03cm] 
     &\approx \braket[\cbig]{z_\text{FV}\!\!\:|U(-T)\!\:U(T)\!\!\:| z_\text{FV}} = \braket{z_\text{FV}\!\!\:| z_\text{FV}} = \delta(0)= P_\text{FV}(0)\, ,
    \end{align}
    once more inferring that we can complete the subspace-projector to a resolution of unity, with contributions arising from $z$ in the TV region assumed to be exponentially suppressed. However, this approximation essentially kills all interesting dynamics in the present case. Nevertheless, the above step seems to be precisely the approach they took, as it results in
    \begin{align}
        P_\text{FV}(0)&=\mathlarger{\int}_{-\infty}^{\infty}\:  K\!\!\:\big(z_\text{FV},z\:\!;T\big)\,\closure{K\!\!\:\big(z_\text{FV},z\:\!;T\big)} \, \text{d}z = \mathlarger{\int}_{-\infty}^{\infty}\:  K\!\!\:\big(z_\text{FV},z\:\!;T\big)\:\! K\!\!\:\big(z,z_\text{FV}\:\!;-T\big) \, \text{d}z \nonumber  \\
        &=\Bigg\{\mathlarger{\int}_{-\infty}^{\infty}\:  K_\mathrm{E}\big(z_\text{FV},z\:\!;\mathcal{T}\big)\:\! K_\mathrm{E}\big(z,z_\text{FV}\:\!;\mathcal{T}'\big) \, \text{d}z \Bigg\}_{\substack{\scalebox{0.75}{$\mathcal{T},\mathcal{T}'\!\!\:>\!\!\:0$}\\[0.05cm] \scalebox{0.75}{$\mathcal{T}\:\!\!\:=\!\!\:+ iT$}\\[0.05cm] \scalebox{0.75}{$\mathcal{T}'\!\!\:=\!\!\:-iT$}}}  \nonumber \\ 
        &= \scalebox{1.3}{\Bigg\{} \mathlarger{\mathlarger{\int}}_{\scalebox{0.75}{$z(-\mathcal{T})\!=\!z_\text{FV}$}}^{\scalebox{0.75}{$z(\mathcal{T}')\!=\!z_\text{FV}$}} \mathcal{D}_\mathrm{E}\llbracket z\rrbracket \, \exp\!\!\:\bigg(\!\!\!\:-\!\!\:\frac{S_\mathrm{E}\llbracket z\rrbracket}{\hbar}\bigg) \scalebox{1.3}{\Bigg\}}_{\substack{\scalebox{0.75}{$\mathcal{T},\mathcal{T}'\!\!\:>\!\!\:0$}\\[0.05cm] \scalebox{0.75}{$\mathcal{T}\:\!\!\:=\!\!\:+ iT$}\\[0.05cm] \scalebox{0.75}{$\mathcal{T}'\!\!\:=\!\!\:-iT$}}} \, , 
    \end{align}
    granting their result modulo the ostensible sign error from equating $\mathcal{T}$ and $\mathcal{T}'$. The above result is simply a consequence of the elementary relation $K_\mathrm{E}\big(z_\text{i},z_\text{f}\:\!;0\big)=\delta(z_\text{i}-z_\text{f})$ and the fact that the arising propagator solely depends on $\mathcal{T}+\mathcal{T'}=0$. While it is indeed reasonable to assume that the survival probability $P_\mathrm{FV}(T)$ remains constant at leading order, the extent to which the given approximation trivializes $P_\mathrm{FV}(T)$ highlights the uncertainty as to whether the similar simplification could previously be applied to $P_\mathrm{TV}(T)$, or whether doing so would likewise render it trivial.
     \item Their equation (4.19) in exposition~\cite{SchwartzPrecisionDecayRate} is incorrect, as the Dirichlet boundary conditions of the path integral explicitly break time translation invariance, rendering $\mathcal{E}_\tau$ $\tau$-dependent. Usually, this would not be a problem, as the underlying time translation symmetry is fully restored in the limit of large $T$. However, for finite $T$, the absence of the exact symmetry is encoded in exponentially small correction terms of the form $e^{-T}$. In the present case, these contributions read $e^{-\mathcal{T}},e^{-\mathcal{T'}}$ and become purely oscillatory upon analytic continuation back to real time, thus they cannot be dropped trivially. The identical issue arises from the fact that Andreassen et al. redefine $T\mapsto T/2$ to bring the total time interval length back to $T$. Again, this step is highly nontrivial in light that the otherwise exponentially small contributions in $\mathcal{T},\mathcal{T}'$ turn oscillatory under the Wick-rotation back to real time. 
\end{itemize}
The second bullet point reveals a last notable concern with their approach. Glossing over the previously stated uncertainty in their argument of supplementing the projectors onto the FV and TV subspaces to a resolution of unity, their derivation can be faithfully reproduced until one arrives at the final expression 
\begin{equation}
	\Gamma_0 \approx -2\, \lim_{T\to\infty}\:\mathrm{Im}\left\{ \frac{\displaystyle{\mathlarger{\mathlarger{\int}}_{\scalebox{0.75}{$z(-\mathcal{T})\!=\!z_\text{FV}$}}^{\scalebox{0.75}{$z(\mathcal{T}')\!=\!z_\text{FV}$}} \mathcal{D}_\mathrm{E}\llbracket z\rrbracket \, \exp\!\!\:\bigg(\!\!\!\:-\!\!\:\frac{S_\mathrm{E}\llbracket z\rrbracket}{\hbar}\bigg) \,  \delta\big(\tau^\star \llbracket z\rrbracket\big)}}{\displaystyle{\mathlarger{\mathlarger{\int}}_{\scalebox{0.75}{$z(-\mathcal{T})\!=\!z_\text{FV}$}}^{\scalebox{0.75}{$z(\mathcal{T}')\!=\!z_\text{FV}$}} \mathcal{D}_\mathrm{E}\llbracket z\rrbracket \, \exp\!\!\:\bigg(\!\!\!\:-\!\!\:\frac{S_\mathrm{E}\llbracket z\rrbracket}{\hbar}\bigg)}} \right\}_{\substack{\scalebox{0.75}{$\mathcal{T},\mathcal{T}'\!\!\:>\!\!\:0$}\\[0.05cm] \scalebox{0.75}{$\mathcal{T}\:\!\!\:=\!\!\:+ iT$}\\[0.05cm] \scalebox{0.75}{$\mathcal{T}'\!\!\:=\!\!\:-iT$}}}\, .
\end{equation}
However, these Euclidean path integrals cannot be evaluated the same way one would go about in usual instanton calculus, hinging crucially on the limit $T\to\infty$ to project onto the ground state. In the present case, the finite-time path integrals have to be evaluated exactly, as only after the Wick-rotation back to real time has been instated do we infer the late-time limit $T\to\infty$. Exponentially small contributions of the form $e^{-\mathcal{T}}$ or $e^{-\mathcal{T}'}$ that one would traditionally ignore from the get-go cannot be dropped in the present case. This renders a convenient evaluation practically infeasible, even in the semiclassical limit $\hbar\to 0^+$. Andreassen et al. sidestep this intricate discussion and simply drop all terms vanishing in the limit $\mathcal{T}\to\infty$ before analytically continuing, which is not permitted without strong supplementary justification. \\

\noindent 
While the presented approach uses some very clever techniques that could come in handy for other questions, in its current state it falls short in providing an alternative way of understanding the instanton method. Researchers that desire to implement the direct method should proceed with caution, ensuring that the previously identified gaps are thoroughly addressed beforehand.

\subsection{Critical perspective on the ``steadyon'' picture by Steingasser \& Kaiser}
\label{sec:5_2_Flaws_Steadyons}

Another approach has been proposed by Steingasser \& Kaiser~\cite{SteingasserRealTimeInstantons}, building upon the ansatz employed by the previously discussed ``direct method'', essentially starting with equation~\eqref{eq:TVProbability_Derivative}. They instate the slight generalization of utilizing an initial state of the form $\Psi_{T=0}(z)=\delta(z-z_\mathrm{i})$, for which $z_\mathrm{i}$ not necessarily agrees with $z_\mathrm{FV}$, adjusting the boundary conditions of the functional integrals and the crossing condition accordingly.\footnote{As stated previously, the $\delta$-distribution should be appropriately regularized in order for the FV probability to be normalizable at $T=0$.} Their overall goal consists in computing the excited-state decay rate pertaining to energy $E=V(z_\mathrm{i})$ solely from the time evolution of the provided initial state, which in itself constitutes a premise that would need a proof and appears at odds with the picture of tunneling reviewed in section~\ref{sec:2_Resonant_State_Introduction}. The key issue is that the given choice for the initial wave function $\Psi_{T=0}(z)=\delta(z-z_\mathrm{i})$ appears ill-suited for capturing the desired excited-state decay rate. Referencing our previous discussion provided at the onset of section~\ref{sec:2_Resonant_State_Introduction}, it is of utmost importance to minimize the sloshing of the initial state within the FV region at the beginning of its evolution. This is required to ensure the prompt onset of the uniform decay phase; otherwise, the rapidly decaying excited state of interest will no longer dominate the time evolution at the intermediate times under investigation, with the FV ground state instead governing the decay dynamics. A more appropriate approach would thus involve initializing with a state that closely resembles an energy eigenstate within the FV basin, essentially skipping the undesirable sloshing phase. In contrast, the peaked initial state employed by Steingasser \& Kaiser, being inherently far from a quasi-stationary resonance state, prolongs the sloshing phase to such an extent that, by the onset of uniform decay, the FV ground state will already dominate the system’s evolution. The authors try to circumvent this problem by clarifying that, despite eventually taking the limit $T\to\infty$, their intention is not to wait for the state to approximately equilibrate. Instead, they conjecture that the desired exponential decay dynamics can identically be captured during the brief temporal periods for which the highly oscillatory state inside the FV region reaches its highest point when approaching the barrier. It is however expected that the underlying complex transient dynamics, dominated by reflections and interference phenomena rather than simple exponential decay, make it virtually impossible to extract the sought-after pure decay rate during these short time intervals. Thus, their central assumption that this particular time evolution allows for capturing of the excited-state decay rate is open to doubt. Glossing over this issue, we turn to their evaluation of the arising real-time path integrals, which, as we shall see, falls somewhat short in providing all the technical steps needed to convincingly support their subsequent claims.\\

\noindent 
Let us focus on the arising points of concern regarding the numerator\\[-0.5cm]
\begin{equation}
  \begin{adjustbox}{width=\linewidth, center}
    \begin{minipage}{\linewidth}
      \begin{align*}
	   \frac{\text{d}P_\text{TV}(T)}{\text{d}T}\approx 2\, \mathrm{Re}\scalebox{1.4}{\Bigg\{} \!\!\left[ \mathlarger{\mathlarger{\int}}_{\scalebox{0.75}{$z(0)\!=\!z_\text{i}$}}^{\scalebox{0.75}{$z(T)\!=\!z_\text{s}$}} \!\!\mathcal{D}\llbracket z\rrbracket \,
		\exp\!\!\:\bigg(\frac{i\:\! S\llbracket z\rrbracket}{\hbar}\bigg) \,  \delta\big(t_\mathrm{s}\llbracket z\rrbracket-T\big)\right] \! \closure{\left[ \mathlarger{\mathlarger{\int}}_{\scalebox{0.75}{$z(0)\!=\!z_\text{i}$}}^{\scalebox{0.75}{$z(T)\!=\!z_\text{s}$}} \!\!\mathcal{D}\llbracket z\rrbracket \,
			\exp\!\!\:\bigg(\frac{i\:\! S\llbracket z\rrbracket}{\hbar}\bigg)\right]}\scalebox{1.4}{\Bigg\}}\, , \\[-0.55cm]
        \end{align*}
    \end{minipage}
  \end{adjustbox}
  \tag{\normalsize \theequation} 
  \label{eq:TVProbability_Derivative_Steingasser}
\end{equation}
the subsequent discussion of the normalization factor $P_\mathrm{FV}(T)$ would give rise to similar questions. In order to compute expression~\eqref{eq:TVProbability_Derivative_Steingasser}, one desires to find the leading-order semiclassical approximation of the two propagators $K\!\!\;\big(z_\text{i},z_\text{s}\:\!;T\big)$ and $\widetilde{K}\!\!\;\big(z_\text{i},z_\text{s}\:\!;T\big)$, with $z_\text{s}$ denoting the classical escape point to energy $E=V(z_\text{i})$. In their work, the authors assert that the traditional propagator and its tilded variant, involving an additional functional $\delta$, possess the identical dominant stationary trajectory, thereby implicitly treating both path integrals as the same object. Let us temporarily set this aspect aside, addressing why a semiclassical evaluation of $\widetilde{K}$ is significantly more involved compared to that of the already demanding real-time propagator $K$ only further down the section. Instead, we first focus on their evaluation of the standard propagator $K\!\!\;\big(z_\text{i},z_\text{s}\:\!;T\big)$, which already poses several questions. Given Dirichlet boundary conditions, one generally finds a countably infinite set of (complex) critical trajectories, satisfying both the Euler--Lagrange equation of motion together with the demanded boundary conditions. Which of those saddle points ultimately contribute to a real-time path integral has remained an open question to this day due to the complicated thimble structure in the complexified function space, constituting an extremely intricate subject as shown by earlier studies, see e.g. Tanizaki \& Koike~\cite{TanizakiLefschetz}. Despite this inherent complexity, Steingasser \& Kaiser make the assertion that, among the infinite set of critical trajectories, the relevant classical motion is given by what they dub the ``steadyon'' (or the ``drag'' when investigating the normalization factor). This configuration essentially constitutes a purely periodic trajectory inside the FV basin with classical energy $V(z_\mathrm{i})$, whose time argument has been Wick-rotated by an infinitesimal amount $\epsilon$ as $t\mapsto e^{-i\epsilon}t$.\footnote{This regulator is required for this nontrivial solution to even show up, as in the strict limit $\epsilon\to 0$, the solution becomes a trivial oscillatory motion inside the FV basin, thus never coming close to the desired end-point $z_\mathrm{s}$. Due to the limit $\epsilon\to 0$ being singular, the introduction of the regulator is indeed justified.} The arising trajectory is seen to eventually cross into the TV region after a time of order $1/\epsilon$, passing close to the classical turning point $z_\mathrm{s}$. Our main difficulty with this solution is the fact that, for general times $T$, the steadyon does not satisfy the correct Dirichlet boundary conditions of the path integral in question.\footnote{Obtaining the critical trajectories faithfully requires the two complex integration constants in guise of the classical energy of the trajectory and a temporal offset, arising from integrating the second-order EOM, to be fixed by the two Dirichlet boundary conditions. While Steingasser \& Kaiser determine the time translation parameter from the condition $z_\mathrm{steadyon}(0)=z_\mathrm{i}$, they simply fix the classical energy of the trajectory to match the energy of the decaying state. Therefore, the second boundary condition $z_\mathrm{steadyon}(T)=z_\mathrm{s}$ is generally not satisfied.} While acknowledging this fact, Steingasser \& Kaiser promote this imprecise assessment of the critical trajectory of the path integral into an inherent feature of their calculation scheme, stating that the desired quantity~\eqref{eq:TVProbability_Derivative_Steingasser} can only be computed for specific, discrete times $T$, which they justify with the oscillation of the initial state inside the FV basin. To meet the required endpoint boundary condition at least approximately, they relate their regulator $\epsilon$ to the length of the time interval $T$, arguing that ``$\epsilon$ is not an entirely free parameter'', while stating that ``for general physical times, no exact solution exists''~\cite{SteingasserRealTimeInstantons}. This assertion is incorrect; instead, a more precise statement would be that for general times $T$ and fixed regulator $\epsilon$, computing and classifying the infinite set of complex solutions to the equations of motion satisfying both boundary conditions proves to be a rather formidable task. Rather than finding the critical trajectories of the path integral in question directly, the authors insist on the existence of a specific class of solutions, retrospectively adjusting the parameters $T$ and $\epsilon$ in order for the desired steadyon solution to constitute an approximate saddle point.\footnote{Their construction of ``non-periodic steadyons'' follows a similar argument. Instead of working with a fixed time interval $T$, they fine-tune $T$ and $\epsilon$ such that a solution starting with an imaginary velocity $\dot{z}_\mathrm{steadyon}(0)\in i\mathbb{R}$ closes in on $z_\mathrm{s}$ at time $T$. When comparing the action of these configurations in their figure 13, they therefore compare trajectories living either on different time intervals $T$ or possess a different regularization parameter $\epsilon$. Hence, we argue that no useful information can be gained from comparing these motions.} This practice gives rise to another potential concern, as regardless of the size of the time interval $T$, the regulator $\epsilon$ must be taken to zero in order to recover the desired real-time propagator. Accordingly, one should not consider the double scaling limit $T\to\infty$, $\epsilon\to0$ with $\epsilon \!\;T=\mathrm{const.}$, as the limit $\epsilon \to 0$ must instead be taken strictly before the limit $T \to \infty$. The authors acknowledge the fact that instating a finite $\epsilon$ inevitably ``implies a finite error'', asserting that their limiting procedure adequately takes care of any undesirable artifacts. However, since combinations of the form $\epsilon \:\! T$ yield non-vanishing contributions that ought to be absent from the correct result, thereby potentially compromising it, this aspect calls for further clarification. Given the reasons discussed above, we question whether the steadyon trajectory truly represents the classical configuration that dominates the real-time propagator. This doubt is further compounded when considering the steadyon motion as the dominant saddle point for the modified propagator $\widetilde{K}$. To this end, note that the introduced $\delta$-functional with its entailed crossing condition is only a sensible object when evaluated on real trajectories.\footnote{Note that the treatment by Andreassen et al.~\cite{SchwartzDirectMethod} never suffered this issue as their critical paths were fully real.} In the present case, a semiclassical evaluation however requires the incorporation of complex trajectories via Picard--Lefschetz theory, for which the crossing condition is ill-defined. The important question which complex paths are to be included in the functional integration and which ones are not is left unaddressed in their work, solely a side remark states that they constrain their steadyon solution to a single period due to the entailed crossing condition. This is insufficient to address this vital point. Since the condition of only receiving contributions from paths that reach $z_\mathrm{s}$ at time $T$ \emph{for the first time} is a nontrivial deformation of the functional integral in the real function space for which the condition applies, its impact of weighting the contributions arising from complex trajectories remains obscure.\\ 

\noindent 
While the steadyon trajectories discussed by Steingasser \& Kaiser are indeed interesting configurations to study, the substantial gaps in their treatment leave it unclear how these classical motions relate to the desired real-time decay dynamics of the provided peaked initial state. Whereas the numerical recovery of the sought-after exponential WKB factor constitutes a promising step, the absence of a treatment of the fluctuation determinant---together with the previously identified ambiguities---makes it difficult to regard the method as a complete first-principles derivation of the excited-state tunneling rate. It would be valuable to see whether their approach can analytically reproduce the known results, including their next-to-leading-order contributions, within the same set of principles.

\subsection{Compilation of other approaches}
\label{sec:5_3_Other_Approaches}

Let us finally address some other recent works that were undertaken into the direction of uncovering the intricacies of the instanton method in relation to real-time computations. Given the vast existing literature, the following list is not intended to be exhaustive.
\begin{itemize}
    \item Early works regarding the topic of instantons in real time, as those by Cherman \& Ünsal~\cite{UnsalRealTimeInstantons} or Ai et al.~\cite{GarbrechtFunctionalMethods}, have concentrated on the ``real-time'' equivalent of bounce motions arising from an analytic continuation of the Euclidean instanton solution. However, the label ``real time'' in that context merely reflects the choice $\theta=0$ in the earlier formulation~\eqref{eq:Ground_State_Energy_Resonant_FullySpelledOut} of the traditional instanton method, and should not be interpreted too literally. Therefore, although it formally involves the evaluation of Lorentzian path integrals, the procedure raised the same unresolved questions about the choice of integration contour as the Euclidean instanton method.
    \item Bender \& Hook~\cite{BenderTunnelingAsAnomaly} focused on classical motions in metastable potentials possessing a complex energy $E$, relating the trajectories arising in the formal limit $\mathrm{Im}(E)\to 0$ to the FV and TV probability of (global) eigenstates of the system. Constituting only a quantitative result, the deeper connection between the different quantities at play has remained obscure.
    \item The works by Tanizaki \& Koike~\cite{TanizakiLefschetz} and Feldbrugge et al.\cite{FeldbruggeRealTimeTunneling} pursued a slightly different motivation, emphasizing the formal structure of real-time path integrals and the characterization of contributing (complexified) saddle points, rather than their connection to Euclidean instanton solutions. Thus, these studies are rooted more in the understanding of Picard--Lefschetz theory than in questions concerning the instanton method itself. 
    \item Another notable attempt by Turok~\cite{TurokRealTimeTunneling} has focused on studying the real-time evolution of the wave function by utilizing the relation 
    \begin{align}
	\Psi\big(z_T,T\big)&=\!\!\:\mathlarger{\int}_{-\infty}^{\infty} \Psi_{T=0}(z_0)\:\! K\!\!\;\scalebox{1.15}{$\big($}z_0,z_T\:\!;T\scalebox{1.15}{$\big)$}\:\mathrm{d}z_0 \, .
    \end{align} 
    Obtaining the critical trajectory that stationarizes both the real-time propagator as well as the endpoint integral over $z_0$ in the dual limit $T\to\infty$, $z_T\to\infty$ resulted in the correct exponential suppression factor of the wave function far outside the FV region. While certainly constituting a big step into the right direction, it is not yet clear how this result connects to the decay of the survival probability $P_\mathrm{FV}(T)$ for increasing $T$, which requires the evaluation of yet another endpoint-integration over $z_T$. Additionally, one would need to keep the time interval $T$ variable in order to study the rate of change, turning the full problem into a formidable task. Strides in this direction have been made using numerical tools, see e.g.~\cite{NishimuraRealTimeTunneling}. 
    \item Blum \& Rosner~\cite{BlumRealTimeTunneling} studied critical trajectories of the real-time propagator $K\!\!\:\big(z_\mathrm{FV},z\:\! ; T\big)$ with $z\geq z_\mathrm{escape}$. Due to assuming the classical energy $E$ of the relevant motion to be small, to leading-order in $E$, they simply arrive at the Wick-rotated versions of time-translated bounce motions discussed in references~\cite{UnsalRealTimeInstantons,GarbrechtFunctionalMethods}.\footnote{Note that the case $z> z_\mathrm{escape}$ arises from an imaginary time translation.} Due to their finite energy, these solutions are regular, however their relation to the time evolution of $P_\mathrm{FV}(T)$ was not clarified. Similarly to Turok's work, there is a sophisticated discussion missing on how these ideas influence the derivation of the full survival probability $P_\mathrm{FV}(T)$ and ultimately the decay rate $\Gamma_0$, as solely investigating the propagator or the wave function only provides a partial picture. Additionally, the previously shared objection to the na\"{\i}ve evaluation of such a real-time propagator given in section~\ref{sec:5_2_Flaws_Steadyons} applies similarly to the present case. While selecting a critical trajectory that reproduces the desired exponential suppression factor may seem effective, doing so without a rigorous justification offers little clarification of the theoretical basis of the instanton method.
    \item A recent work by Harada et al.~\cite{HaradaWKB} reexamines the common free energy method for obtaining decay rates at finite (inverse) temperature $\beta$, claiming that non-periodic shot solutions constitute relevant saddle points in the semiclassical evaluation of the partition function. They state that the associated functional integral solely runs over paths $z(t)$ possessing periodic endpoints $z(0)=z(\hbar\beta)$, which however does not rule out paths which are non-periodic in their velocity, i.e. satisfy $\dot{z}(0)\neq \dot{z}(\hbar\beta)$. Under this premise, they argue that classical, non-periodic shot-like solutions also constitute critical points of the partition function, which had been dismissed in previous considerations. While their initial statement regarding periodicity is clearly correct, their drawn conclusion that shot-like solutions constitute admissible critical points of the functional is not. When carefully deriving the Euler--Lagrange equation in the present case, the required integration by parts spawns additional boundary terms. Demanding stationarity of the functional in question not only requires the typical equation of motion to be fulfilled, but also the emerging boundary terms to vanish, leading to a so-called \emph{transversality condition} arising from the fact that the endpoints are not fixed~\cite{CalcOfVariations}. In the desired case of periodic boundary conditions $z(0)= z(\hbar\beta)$, this supplementary constraint on the critical trajectory $z_\mathrm{crit}(t)$ takes the simple form $\dot{z}_\mathrm{crit}(0)= \dot{z}_\mathrm{crit}(\hbar\beta)$, see e.g the lucid statement surrounding equation (2.16) in reference~\cite{LapedesComplexPI}, ruling out relevant contributions from non-periodic shot-like solutions. Instead, the only relevant real critical configurations of the partition function are the trivial solutions $z_\mathrm{FV}(t)=z_\mathrm{FV}$ and $z_\mathrm{TV}(t)=z_\mathrm{TV}$ as well as fully periodic bounce motions. Thus, the traditional procedure of ignoring non-constant, shot-like trajectories has always been the correct approach. 
    
    \noindent 
    Yet another criticism concerns their comparison of the zero-temperature decay rate results arising from the free energy method to the associated WKB result for the ground-state decay rate. The NLO WKB result provided by them is incorrect, as it arises by wrongfully utilizing linear turning point formulas when matching the WKB solutions around the quadratic FV turning point $z_\mathrm{FV}$.\footnote{Note that their provided representation of the NLO WKB decay rate is taken from Andreassen et al.~\cite{SchwartzPrecisionDecayRate}. However, Andreassen et al. themselves explicitly acknowledge the incompleteness of this result directly below the expression---a caveat that was apparently overlooked by Harada et al.} Further clarifications on this misconception for the similar case of quantum-mechanical level splittings have been provided by Garg~\cite{GargTunnelingRevisited}. Indeed, a correct treatment would find the decay rate results arising from Callan \& Coleman's instanton method and WKB to fully agree, see e.g. appendix A of our previous work~\cite{WagnerExcitedStateTunneling} and references therein. Similarly, the free energy method also provides the correct ground-state decay rate in the limit $\beta\to \infty$, which their computation cannot recover.  

    Despite these shortcomings, it is worth noting that in a follow-up paper~\cite{YinTunneling}, Yin correctly conjectures the connection between the instanton method and Gamow--Siegert boundary conditions.
\end{itemize}
We conclude by briefly mentioning additional noteworthy works that touch on related aspects, albeit with less direct relevance to the traditional instanton method, which are thus not discussed in detail. An early paper by Lapedes \& Mottola~\cite{LapedesComplexPI} investigated the extraction of resonant energy states using functional techniques, strongly emphasizing the relevant outgoing Gamow--Siegert boundary conditions in the process. The authors similarly deform the initial metastable potential to pronounce this feature of resonant states, subsequently picking only a subset of critical trajectories for inclusion in their path integral calculation. While this work yields an alternative procedure to compute decay rates, to the best of our knowledge, it does not relate the different existing procedures in an illuminating way. Other works by e.g. Carlitz \& Nicole~\cite{CarlitzClassicalPaths} or Weiss \& Haeffner~\cite{WeissHaeffnerMetastableDecayFiniteTemp} have focused on deforming the $T$-integration contour when dealing with the integral
kernel of the quantum-mechanical resolvent, i.e. the Fourier-transformed real-time propagator, providing yet another alternative for arriving at the sought-after decay rates. Once more, the procedure's relation to other methods is unclear. McLaughlin~\cite{McLaughlinComplexTimeMethod} generalized the conventional path integral treatment from a purely Wick-rotated time contour to arbitrary temporal contours, utilizing the emerging functional integrals to compute barrier penetration rates. Despite following a rather similar idea to our work by generalizing the conventional path integral formulations in quantum mechanics to more demanding situations, his work again seems to have relatively little overlap with the ordinary instanton method.

\newpage
\section{Conclusion}
\label{sec:6_Conclusion}

In this work, we have provided a detailed account of how the traditional instanton method, as formulated by Callan \& Coleman, relates to the pivotal concept of resonant states. These long-lived quasi-bound states mimic the approximate real-time behavior of a wave function during the intermediate temporal regime in which the quantum-mechanical decay remains uniform enough to be meaningfully studied. As such, they form the backbone of virtually all decay rate calculations employing the traditional wave function picture. Their characteristic outgoing Gamow--Siegert boundary conditions are imprinted in the fact that the associated eigenvalue problem is not defined on the real axis, but rather in angular sectors of the complex plane, bordering the real line in all spatial directions for which barrier penetration effects are relevant. To efficiently address this broader class of boundary conditions, we have extended the standard notion of the quantum-mechanical propagator to encompass such generalized eigenvalue problems, leading to the universally applicable master formulas~\eqref{eq:Final_Relation_Propagator} and~\eqref{eq:Final_Relation_Trace}. \\ 

\noindent 
Straightforwardly applying these fundamental relations to the case of determining the sought-after resonant ground state energy, the functional contour deformation prescribed by Callan \& Coleman arises naturally from the fact that when dealing with unbounded potentials admitting resonant states, the real function space $\mathcal{C}\big([0,T]\big)$ is not an admissible integration cycle. Instead, the proper contour $\mathcal{C}\big([0,T],\Gamma\big)$ is intrinsically distorted into the upper complex half-plane, thereby picking up the vital imaginary contributions from the critical bounce saddle. In cases where the original potential features a global minimum alongside the metastable FV region, we have shown that shot-like solutions are indeed excluded from the decay rate calculation, a consequence of the requirement to destabilize the potential before being able to extract the desired resonant eigenenergies. The provided explanation offers new insight into the inner workings of the instanton method, which previously rested on heuristic arguments and had not yet been placed on firm mathematical footing. As we have argued in section~\ref{sec:5_Assessment_Earlier_Attempts}, prior attempts to elucidate the instanton procedure have fallen short of providing a compelling explanation. In contrast, our approach offers a clear and coherent account of the method. Despite these conceptual advancements, the question of how to rigorously establish the full thimble decomposition of the generalized contour $\mathcal{C}\big([0,T],\Gamma\big)$ in the complexified function space $\mathcal{C}^\mathbb{C}\big([0,T]\big)$ remains an open problem. This particularly involves the question of how multi-bounce configurations enter the na\"{\i}ve thimble decomposition illustrated in figure~\ref{fig:Tunneling_Thimble_Structure_unstable}. We hope that ongoing advances in Picard--Lefschetz theory will ultimately shed light on this pending issue.\\ 

\noindent 
Apart from the provided insights into the instanton method, the developed techniques could offer a deeper understanding of other areas of theoretical physics relying on generalized Schrödinger-type eigenvalue problems. One particularly promising application of the framework is its natural treatment of $\mathcal{PT}$-symmetric eigenvalue problems, enabling the extraction of spectral information using purely functional methods. We anticipate that this riveting direction will be fruitfully explored in future work.

\vfill

\subsection*{Acknowledgments}
N.W. owes special thanks to Soo-Jong Rey for preliminary discussions that helped catalyze the ideas developed in this work. The authors furthermore acknowledge Thomas Steingasser for insightful discussions regarding the steadyon method. N.W. is supported by the German Academic Scholarship Foundation, the Marianne-Plehn-Program of the Elite Network of Bavaria, as well as the International Max Planck Research School on Elementary Particle Physics (IMPRS EPP), hosted by the Max Planck Institute for Physics. N.W. extends his gratitute to the Mainz Institute for Theoretical Physics (MITP) of the Cluster of Excellence $\text{PRISMA}^+$ (Project ID 39083149) for its hospitality during the inspiring early stages of this work.

\newpage 
\appendix 

\section{Specialized (niche) path integral techniques}
\label{sec:A_PathIntegralTechniques}

This appendix chapter revisits lesser-known but indispensable details in the derivation of the path integral representation, which bear material consequences when dealing with non-standard Hamiltonians, such as the one portrayed in equation~\eqref{eq:NonStandard_Hamiltonian}. Whereas the subsequently presented techniques can be treated as peripheral for the vast majority of cases, they are crucial in the theory of stochastic calculus~\cite{ItoIntegral,StratonovichIntegral,LangouchePathIntegralTechniques,LangoucheFunctionalIntegration} and for path integrals on curved manifolds~\cite{deWittCurvedSpaces,McLaughlinSchulmanCurvedSpaces,HirshfeldCovariantPI,WeissGravPathIntegral}, for which the arising caveats have been studied extensively. Identical techniques are required for discussing point canonical transformations in path integrals, see e.g.~\cite{deWittPointTransformations,GervaisPointCanonicalTrafos}. Since most HEP-theorists presumably have had relatively little contact with these methods, we will provide a rather detailed analysis in order for the work to be self-contained.\\

\noindent 
When dealing with non-standard Hamiltonians, possessing kinetic terms with explicit position-dependence, both the derivation of the intermediate discretized path integral as well as attaining its associated continuum limit are more elaborate than it might appear at first glance. For brevity, let us focus on the desired single-dimensional case, for which we want to obtain the path integral representation of the (partially Wick-rotated) amplitude $\smash{\braket[\rbig]{x_\mathrm{f}\!\!\:| \:\! \widehat{U}_\theta(T) \!\!\:|x_\mathrm{i}}}$. In the section at hand, we abbreviate the analytically continued time evolution operator by $\smash{\widehat{U}_\theta(t)\coloneqq e^{-ie^{-i\theta}\widehat{H}t}}$, while deviating from the main body of text in the use of the real position variable $x$. We will subsequently examine the most general case involving a Hamiltonian that is quadratic in the momenta\footnote{As introduced in the main body of text, we highlight operators like the Hamiltonian $\widehat{H}$ with hats, whereas for functions of operators as e.g. $H\big(\hat{x},\hat{p}\big)$ the hat is waived for the function itself.}
\begin{align}
	\widehat{H}=H\big(\hat{x},\hat{p}\big) = f(\hat{x}) \,\hat{p}^2+ g(\hat{x}) \,\hat{p}+ h(\hat{x})\, ,
    \label{eq:GeneralQuadraticHamiltonian}
\end{align} 
assuming the largely arbitrary functions $f$, $g$ and $h$ to be real-valued. All subsequent notions can be easily generalized to encompass complex coefficient functions, solely requiring convergence of the arising integrals. As usual, the path integral derivation hinges on the position and momentum completeness relations~\eqref{eq:CompletenessRelations_PosMom}. Inserting sets of position eigenstates into a time-sliced version of the time evolution operator, one arrives at the intermediate expression
\begin{align}
	K_\theta\big(x_\mathrm{i},x_\mathrm{f}\:\! ; T\big) \coloneqq \braket[\rbig]{x_\mathrm{f}\!\!\:| \:\! \widehat{U}_\theta(T) \!\!\:|x_\mathrm{i}} = \!\lim_{N \to \infty} \mathlarger{\mathlarger{\int}}_{-\infty}^{\infty} \, \mathlarger{\prod}_{k=1}^{N-1} \:\!\mathrm{d}x_{k} \; \Bigg\{\prod_{\ell=1}^N K_\theta\big(x_{\ell-1},x_{\ell}\:\! ; \Delta t\big)\Bigg\}_{\!\substack{\Delta t\:\!=\:\!T/N\\[0.05cm] \,x_0\:\!=\:\!x_\mathrm{i}\;\;\;\, \\ x_N\:\!=\:\!x_\mathrm{f}\;\;\;\,}} \, .
    \label{eq:SplitPathIntegral}
\end{align}
Note that we do not dress the propagator with additional factors here as was done in the main body of text, see equation~\eqref{eq:Definition_Propagator}. The arising infinitesimal propagators $\braket[\rbig]{x\!\!\:| \:\!\widehat{U}_\theta(\Delta t) \!\!\:|x-\Delta x}$ (defining $\Delta t \coloneqq T/N>0$) then constitute the starting point for the path integral treatment. To this end, let us proceed by studying the behavior of the infinitesimal propagators in the limit $\Delta t\to 0^+$, initially by using a very na\"{\i}ve approximation scheme outlined in section~\ref{sec:NaiveEstimate}. Afterwards, we will confirm that a rigorous derivation indeed reproduces the identical result obtained by the previously portrayed flawed argument, spanning subsections~\ref{sec:RigorousEstimate} to~\ref{sec:CollectingResults}. Keeping notational clutter at bay, we chose to set $\hbar=1$ throughout the given section.

\subsection{Na\"{\i}ve derivation of the infinitesimal propagator} 
\label{sec:NaiveEstimate}

A frequently portrayed method to arrive at the leading-order expression of the infinitesimal propagator is to expand the exponential time evolution operator to leading-order in $\Delta t$, followed by evaluating the leftover matrix element and re-exponentiating the expression. 
Inserting a full set of momentum eigenstates, this would amount to 
\begin{align}
	K_\theta\big(x-\Delta x,x\:\!;\Delta t\big)\, &=\mathlarger{\mathlarger{\int}}_{-\infty}^{\infty}\frac{\mathrm{d}p}{2\pi} \: \braket[\cBig]{x|\,\exp\!\Big[\!-\!\!\:ie^{-i\theta}\Delta t\, H\!\!\:\big(\hat{x},\hat{p}\big)\Big]\!\!\:|\,p} \braket[\rbig]{p\!\!\:|x-\Delta x} \nonumber \\[0.05cm] 
    &=\mathlarger{\mathlarger{\int}}_{-\infty}^{\infty}\frac{\mathrm{d}p}{2\pi} \: \braket[\cBig]{x|\,1-ie^{-i\theta}\Delta t\, H\!\!\:\big(\hat{x},\hat{p}\big)+\mathcal{O}\big(\Delta t^2\big)\!\!\:|\,p} \braket[\rbig]{p\!\!\:|x-\Delta x} \nonumber \\[0.05cm] 
    &=\mathlarger{\mathlarger{\int}}_{-\infty}^{\infty}\frac{\mathrm{d}p}{2\pi} \: \Big[1-ie^{-i\theta}\Delta t\, H\!\!\:\big(x,p\big)+\mathcal{O}\big(\Delta t^2\big)\Big] \exp\!\big(ip\:\! \Delta x\big)\nonumber \\\
    &= \mathlarger{\mathlarger{\int}}_{-\infty}^{\infty}\frac{\mathrm{d}p}{2\pi} \: \bigg\{\exp\!\Big[\!-\!\!\:ie^{-i\theta}\Delta t\, H\!\!\:\big(x,p\big)+ip\:\! \Delta x\Big] +\mathcal{O}\big(\Delta t^2\big)\bigg\}\, .
    \label{eq:InfinitesimalPropagatorMomentumInsertion}
\end{align} 
While this expansion is correct up to the given point, it is incorrect to infer that the $\mathcal{O}\big(\Delta t^2\big)$ terms entering the integrand also yield overall contributions to order $\Delta t^2$ after performing the leftover momentum integral, essentially failing due to the fact that the momentum-operator is unbounded. In the following subsection~\ref{sec:RigorousEstimate} we will see that the momentum scale from which the dominant contributions of the integral arise is $p\sim \Delta t^{-1/2}$, thus $p$ itself possesses a nontrivial $\Delta t$-counting. Ignoring this caveat for now, a na\"{\i}ve approximation then drops all $\mathcal{O}\big(\Delta t^2\big)$ already at the level of the integrand, thus concluding with the (over-)simplified result 
\begin{align}
    \!\!K_\theta^{(\text{na\"{\i}ve})}\!\!\:\big(x\!\!\:-\!\!\:\Delta x,x\:\!;\Delta t\big)&=\cbig[4\pi ie^{-i\theta}f(x) \Delta t\cbig]^{\!\!\: -\frac{1}{2}}\: \exp\Bigg\{\frac{ie^{i\theta}\Delta t}{4f(x)}\bigg[\frac{\Delta x}{\Delta t}-e^{-i\theta}g(x)\bigg]^2 \label{eq:NaiveInfPropagator} \\[-0.1cm] 
    &\qquad\qquad\qquad\qquad\qquad\qquad\qquad\qquad\quad\;\; -ie^{-i\theta}\Delta t\, h(x)\Bigg\} + \mathcal{O}\big(\Delta t^2\big) \, , \nonumber
\end{align}
where we inserted the generic Hamiltonian~\eqref{eq:GeneralQuadraticHamiltonian}. While the instated simplification is not admissible per se, leading to an erroneous short-time propagator, when inserting the approximation~\eqref{eq:NaiveInfPropagator} into relation~\eqref{eq:SplitPathIntegral}, the limit $N\to \infty$ takes care of the initially flawed terms, resulting in correct expressions for the full, macroscopic propagator $K_\theta\big(x_\mathrm{i},x_\mathrm{f}\:\! ; T\big)$. The subsequent sections~\ref{sec:RigorousEstimate} to~\ref{sec:CollectingResults} detail how this miraculous result emerges from a rigorous derivation.

\subsection{Rigorous derivation of the infinitesimal propagator}
\label{sec:RigorousEstimate}

Whereas in the conventional quantum-mechanical case $f(x)=\text{const.}$ and $g(x)=0$, the previous simplification~\eqref{eq:NaiveInfPropagator} can be justified by invoking a Trotter decomposition~\cite{TrotterProduct} to split the kinetic and potential terms into two separate exponential pieces, allowing for a trivial evaluation, the general Hamiltonian~\eqref{eq:GeneralQuadraticHamiltonian} with its position-dependent kinetic terms demands a more sophisticated approach. An elegant method of studying the infinitesimal propagator to any desired order is provided by the $\Delta t^{1/2}\,$-scaling method conceptualized by Graham~\cite{GrahamPathIntegralMethods1,GrahamPathIntegralMethods2}, for further expositions of the method see e.g.~\cite{WeissGravPathIntegral,LangouchePathIntegralTechniques}. Given a Hamiltonian $H\big(\hat{x},\hat{p}\big)$ quadratic in the momenta, the broken line path entering the discretized path integral obeys the Brownian motion-like scaling relation $\Delta x^2\sim \Delta t$, which was already pointed out by Feynman in his seminal work introducing path integration~\cite{FeynmanPathIntegrals}. To see this, one exploits the fact that after performing the Gaussian momentum integrals, one ends up with terms of the schematic form $e^{ie^{i\theta}\Delta t \!\:(\Delta x/\Delta t)^2}$, whose dominant contributions to the intermediate position integrals arise when the exponent is of order unity.\footnote{The identical argument applies to the conventional real-time case, leading to the purely oscillatory behavior $\smash{e^{i\Delta t \!\:(\Delta x/\Delta t)^2}}$. Invoking the method of stationary phase again reveals that the dominant contributions arise from the integration region for which the characteristic size relation $\Delta x^2\sim \Delta t$ holds.} When splitting the initial amplitude into $N\gg 1$ short-time propagators $K_\theta\big(x\!\!\:-\!\!\:\Delta x,x\:\!;\Delta t\big)$, one needs to compute the individual pieces up to order $\Delta x^2\sim \Delta t \sim 1/N$, with terms of subleading order vanishing when subsequently taking the continuum limit $N\to\infty$.  \\

\noindent 
Once more, the first step consists in inserting a full set of momentum eigenstates such that each infinitesimal propagator takes the form portrayed in the first line of equation~\eqref{eq:InfinitesimalPropagatorMomentumInsertion}, which we now want to expand in powers of $\Delta t \ll 1$ properly. With the Hamiltonian being quadratic in $p$, the emerging Gaussian momentum integral receives its dominant contributions from the parametric region $p\sim \Delta t^{-1/2}$, such that it is convenient to rescale both the momentum operator $\hat{p}$ as well as $p$ itself by this characteristic scale~\cite{GrahamPathIntegralMethods1,GrahamPathIntegralMethods2}. This makes the scaling manifest and renders the contributing momenta to be of order unity, allowing for an illuminating power counting. Denoting the rescaled momentum by $k$ and the associated rescaled momentum operator by $\hat{k}$, we arrive at the expression 
\begin{align}
	K_\theta\big(x-\Delta x,x\:\!;\Delta t\big)\, =\Delta t^{-\frac{1}{2}}\mathlarger{\mathlarger{\int}}_{-\infty}^{\infty}\frac{\mathrm{d}k}{2\pi} \: \braket[\cBig]{x|\,\exp\!\Big[\!-\!\!\:ie^{-i\theta} H_\mathrm{res}\big(\hat{x},\hat{k}\big)\Big]\!\!\:|\,k}_{\!\mathrm{res}} \braket[\rbig]{k\!\!\:|x-\Delta x}_\mathrm{res} \, ,
    \label{eq:RescaledPropagatorIntegral}
\end{align} 
defining the rescaled Hamiltonian $\widehat{H}_\mathrm{res}$ as 
\begin{align}
	\widehat{H}_\mathrm{res}=H_\mathrm{res}\big(\hat{x},\hat{k}\big) = f(\hat{x}) \,\hat{k}^2+ \Delta t^{\frac{1}{2}}\:\! g(\hat{x}) \,\hat{k}+ \Delta t \, h(\hat{x})\, .
\end{align}
Beware that the new momentum operator $\hat{k}$ satisfies the altered commutation relation $[\hat{x},\hat{k}]=i \Delta t^{1/2}$, leading to the relevant equality $\smash{\braket{x\!\!\:|k}_\mathrm{res}=e^{i \Delta t^{-\!\!\;1\!\!\:/\!\!\;2} kx}}$. We are now fit to expand the exponential operator in $\Delta t$, keeping only the terms up to linear order. Since each commutator introduces additional factors of $\Delta t^{1/2}$, this expansion naturally truncates, granting us the ability to approximate the arising integral~\eqref{eq:RescaledPropagatorIntegral} accordingly. Utilizing the definition of the operator exponential as a power series, the overarching goal consists in obtaining the first orders in the normal ordered expansion of each monomial, written in the form 
\begin{align}
	\cbig(\widehat{H}_\mathrm{res}\cbig)^{\!\!\: n}=\Big[H_\mathrm{res}\big(\hat{x},\hat{k}\big)\Big]^n \eqqcolon H_n^{(0)}\big(\hat{x},\hat{k}\big) + \Delta t^{\frac{1}{2}}H_n^{(1)}\big(\hat{x},\hat{k}\big) + \Delta t\,H_n^{(2)}\big(\hat{x},\hat{k}\big) + \mathcal{O}\cbig(\Delta t^{\frac{3}{2}}\cbig)\, .
\end{align}
In order to determine the exact form of the operators $\widehat{H}_n^{(0,1,2)}$, one introduces recurrence relations between these operators and their pendants to lower $n$. In a final step, we resum each order of the exponential series to obtain a convenient normal ordered expression for the rescaled time evolution operator $\smash{\widehat{U}_{\theta,\mathrm{res}}\coloneqq e^{-ie^{-i\theta} \widehat{H}_\mathrm{res}}}$ to the desired order. As explained, let us start by finding recurrence relations between the different coefficients. We can relate the operators $\widehat{H}_n^{(0,1,2)}$ to $\widehat{H}_{n-1}^{(0,1,2)}$ by utilizing the equality
\begin{align}
	\cbig(\widehat{H}_\mathrm{res}\cbig)^{\!\!\: n}=\widehat{H}_\mathrm{res}\cbig(\widehat{H}_\mathrm{res}\cbig)^{\!\!\: n-1} &\doteq \widehat{H}^{(0)}_n+ \Delta t^{\frac{1}{2}}\:\! \widehat{H}^{(1)}_n + \Delta t\,\widehat{H}^{(2)}_n \nonumber \\ 
    &\doteq \Big[f(\hat{x}) \,\hat{k}^2+ \Delta t^{\frac{1}{2}}g(\hat{x}) \,\hat{k}+ \Delta t\, h(\hat{x})\Big] \Big[\widehat{H}^{(0)}_{n-1}+ \Delta t^{\frac{1}{2}}\:\! \widehat{H}^{(1)}_{n-1} + \Delta t\,\widehat{H}^{(2)}_{n-1}\Big] \nonumber \\
	&\doteq  f(\hat{x}) \cbig[\hat{k}^2, \widehat{H}^{(0)}_{n-1}\cbig] + \,\Delta t^{\frac{1}{2}}\:\! g(\hat{x}) \cbig[\hat{k},\widehat{H}^{(0)}_{n-1}\cbig] + \, \Delta t^{\frac{1}{2}} \:\! f(\hat{x}) \cbig[\hat{k}^2,\widehat{H}^{(1)}_{n-1} \cbig] \nonumber \\
	&\quad + f(\hat{x})\:\! \widehat{H}^{(0)}_{n-1}\,\hat{k}^2 + \Delta t^{\frac{1}{2}}\:\! g(\hat{x}) \:\! \widehat{H}^{(0)}_{n-1}\,\hat{k} + \Delta t^{\frac{1}{2}}\:\! f(\hat{x}) \:\! \widehat{H}^{(1)}_{n-1}\,\hat{k}^2\nonumber \\
	&\quad + \Delta t \, h(\hat{x}) \:\! \widehat{H}^{(0)}_{n-1} + \Delta t \, g(\hat{x}) \:\!\widehat{H}^{(1)}_{n-1}\, \hat{k} + \Delta t\, f(\hat{x})\:\! \widehat{H}^{(2)}_{n-1}\, \hat{k}^2 \, ,
\end{align}
readily dropping all terms of order $\Delta t^{3/2}$, for which we adopt DeWitt's notation of labeling equality in the path integral sense with the symbol $\doteq$~\cite{deWittCurvedSpaces}.\footnote{Since the higher-order contributions in $\Delta t$ will eventually drop out when taking the continuum limit, both sides can be truncated at the given order, leaving the emerging discretized path integral invariant. The symbol $\doteq$ hereby symbolizes equality after having taken the respective continuum limit $N\to \infty$ ($\Delta t\to 0^+$), manifesting the subleading terms to drop out.} Note that we also ignored the terms proportional to $\Delta t$ which entail an additional commutator, as each commutator constitutes at least a factor of $\Delta t^{1/2}$. The remaining commutators can be computed by utilizing the commutation relations
\begin{subequations}
\begin{align}
	\Big[\hat{k}, O\big(\hat{x},\hat{k}\big)\Big]&= -i\:\!\Delta t^{\frac{1}{2}} \,\frac{\partial}{\partial \hat{x}}\, O\big(\hat{x},\hat{k}\big) \, , \\
	\Big[\hat{k}^2, O\big(\hat{x},\hat{k}\big) \Big]&=-2i\:\!\Delta t^{\frac{1}{2}} \,\frac{\partial}{\partial \hat{x}}\, O\big(\hat{x},\hat{k}\big)\, \hat{k}-\Delta t\,\frac{\partial^2}{\partial \hat{x}^2}\,  O\big(\hat{x},\hat{k}\big)\, ,
\end{align}
\end{subequations}
with the terms readily brought into normal order, assuming the initial operator $\widehat{O}$ to be normal ordered itself. Collecting the terms contributing to each order leads to the desired recurrence relations
\begin{subequations}
\begin{align}
	\widehat{H}^{(0)}_n&=f(\hat{x})\:\! \widehat{H}^{(0)}_{n-1}\,\hat{k}^2 \, ,\\
	\widehat{H}^{(1)}_n&=f(\hat{x})\:\! \widehat{H}^{(1)}_{n-1}\,\hat{k}^2+g(\hat{x})\:\! \widehat{H}^{(0)}_{n-1}\,\hat{k}-2i f(\hat{x})\:\!\frac{\partial \widehat{H}^{(0)}_{n-1}}{\partial \hat{x}}\, \hat{k} \, , \\[0.15cm]
	\widehat{H}^{(2)}_n&= f(\hat{x})\:\! \widehat{H}^{(2)}_{n-1}\,\hat{k}^2+ g(\hat{x})\:\! \widehat{H}^{(1)}_{n-1}\,\hat{k} + h(\hat{x})\:\! \widehat{H}^{(0)}_{n-1}  \nonumber \\ 
	&\quad -f(\hat{x})\:\!\frac{\partial^2 \widehat{H}^{(0)}_{n-1}}{\partial \hat{x}^2}-i\:\! g(\hat{x})\:\!\frac{\partial \widehat{H}^{(0)}_{n-1}}{\partial \hat{x}}-2i f(\hat{x})\:\!\frac{\partial \widehat{H}^{(1)}_{n-1}}{\partial \hat{x}}\, \hat{k}\, ,
\end{align}
\end{subequations}
which can be solved consecutively, using the initial values $\widehat{H}^{(0)}_0=1, \widehat{H}^{(1)}_0=0$ and $\widehat{H}^{(2)}_0=0$. After some cumbersome algebra, one arrives at the final results
\begin{subequations}
\begin{align}
    \widehat{H}^{(0)}_n &= f(\hat{x})^n \,\hat{k}^{2n}\, , \\[0.15cm]
    \widehat{H}^{(1)}_n&=n\:\!f(\hat{x})^{n-1}\Big[g(\hat{x})-(n-1)i f'(\hat{x})\Big]\:\!\hat{k}^{2n-1}\, , \\[0.15cm]
    \widehat{H}^{(2)}_n&=n\:\! f(\hat{x})^{n-2}\Bigg\{f(\hat{x})h(\hat{x})+\frac{(n-1)}{2}\bigg[g(\hat{x})^2-(2n-3)\,if'(\hat{x}) g(\hat{x}) \nonumber\\[-0.15cm] 
	&\qquad\qquad\qquad  - \frac{4n-5}{3}\,f(\hat{x})f''(\hat{x})-(n-1)(n-2)\, f'(\hat{x})^2 -2i f(\hat{x}) g'(\hat{x}) \bigg]\!\!\:\Bigg\}\,\hat{k}^{2n-2}\, , 
\end{align}
\end{subequations}
which are easily verified by plugging the solutions back into the recurrence relation. 
At this point we may resum the expressions, with the full (analytically continued) time evolution operator $\widehat{U}_{\theta,\mathrm{res}}$ represented as a power series. Once more skipping the fairly straightforward algebra, one obtains the desired result
\begin{align}
	\!\!\widehat{U}_{\theta,\mathrm{res}}^{(0)}&\coloneqq \mathlarger{\sum}_{n=0}^{\infty} \, \frac{\big(\!-\!\!\:ie^{-i\theta}\big)^{\!\!\:n}}{n!}\,\widehat{H}^{(0)}_n =\normordbig \exp\cbig[-ie^{-i\theta}f(\hat{x})\:\!\hat{k}^2\cbig] \normordbig \; , \\[0.15cm] 
	\!\!\widehat{U}_{\theta,\mathrm{res}}^{(1)}&\coloneqq\mathlarger{\sum}_{n=0}^{\infty} \, \frac{\big(\!-\!\!\:ie^{-i\theta}\big)^{\!\!\:n}}{n!}\:\!\widehat{H}^{(1)}_n
	=-ie^{-i\theta}\!\: \normordbig \Big[g(\hat{x})\,\hat{k}-e^{-i\theta} f(\hat{x})f'(\hat{x})\,\hat{k}^3\Big]\exp\cbig[-ie^{-i\theta}f(\hat{x})\,\hat{k}^2\cbig]\normordbig\, , \\[0.15cm] 
	\!\!\widehat{U}_{\theta,\mathrm{res}}^{(2)}&\coloneqq\mathlarger{\sum}_{n=0}^{\infty} \, \frac{\big(\!-\!\!\:ie^{-i\theta}\big)^{\!\!\:n}}{n!}\:\!\widehat{H}^{(2)}_n \nonumber \\ 
    &= -ie^{-i\theta}\, \normordbig \Bigg\{h(\hat{x})-\frac{ie^{-i\theta}}{2}\,g(\hat{x})^2\hat{k}^2-\frac{e^{-i\theta}}{2}\,f'(\hat{x})g(\hat{x})\cbig[1-2ie^{-i\theta}f(\hat{x})\hat{k}^2\cbig]\hat{k}^2 \nonumber \\[-0.15cm]
    &\qquad\qquad\qquad\qquad\!\!\!-e^{-i\theta} f(\hat{x}) g'(\hat{x}) \hat{k}^2 + \frac{ie^{-i\theta}}{6}\,f(\hat{x})f''(\hat{x})\cbig[3-4ie^{-i\theta}f(\hat{x})\hat{k}^2\cbig]\hat{k}^2 \nonumber \\ 
    &\qquad\qquad\qquad\qquad\!\!\! + \frac{e^{-2i\theta}}{2}\,f(\hat{x})f'(\hat{x})^2\cbig[2-ie^{-i\theta}f(\hat{x})\hat{k}^2\cbig]\hat{k}^4  \Bigg\} \exp\cbig[-ie^{-i\theta}f(\hat{x})\,\hat{k}^2\cbig] \normordbig \; . 
\end{align}
Note that we constructed the operators $\widehat{H}_n^{(0,1,2)}$ such that they are readily in normal order, thus the same applies to the time evolution operators $\smash{\widehat{U}_{\theta,\mathrm{res}}^{(0,1,2)}}$, indicated by colons. We are now fit to perform the Gaussian momentum integral explicitly, arriving at the full expression for the propagator to the desired order. Up to irrelevant contributions, we obtain
\begin{align}
    \!K_\theta\big(x\!\!\:-\!\!\:\Delta x,x\:\!;\Delta t\big) &\doteq \Delta t^{-\frac{1}{2}}\!\!\:\mathlarger{\mathlarger{\int}}_{-\infty}^{\infty}\frac{\mathrm{d}k}{2\pi} \: \cBig\langle x\,\cBig\vert\, U_{\theta,\mathrm{res}}^{(0)}\big(\hat{x},\hat{k}\big) +\Delta t^{\frac{1}{2}}\:\! U_{\theta,\mathrm{res}}^{(1)}\big(\hat{x},\hat{k}\big)  \\[-0.45cm] 
    &\phantom{\doteq \Delta t^{-\frac{1}{2}}\!\!\:\mathlarger{\mathlarger{\int}}_{-\infty}^{\infty}\frac{\mathrm{d}k}{2\pi} \: \cBig\langle x\,\cBig\vert\, U_{\theta,\mathrm{res}}^{(0)}\big(\hat{x},\hat{k}\big)\;} + \Delta t \,U_{\theta,\mathrm{res}}^{(2)}\big(\hat{x},\hat{k}\big) \,\cBig\vert\, k\cBig\rangle_{\!\!\!\:\mathrm{res}} \braket[\rbig]{k\!\!\:|x-\Delta x}_\mathrm{res} \nonumber\\[0.05cm]
	&=\Delta t^{-\frac{1}{2}}\!\!\:\mathlarger{\mathlarger{\int}}_{-\infty}^{\infty}\frac{\mathrm{d}k}{2\pi} \:\! \Big[U_{\theta,\mathrm{res}}^{(0)}\big(x,k\big)\!\!\:+\!\!\:\Delta t^{\frac{1}{2}}\:\! U_{\theta,\mathrm{res}}^{(1)}\big(x,k\big)\!\!\:+\!\!\: \Delta t \,U_{\theta,\mathrm{res}}^{(2)}\big(x,k\big)\Big] \exp\!\bigg(\frac{ik\:\!\Delta x}{\Delta t^{\frac{1}{2}}}\bigg)  , \nonumber
\end{align}  
where we recall $\Delta x\sim \Delta t^{1/2}$, rendering the combination in the exponent to be of order unity. With the $U_{\theta,\mathrm{res}}^{(i)}\big(x,k\big)$ being independent of $\Delta t$, this retrospectively makes it apparent why the expansion up to the portrayed order suffices. Recall that the normal ordering procedure ensured that the matrix element involving the evolution operators $\widehat{U}_{\theta,\mathrm{res}}^{(0,1,2)}$ could be evaluated trivially. Performing the leftover momentum integral, the propagator takes the schematic form,
\begin{align}
    K_\theta\big(x-\Delta x,x\:\!;\Delta t\big) = \cbig[4\pi ie^{-i\theta}f(x) \Delta t\cbig]^{\!\!\:-\frac{1}{2}}\, \exp\!\!\!\;\bigg[\frac{ie^{i\theta}\Delta x^2}{4f(x) \Delta t}\bigg] \: \mathlarger{\sum}_{\substack{i,j\!\:=\!\:0\\ j\!\:\leq\!\: 2i}}^\infty \,F_{i,j}(x) \, \Delta t^{\!\:i}\:\!\bigg(\frac{\Delta x}{\Delta t}\bigg)^{\!\!\!\: j} \, ,
    \label{eq:SchematicForm_InfPropagator}
\end{align}
where the coefficient functions to the desired order are given by 
\begin{subequations}
\begin{align}
    F_{0,0}(x)&=1\, ,  & & & F_{i,2i}(x)&=0 \quad \text{for } i\geq 1\, ,\\[0.15cm]  
    F_{1,1}(x)&=\frac{3f'(x)-2ig(x)}{4f(x)}\, , & F_{2,3}(x)&=\frac{ie^{i\theta}f'(x)}{8f(x)^2}\, , & F_{i,2i-1}(x)&=0 \quad \text{for } i\geq 3 \, ,  
\end{align}
\label{eq:CoefficientFunctionsComplete1}
\end{subequations}
and 
\begin{align}
    F_{1,0}(x)&=ie^{-i\theta}\Bigg\{\frac{4f(x)f''(x)-3f'(x)^2-8if(x)g'(x)+8if'(x)g(x)+4g(x)^2}{16f(x)}-h(x)\Bigg\} \, , \nonumber \\[0.15cm] 
    F_{2,2}(x)&=\frac{-12f(x)f''(x)+21f'(x)^2+8if(x)g'(x)-20if'(x)g(x)-4g(x)^2}{32f(x)^2} \, , \nonumber  \\[0.15cm] 
    F_{3,4}(x)&=-ie^{i\theta}\Bigg\{\frac{8f(x)f''(x)-33f'(x)^2+12i f'(x)g(x)}{192f(x)^3}\Bigg\} \, , \nonumber  \\[0.15cm]
    F_{4,6}(x)&=-\frac{e^{2i\theta}f'(x)^2}{128 f(x)^4}\, ,  \qquad\qquad\qquad\qquad\qquad\qquad F_{i,2i-2}(x)=0\quad \text{for } i\geq 5 \, .
    \label{eq:CoefficientFunctionsComplete2}
\end{align}
\noindent 
All terms with $j\leq 2i-3$ can be neglected, as their overall order is at least $\mathcal{O}\big(\Delta t^{3/2}\big)$, thus dropping out when taking the continuum limit $N\to\infty$ ($\Delta t \to 0$). It is immanent that the provided result is yet too complicated to be put to practical use. Notably, it does not at all resemble the na\"{\i}ve result~\eqref{eq:NaiveInfPropagator}, which instead would amount to 
\begin{subequations}
\begin{align}
    F_{0,0}^{(\text{na\"{\i}ve})}(x)&=1\, , & F_{1,1}^{(\text{na\"{\i}ve})}(x)&=-\frac{ig(x)}{2f(x)}\, ,  \\[0.15cm] 
    F_{1,0}^{(\text{na\"{\i}ve})}(x)&=ie^{-i\theta}\cbigg\{\frac{g(x)^2}{4f(x)}-h(x)\!\!\:\cbigg\} \, , & F_{2,2}^{(\text{na\"{\i}ve})}(x)&=-\frac{g(x)^2}{8f(x)^2} \, ,  
\end{align}
\end{subequations}
with $F_{i,2i-j}^{(\text{na\"{\i}ve})}(x)=0$ for $i\geq j+1$ and $j\in \big\{1,2,3\big\}$. In that result, all contributions involving derivatives of $f$ and $g$ are completely absent. Fortunately, one can invoke simplifications to the full result~\eqref{eq:SchematicForm_InfPropagator},~\eqref{eq:CoefficientFunctionsComplete1} and~\eqref{eq:CoefficientFunctionsComplete2}, greatly shrinking down the size of the arising terms by utilizing so-called substitution rules, see e.g.~\cite{LangoucheFunctionalIntegration,SubstitutionRules,dePireyPathIntegralMethods}.

\subsection{Substitution rules}
\label{sec:SubstitutionRules}

After having preformed the daunting expansion procedure, we found the full propagator given by 
\begin{align}
	\!\!K_\theta\big(x_\mathrm{i},x_\mathrm{f}\:\! ; T\big) = \!\lim_{N \to \infty} \mathlarger{\mathlarger{\int}}_{-\infty}^{\infty} \, \mathlarger{\prod}_{k=1}^{N-1} \:\!\mathrm{d}x_{k} \; \cBigg\{\prod_{\ell=1}^N \,\exp\!\!\;\cbigg[\frac{ie^{i\theta}\big(\Delta x_\ell\big)^{\!\!\: 2}\:\!}{4f(x_{\ell}) \Delta t}\cbigg] \: \mathlarger{\sum}_{j\!\:=\!\:0}^\infty \,G_{j}(x_{\ell}) \, \big(\Delta x_\ell\big)^j\cBigg\}_{\!\substack{\Delta t\:\!=\:\!T/N\\[0.05cm] \,x_0\:\!=\:\!x_\mathrm{i}\;\;\;\, \\ x_N\:\!=\:\!x_\mathrm{f}\;\;\;\,}} \, ,
    \label{eq:DiscretizedPathIntegral}
\end{align}
with each infinitesimal propagator taking the schematic form provided in equation~\eqref{eq:SchematicForm_InfPropagator}, replacing $\Delta x$ with the shorthand $\Delta  x_\ell\coloneqq x_\ell -x_{\ell-1}$. We also abbreviated the combination
\begin{align}
    G_{j}(x_\ell)\coloneqq \cbig[4\pi ie^{-i\theta}f(x_\ell) \Delta t\cbig]^{\!\!\:-\frac{1}{2}}\mathlarger{\sum}_{i=\lceil j/2\rceil}^{\infty} \Delta t^{i-j}\:\! F_{i,j}(x_\ell) \, ,
    \label{eq:Definition_G_Function}
\end{align}
making the resulting expressions more bearable for further manipulation, as for all subsequent arguments solely the $\Delta x_\ell$-dependence will be of relevance. While in the formal continuum limit $N\to \infty$, quantities such as $\Delta t \to  \mathrm{d}t, \Delta x_\ell\to \dot{x}(t_\ell) \mathrm{d}t$ and $\Delta x_\ell^2/\Delta t\rightarrow \dot{x}(t_\ell)^2 \mathrm{d}t$ can be straightforwardly converted (if considerable care is taken), the remaining combinations $\smash{\Delta t^{i-j} \big(\Delta x_\ell\big)^{j}}$ multiplying the $F_{i,j}$ terms do not possess a comparably simple equivalent. This is where so-called substitution rules come into play, as they allow us to infer the replacements 
\begin{subequations}
    \begin{align}
    \big(\Delta x_\ell\big)^{\!\!\: 2} &\doteq 2ie^{-i\theta} f(x_\ell) \:\! \Delta t\, , \\
    \Delta t^{-1}\big(\Delta x_\ell\big)^{\!\!\: 3} &\doteq 6ie^{-i\theta} f(x_\ell) \:\! \Delta x_\ell \, , \\
    \Delta t^{-2}\big(\Delta x_\ell\big)^{\!\!\: 4} &\doteq -12e^{-2i\theta} f(x_\ell)^2 \:\! \Delta t\, , \\
    \Delta t^{-3}\big(\Delta x_\ell\big)^{\!\!\: 6}&\doteq -120ie^{-3i\theta} f(x_\ell)^3 \:\! \Delta t\, ,
\end{align}
\label{eq:SubstitutionRules}%
\end{subequations}
which when instated do not alter the value of the total discretized path integral, as the deviations only enter higher-order corrections that eventually drop out in the continuum limit $N\to \infty$. The general idea behind these substitution rules~\eqref{eq:SubstitutionRules} is strikingly straightforward, as it simply boils down to performing Gaussian integrations. As stated in the onset of section~\ref{sec:RigorousEstimate}, notice that the leftover integrations over the intermediate position variables $x_k$ receive their dominant contributions from the parametric region for which $\big(\Delta x_\ell\big)^{\!\!\: 2}\lesssim \Delta t$, since contributions outside this parametric range are exponentially suppressed in the limit $\Delta t\to 0^+$.\footnote{Beware that for all expressions to be well-defined, we require $\mathrm{Re}\big[ie^{-i\theta}f(x_\ell)^{-1}\big] < 0$, which we assume to be met at all times. In the real-time case $\theta=0$, for which the real part vanishes, the argument would similarly proceed by invoking the method of stationary phase (whereas for general $\theta$, the argument hinges on the method of steepest descent).} Expanding the non-exponential piece $G_j(x_\ell)$ around $\Delta x_\ell=0$ and consistently truncating the expansion at order $\smash{\Delta t\sim \big(\Delta x_\ell\big)^{\!\!\: 2}}$, one finds that one can substitute higher-order terms in $\Delta x_\ell$ by lower-order pendants as specified in equation~\eqref{eq:SubstitutionRules}, leaving the value of the integrals intact to the desired order.\\ 

\noindent 
Before making the stated notions explicit for the high-dimensional integral~\eqref{eq:DiscretizedPathIntegral}, let us briefly show how this pans out in case of a single-dimensional Laplace-type integral. Given an arbitrary smooth function $f$ that does not depend on the small parameter $\Delta t$, in the limit $\Delta t\to 0^+$, one finds the asymptotic relations
\begin{align}
    I_{2n}(\Delta t)&=\big(4\pi\Delta t\big)^{-\frac{1}{2}}\mathlarger{\int}_{-\infty}^\infty \mathrm{d}(\Delta x) \; f(\Delta x) \exp\!\bigg(\!\!-\!\!\:\frac{\Delta x^2}{4\Delta t}\bigg)\frac{\Delta x^{2n}}{\Delta t^n}\nonumber \\
    &\sim \frac{(2n)!}{n!}\,f(0)+ \frac{(2n+2)!}{2(n+1)!}\, f''(0)\, \Delta t + \mathcal{O}(\Delta t^2)\, , \label{eq:SingleDimIntegral_Example}\\[0.2cm]
    \frac{(2n)!}{n!}\,I_0(\Delta t)&=\big(4\pi\Delta t\big)^{-\frac{1}{2}}\:\frac{(2n)!}{n!}\,\mathlarger{\int}_{-\infty}^\infty \mathrm{d}(\Delta x) \; f(\Delta x) \exp\!\bigg(\!\!-\!\!\:\frac{\Delta x^2}{4\Delta t}\bigg)\nonumber \\
    &\sim \frac{(2n)!}{n!}\,f(0)+\frac{(2n)!}{n!}\,f''(0)\, \Delta t+ \mathcal{O}(\Delta t^2)\, .
\end{align}
Thus, in case one is solely interested in the leading-order behavior, one can simply replace the factor $\Delta x^{2n}\Delta t^{-n}$ inside the original Laplace-integral $I_{2n}$ by $(2n)!/n!$ without altering the result to the desired order. This furthermore assumes the function $f$ not to vanish at the origin, which is subliminally assumed in the following.\footnote{Note that $f(0)\neq 0$ is a generic behavior, being violated only on a set of vanishing Lebesgue measure in case one deals with a multi-dimensional integral as is the case for the discretized path integral~\eqref{eq:DiscretizedPathIntegral}. Therefore we can always assume this condition to be met.}\\

\noindent 
While the argument proceeds in the identical manner for the discretized path integral~\eqref{eq:DiscretizedPathIntegral}, the explicit reasoning will be slightly cluttered by the additional notational overhead. The previous definition of $G_j$ as introduced in equation~\eqref{eq:Definition_G_Function} constituted a first step in keeping the expressions at bay. The most lucid way to obtain the correct substitution rules~\eqref{eq:SubstitutionRules} is to change the integration variables from $\big\{x_\ell\big\}$ to $\big\{\Delta x_\ell\big\}$, $\ell\in 1,\ldots, N-1,$ invoking the relations
\begin{align}
	x_\ell=x_N - \!\!\!\:\sum_{k=\ell+1}^N \!\!\!\:\Delta x_k \, .
    \label{eq:UsefulCoordinateChange}
\end{align}
The transformation bestows a Jacobian factor of $(-1)^{N-1}$ upon us, which will be irrelevant for all subsequent arguments.\footnote{The coordinate transformation has the sole use of making the to-be-determined substitution rules most apparent. After instating them, one would transform back to the previous coordinate basis $\big\{x_\ell\big\}$.} Note that while we could have instead transformed the coordinates via the seemingly simpler relation 
\begin{align}
	x_\ell=x_0 + \sum_{k=1}^\ell \,\Delta x_k\, ,
    \label{eq:SimplifiedCoordinateChange}
\end{align}
this would grant additional complications downstream due to our initial choice of using a post-point discretization scheme. We will find that the parametrization~\eqref{eq:UsefulCoordinateChange} is favored in our case as it renders the factor $f(x_k)$ independent of $\Delta x_k$, greatly simplifying the subsequent treatment. Similarly, in case we would have chosen a pre-point discretization scheme, the transformation~\eqref{eq:SimplifiedCoordinateChange} would have been preferential due to making $f(x_{k-1})$ manifestly $\Delta x_k$-independent. While we for now stick to the transformation~\eqref{eq:UsefulCoordinateChange}, we will later briefly comment on the what would have changed if we employed the alternative parametrization~\eqref{eq:SimplifiedCoordinateChange} in the present case, as we feel the arising difficulties and their resolution to be instructive. \\ 

\noindent 
Enforcing the coordinate change~\eqref{eq:UsefulCoordinateChange} and solely focusing on the $\smash{\big(\Delta x_k\big)^{j}}$-contributions to a single $\Delta x_k$-integral, one requires to compute the Gaussian integral 
\begin{align}
	\mathcal{I}_j\coloneqq \mathlarger{\mathlarger{\int}}_{-\infty}^{\infty} \,\mathrm{d}\big(\Delta x_{k}\big) \:\mathcal{G}_j(\Delta x_k) \:\!\exp\!\!\;\cbigg[\frac{ie^{i\theta}\big(\Delta x_k\big)^{\!\!\: 2}\:\!}{4f(x_{k}) \Delta t}\cbigg] \big(\Delta x_k\big)^j \:\! ,
    \label{eq:DiscretizedPathIntegral_SingleInt}
\end{align}
where all the complexity is hidden in the newly defined function 
\begin{align}
    \mathcal{G}_j(\Delta x_k) &=G_j(x_k)\,\prod_{\substack{\ell=1\\ \ell\neq k}}^N \left\{\exp\!\!\;\cbigg[\frac{ie^{i\theta}\big(\Delta x_\ell\big)^{\!\!\: 2}\:\!}{4f(x_{\ell}) \Delta t}\cbigg] \, \mathlarger{\sum}_{i\!\:=\!\:0}^\infty \;G_{i}(x_{\ell}) \:\! \big(\Delta x_\ell\big)^{\!\!\:i}\right\}\, ,
\end{align}
leaving the dependence on $\Delta x_{\ell\neq k}$ and $\Delta t$ in both $\mathcal{I}_j$ and $\mathcal{G}_j$ implicit. Beware that the functions $f(x_k)$ and $G_j(x_k)$ now implicitly depend on all $\Delta x_{\ell>k}$, which is to be understood. Importantly, as stated previously, $f(x_k)$ inside the exponent of equation~\eqref{eq:DiscretizedPathIntegral_SingleInt} is independent of $\Delta  x_k$, thus, we can treat the integral on the identical footing as the single-dimensional example~\eqref{eq:SingleDimIntegral_Example}. While the auxiliary function $\mathcal{G}_j$ still entails a nontrivial $\Delta t$-behavior, this is of no concern here as it is always regulated by contributions involving $\Delta x^2\sim \Delta t$.\footnote{One could clarify the argument even more by rescaling all integration variables $\Delta x_k$ by the common scaling factor $\Delta t^{1/2}$, as was previously done for the momentum integral~\eqref{eq:RescaledPropagatorIntegral}. This would once more manifest the parametric region from which the integrals receive their dominant contributions. We will waive this step as it would introduce yet another layer of notational clutter.} The leading-order behavior of the integral in question is then found to be  
\begin{align}
	\mathcal{I}_j\sim \left\{\begin{matrix}
	    \displaystyle{\Gamma\cBig(\frac{j+1}{2}\cBig)\cbig[4ie^{-i\theta}f(x_{k})\Delta t\cbig]^{\!\frac{j+1}{2}}\,\mathcal{G}_j\big(\Delta x_k\!=\!0\big)\:\! ,\qquad\,} & j \;\mathrm{even}\, , \\[0.4cm] 
        \displaystyle{\Gamma\cBig(\frac{j+2}{2}\cBig)\cbig[4ie^{-i\theta}f(x_{k})\Delta t\cbig]^{\!\frac{j+2}{2}}\:\frac{\partial\mathcal{G}_j(\Delta x_k)}{\partial (\Delta x_k)}\bigg\vert_{\Delta x_k=0} \:\! ,} & j \;\mathrm{odd}\, ,
	\end{matrix} \right.
    \label{eq:DiscretizedPathIntegral_SingleInt_Asymptotics}
\end{align}
receiving corrections of order $\Delta t$ and higher, as only even powers in $\Delta x_k$ survive the Gaussian integration. Since for all $j\geq 2$ terms that we want to substitute away, the leading-order behavior is sufficient to capture the individual $\Delta x_k$-integral to order $\Delta t$, we can infer the replacements 
\begin{align}
    \big(\Delta x_k\big)^{j} &\doteq \left\{\begin{matrix}
	    \displaystyle{\!\!\!\!\,\Gamma\cBig(\frac{j+1}{2}\cBig)\:\!\Gamma\cBig(\frac{1}{2}\cBig)^{\!\!-1}} \:\!\cbig[4ie^{-i\theta}f(x_{k})\Delta t\cbig]^{\!\frac{j}{2}}\, , \phantom{\Delta x_k}& j \; \mathrm{even}\, , \\[0.4cm] 
	    \displaystyle{\,\Gamma\cBig(\frac{j+2}{2}\cBig)\:\!\Gamma\cBig(\frac{3}{2}\cBig)^{\!\!-1}} \:\!\cbig[4ie^{-i\theta}f(x_{k})\Delta t\cbig]^{\!\frac{j-1}{2}} \Delta x_k\, , & j \; \mathrm{odd}\, ,
        \end{matrix}\right.
        \label{eq:SubstRules_Full}
\end{align}
leaving~\eqref{eq:DiscretizedPathIntegral_SingleInt} invariant to leading-order. This then directly leads to the substitution rules quoted earlier, see equation~\eqref{eq:SubstitutionRules}. \\

\noindent 
Let us take a slight detour to explain what would have happened in case we would have utilized the coordinate transformation~\eqref{eq:SimplifiedCoordinateChange}, making the term $f(x_k)$ inside the exponent $\Delta x_k$-dependent. In that case, the leading-order behavior of the integral would change, both in the even and odd $j$ case. For even $j$, one would obtain the relation 
\begin{align}
	\mathcal{I}_j\sim \Gamma\cBig(\frac{j+1}{2}\cBig)\cbig[4ie^{-i\theta}f(x_{k-1})\Delta t\cbig]^{\!\frac{j+1}{2}}\mathcal{G}_j\big(\Delta x_k\!=\!0\big)\, ,
    \label{eq:DiscretizedPathIntegral_SingleInt_Asymptotics2}
\end{align}
being identical to the one previously found with the exception that $f$ is evaluated at $x_{k-1}$ instead of $x_k$ since to leading-order, $\Delta x_k$ is set to 0 inside the expression $f(x_k)$. The substitution rule for even $j$ would change accordingly, being given by 
\begin{align}
    \big(\Delta x_k\big)^{j} &\doteq \Gamma\cBig(\frac{j+1}{2}\cBig)\:\!\Gamma\cBig(\frac{1}{2}\cBig)^{\!\!-1} \:\!\cbig[4ie^{-i\theta}f(x_{k-1})\Delta t\cbig]^{\!\frac{j}{2}}\, ,
\end{align}
which is however identical to the previously found expression~\eqref{eq:SubstRules_Full} due to $f(x_k)$ and $f(x_{k-1})$ differing only at order $\Delta x_k$. For the even $j$ terms, these corrections are negligible, as their overall contributions to the infinitesimal propagator~\eqref{eq:SchematicForm_InfPropagator} are already of order $\Delta t$. The odd $j$ case is more involved, as one now also needs to expand the function $f(x_k)$ inside the exponent around $\Delta x_k=0$. Assuming odd $j$, the leading-order behavior of $\mathcal{I}_j$ is given by  
\begin{align}
	\mathcal{I}_{j}&\sim \mathlarger{\mathlarger{\int}}_{-\infty}^{\infty} \,\mathrm{d}\big(\Delta x_{k}\big)\cbigg[\mathcal{G}'_j(0)-\frac{ie^{i\theta}f'(x_{k-1}) \big(\Delta x_k\big)^{\!\!\: 2}}{4f(x_{k-1})^2\Delta t}\,\mathcal{G}_j(0)\cbigg] \:\!\exp\!\!\;\cbigg[\frac{ie^{i\theta}\big(\Delta x_k\big)^{\!\!\: 2}\:\!}{4f(x_{k-1}) \Delta t}\cbigg] \big(\Delta x_k\big)^{j+1}  \nonumber \\
    &= \big[4ie^{-i\theta}f(x_{k-1})\Delta t\big]^{\frac{j+2}{2}}\,\Gamma\cBig(\frac{j+2}{2}\cBig)\bigg[\mathcal{G}'_j(0)+\frac{j+2}{2}\frac{f'(x_{k-1})}{f(x_{k-1})}\,\mathcal{G}_j(0)\bigg]\, ,
    \label{eq:Integral_Ij_1}
\end{align}
receiving contributions not only from $\mathcal{G}_j'(0)$, but also inheriting derivative terms involving $f'$, which were previously absent in relation~\eqref{eq:DiscretizedPathIntegral_SingleInt_Asymptotics}. Once more, these solely arise due to the fact that now $f(x_k)$ depends nontrivially on $\Delta x_k$. Utilizing the educated ansatz 
\begin{align}
    \big(\Delta x_k\big)^{j}\doteq \chi_j(x_k)\Delta t^{\frac{j-1}{2}}\Delta x_k
    \label{eq:AnsatzSubstitutionRule}
\end{align}
for the desired substitution rule, with $\chi_j(x_k)$ constituting a to-be-determined function, the integral would receive the leading-order contribution\\[-0.5cm]
\begin{equation}
  \begin{adjustbox}{width=\linewidth, center}
    \begin{minipage}{\linewidth}
      \begin{align*}
	   \mathcal{I}_{j,\,\mathrm{subst.}}&\sim \Delta t^{\frac{j-1}{2}}\!\mathlarger{\mathlarger{\int}}_{-\infty}^{\infty} \,\mathrm{d}\big(\Delta x_{k}\big)\cbigg[\chi_j(x_{k-1})\mathcal{G}'_j(0)+\chi_j'(x_{k-1})\mathcal{G}_j(0)-\frac{ie^{i\theta}f'(x_{k-1})\big(\Delta x_k\big)^{\!\!\: 2}}{4f(x_{k-1})^2\Delta t}\,\chi_j(x_{k-1})\mathcal{G}_j(0)\cbigg] \nonumber \\[-0.05cm] &\qquad\qquad\qquad\qquad\qquad\qquad\qquad\qquad\qquad\qquad\qquad\qquad\qquad\qquad \times \exp\!\!\;\cbigg[\frac{ie^{i\theta}\big(\Delta x_k\big)^{\!\!\: 2}\:\!}{4f(x_{k-1}) \Delta t}\cbigg] \big(\Delta x_k\big)^{2}  \nonumber \\[0.15cm]
        &= \Delta t^{\frac{j-1}{2}}\big[4ie^{-i\theta}f(x_{k-1})\Delta t\big]^{\frac{3}{2}}\,\Gamma\cBig(\frac{3}{2}\cBig)\bigg[\chi_j(x_{k-1})\mathcal{G}'_j(0)+\chi_j'(x_{k-1})\mathcal{G}_j(0)+\frac{3}{2}\frac{f'(x_{k-1})}{f(x_{k-1})}\,\chi_j(x_{k-1})\mathcal{G}_j(0)\bigg] , \\[-0.55cm]
        \end{align*}
    \end{minipage}
  \end{adjustbox}
  \tag{\normalsize \theequation} 
  \label{eq:Integral_Ij_2}
\end{equation}
where now also $\chi_j(x_k)$ needed to be expanded around $\Delta x_{k}=0$. Demanding equality between both expressions~\eqref{eq:Integral_Ij_1} and~\eqref{eq:Integral_Ij_2} precisely yields
\begin{align}
    \chi_j(x_{k-1})=\Gamma\cBig(\frac{j+2}{2}\cBig)\,\Gamma\cBig(\frac{3}{2}\cBig)^{\!-1}\big[4ie^{-i\theta}f(x_{k-1})\big]^{\frac{j-1}{2}}\, ,
\end{align}
once more leading to the correct substitution rule~\eqref{eq:SubstRules_Full} due to $\chi_j$ entering the previous ansatz~\eqref{eq:AnsatzSubstitutionRule} being evaluated at $x_k$. While this secondary derivation demanded additional care, it applies more broadly to the case of a general $\alpha$-prescription for evaluating the infinitesimal propagators at $\alpha x_k+(1-\alpha)x_{k-1}$ instead of committing to a pre- or post-point discretization scheme.

\subsection{Collecting the results}
\label{sec:CollectingResults}

We are now able to heavily simplify the previous result. Instating the substitution rules~\eqref{eq:SubstitutionRules}, the previously obtained short-time propagator takes the form\\[-0.5cm]
\begin{equation}
  \begin{adjustbox}{width=\linewidth, center}
    \begin{minipage}{\linewidth}
      \begin{align*}
        K_\theta\big(x-\Delta x,x\:\!;\Delta t\big) &= \cbig[4\pi ie^{-i\theta}f(x) \Delta t\cbig]^{\!\!\:-\frac{1}{2}}\, \exp\!\!\!\;\bigg[\frac{ie^{i\theta}\Delta x^2}{4f(x) \Delta t}\bigg] \, \Bigg\{ 1+F_{1,1}(x)\!\:\Delta x +F_{1,0}(x)\!\:\Delta t \nonumber \\[-0.3cm] 
        &\hspace{6.85cm}+ F_{2,2}(x)\!\:\Delta x^2+ F_{2,3}(x)\!\:\frac{\Delta x^3}{\Delta t}\nonumber \\ 
        &\hspace{6.85cm}+ F_{3,4}(x)\!\:\frac{\Delta x^4}{\Delta t}+ F_{4,6}(x)\!\:\frac{\Delta x^6}{\Delta t^2} +\mathcal{O}\cbig(\Delta t^{\frac{3}{2}}\cbig)\Bigg\} \nonumber \\ 
        &\doteq \big[\ldots\big]\, \Bigg\{ 1+\cBig[F_{1,1}(x)+6ie^{-i\theta} f(x)F_{2,3}(x)\cBig]\Delta x+F_{2,2}^{(\text{na\"{\i}ve})}\:\!\Delta x^2 \nonumber \\[-0.4cm] 
        &\phantom{\doteq \big[\ldots\big]\, \Bigg\{ 1\;}+\cBig[F_{1,0}(x) +2ie^{-i\theta} f(x)\Big(F_{2,2}(x)-F_{2,2}^{(\text{na\"{\i}ve})}(x)\Big)\nonumber \\[-0.45cm]
        &\phantom{\doteq \big[\ldots\big]\, \Bigg\{ 1+\cBig[F_{1,0}(x)\;}-12e^{-2i\theta} f(x)^2 F_{3,4}(x)-120ie^{-3i\theta} f(x)^3 F_{4,6}(x)\cBig]\Delta t\Bigg\} \nonumber \\[-0.15cm]
        &= \big[\ldots\big]\, \bigg\{ 1+F_{1,1}^{(\text{na\"{\i}ve})}(x)\:\!\Delta x+F_{1,0}^{(\text{na\"{\i}ve})}(x)\:\!\Delta t+F_{2,2}^{(\text{na\"{\i}ve})}\:\!\Delta x^2\bigg\} \nonumber \\
        &=K_\theta^{(\text{na\"{\i}ve})}\big(x-\Delta x,x\:\!;\Delta t\big)\, ,\\[-1.1cm]
      \end{align*}
    \end{minipage}
  \end{adjustbox}
  \tag{\normalsize \theequation} 
  \vspace{0.25cm}
\end{equation}
utilizing the previously determined coefficient functions $F_{i,j}(x)$ as portrayed in equations~\eqref{eq:CoefficientFunctionsComplete1} and~\eqref{eq:CoefficientFunctionsComplete2}. One finds that the square bracketed combinations perfectly conspire such that all derivative contributions involving $f'(x), f''(x)$ and $g'(x)$ drop out, leaving us with the na\"{\i}ve propagator~\eqref{eq:NaiveInfPropagator} computed at the onset of the chapter. Concluding, one finds that while the infinitesimal propagators do not coincide, for our purposes of utilizing them inside the discretized path integral expressions, all deviations only enter higher-order terms that vanish when taking the continuum limit $N\to\infty$.

\section{Independence of the chosen scalar product}
\label{sec:B_ScalarProduct_Discussion}

There arises the question why we utilized the standard scalar product for the derivation of both the path integral as well as the spectral representation of the propagator. As we will explicitly show, this deliberate choice is fully arbitrary, with the standard scalar product constituting the simplest admissible option. Even though the propagator of the so-defined theory will depend on this aforementioned choice, the dependence is trivial and does not interfere with any of the drawn conclusions. \\

\noindent 
Assuming the contour $\Gamma$ together with its parametrization $\gamma(s)$ has been provided, let us define the general $\mu$-scalar product via
\begin{align}
	\braket[\rbig]{\psi\!\!\:|\phi}_{\!\mu} \coloneqq \mathlarger{\int}_{-\infty}^{\infty} \mu(s) \:\! \closure{\psi(s)}\:\! \phi(s)\,\mathrm{d}s\, ,
    \label{eq:MuScalarProduct}
\end{align}
with $\mu(s)>0$ constituting an arbitrary, smooth, real-valued weight function. Before studying in detail how this altered notion influences the position and momentum completeness relations required for the later path integral derivation, let us first investigate the spectral representation. To render all subsequent expressions most convenient, we again modify the definition of the conventional propagator by introducing an additional factor $\mu(s_\mathrm{i})$, setting
\begin{align}
	K_{\gamma,\mu,\theta}\big(s_\mathrm{i},s_\mathrm{f}\:\!;T\big) &\coloneqq \frac{\mu(s_\mathrm{i})}{\gamma'(s_\mathrm{i})}\,\braket[\rbigg]{s_\mathrm{f} | \,\exp\!\bigg(\!\!-\!\frac{i e^{-i\theta} \widehat{H}_\gamma T}{\hbar}\bigg)\;\!\! |\:\! s_\mathrm{i} }_{\!\!\!\!\:\mu}\: .
	\label{eq:Definition_Propagator_MuScalarProd}
\end{align}

\subsection{Spectral representation}

While the effective Hamiltonian $\widehat{H}_\gamma$ on the complex contour $\Gamma$ is still given by 
\begin{align}
	\widehat{H}_\gamma = -\frac{\hbar^2}{2m}\frac{1}{\gamma'(s)} \frac{\mathrm d}{\mathrm{d}s}\bigg[\frac{1}{\gamma'(s)}\, \frac{\mathrm d}{\mathrm{d}s}\bigg] + V\!\!\:\big[\gamma(s)\big]\, ,
    \label{eq:HamiltonianGamma2}
\end{align}
the Hermitian conjugate changes in case $\mu(s)$ is a non-constant function. A simple computation analogous to the one presented in equation~\eqref{eq:Comp_HermitianConjugate} concludes with 
\begin{align}
	\widehat{H}_\gamma^\dagger &= -\frac{\hbar^2}{2m}\frac{1}{\closure{\mu(s)}}\frac{\mathrm d}{\mathrm{d}s}\cbigg[\frac{1}{\closure{\gamma'(s)}} \frac{\mathrm d}{\mathrm{d}s}\frac{\closure{\mu(s)}}{\closure{\gamma'(s)}}\cbigg] + \,\closure{V\!\!\:\big[\gamma(s)\big]} \, . 
	\label{eq:ConjugateHamiltonian_MuScalarProd}
\end{align}
This in turn alters the relation between the elements of the bi-orthogonal basis associated to the Hamiltonian, which instead of equation~\eqref{eq:RelationDualEigenfunctions} now satisfy
\begin{align}
	\closure{\phi_\ell(s)}=\frac{\gamma'(s)\psi_\ell(s)}{\mu(s)}\, .
\end{align}
Inserting the resolution of unity~\eqref{eq:BiorthogonalResolutionOfUnity} using the bi-orthogonal eigenfunctions grants the identity 
\begin{align}
	K_{\gamma,\mu,\theta}\big(s_\mathrm{i},s_\mathrm{f}\:\!;T\big) &=\frac{\mu(s_\mathrm{i})}{\gamma'(s_\mathrm{i})}\: \mathlarger{\mathlarger{\sum}}_{\ell=0}^\infty \; \frac{\closure{\phi_\ell(s_\mathrm{i})}\,\psi_\ell(s_\mathrm{f})}{\braket[\rbig]{\phi_\ell\!\!\:|\psi_\ell}_{\!\!\:\mu}} \: \exp\!\bigg(\!\!-\!\frac{i e^{-i\theta} E_\ell T}{\hbar}\bigg) \nonumber\\
	&= \frac{\mu(s_\mathrm{i})}{\gamma'(s_\mathrm{i})}\frac{\gamma'(s_\mathrm{i})}{\mu(s_\mathrm{i})}\: \mathlarger{\mathlarger{\sum}}_{\ell=0}^\infty \;\psi_\ell(s_\mathrm{i})\,\psi_\ell(s_\mathrm{f}) \,\Bigg\{\mathlarger{\int}_{-\infty}^{\infty}\:\! \mu(s) \:\! \frac{\gamma'(s)}{\mu(s)}\, \psi_\ell(s)^2\,\mathrm{d}s\Bigg\}^{\!\!\!\:-1} \exp\!\bigg(\!\!-\!\frac{i e^{-i\theta} E_\ell T}{\hbar}\bigg) \nonumber \\ 
    &= K_{\gamma,\mu=1,\theta}\big(s_\mathrm{i},s_\mathrm{f}\:\!;T\big) \, .
	\label{eq:Propagator_SpectralRepresentation_MuScalarProd}
\end{align}
Note that we still defined the position representation of the wave function via $\psi(s)=\braket{s\!\!\:|\psi}_{\!\!\:\mu}$, absorbing additional factors of $\mu(s)$ in the normalization of the position eigenstates.\footnote{We could have similarly chosen to set $\smash{\psi(s)=\sqrt{\mu(s)}\:\!\braket{s\!\!\:|\psi}_{\!\!\:\mu}}$, effectively redefining the position eigenstates $\smash{\ket{s}\mapsto \sqrt{\mu(s)}\,\ket{s}}$. While this prescription would be advantageous for defining the associated canonical momentum operators, it is instructive to take the shown route.} The above result~\eqref{eq:Propagator_SpectralRepresentation_MuScalarProd} shows that in the spectral representation, additional factors of $\mu(s)$ enter trivially, canceling in the relevant normalization factor. By introducing the additional factor $\mu(s_\mathrm{i})$ in the definition~\eqref{eq:Definition_Propagator_MuScalarProd} of the propagator, $K_{\gamma,\mu,\theta}$ is indeed found to be independent of the choice of $\mu$. Note that the trace of the analytically continued time evolution operator involves one additional position-integral, bestowing an additional weight factor $\mu(s)$ upon us, thus leaving the previously obtained spectral representation~\eqref{eq:Final_Relation_Trace} intact.

\subsection{Path integral representation}

The key difference in attaining the path integral representation, given the nontrivial measure $\mu(s)$ compared to using the standard scalar product~\eqref{eq:StandardScalarProduct}, are the altered completeness relations for position and momentum eigenstates. Demanding the position eigenstates to still satisfy $\hat{x}\ket{s}=s\ket{s}$ as well as the (position space) wave function to be given by $\psi(s)=\braket{s\!\!\:|\psi}_{\!\!\:\mu}$ yields the position completeness relation
\begin{align}
	\mathds{1} = \mathlarger{\int}_{-\infty}^{\infty} \mu(s)\!\: \ket{s} \bra{s}\,\mathrm{d}s \, .
\end{align} 
This effectively also changes the normalization of the position eigenstates to $\braket{s\!\!\:|s'}_{\!\!\:\mu}=\mu(s)^{-1}\delta(s-s')$. The momentum operator is more problematic, as $\hat{p}_\gamma=-i\hbar \,\mathrm{d}/\mathrm{d}s$ is not Hermitian with respect to the $\mu$-scalar product~\eqref{eq:MuScalarProduct}, requiring us to introduce
\begin{align}
    \hat{p}_{\gamma,\mu}=-i\hbar \, \mu(s)^{-\frac{1}{2}}\,\frac{\mathrm{d}}{\mathrm{d}s} \,\mu(s)^{\frac{1}{2}}\, , \qquad \text{satisfying } \braket[\rbig]{\psi\!\!\:|\!\:\hat{p}_{\gamma,\mu}\:\!\phi} = \braket[\rbig]{\hat{p}_{\gamma,\mu}\:\!\psi\!\!\:|\phi}\, .
    \label{eq:MuMomentumOperator}
\end{align}
Note that the commutator relation between $\hat{x}_\gamma$ and $\hat{p}_{\gamma,\mu}$ is identical to the usual one, satisfying $\big[\hat{x}_\gamma,\hat{p}_{\gamma,\mu}\big]=i\hbar$, however the projection between position and momentum eigenstates changes to 
\begin{align}
    \braket{s\!\!\:|p}_{\!\!\:\mu}=\mu(s)^{-\frac{1}{2}} e^{ips/\hbar}\, .
    \label{eq:Projection_Altered}
\end{align}
This difference arises due to the momentum eigenstates being required to satisfy the conventional relation $\hat{p}_{\gamma,\mu}\:\!\ket{p}=p\:\!\ket{p}$, however utilizing the altered Hermitian momentum operator~\eqref{eq:MuMomentumOperator}. The momentum completeness relation then stays intact, reading 
\begin{align}
	\mathds{1} = \frac{1}{2\pi\hbar}\mathlarger{\int}_{-\infty}^{\infty} \ket{p} \bra{p}\,\mathrm{d}p\, .
\end{align}
Representing the Hamiltonian~\eqref{eq:HamiltonianGamma2} in terms of the canonical operators $\hat{x}_\gamma$ and $\hat{p}_{\gamma,\mu}$ leads to 
\begin{align}
	\widehat{H}_{\gamma,\mu}
	&=\phantom{+}\frac{1}{2m}\frac{1}{\gamma'(\hat{x}_\gamma)^2}\,\mu(\hat{x}_\gamma)^{\frac{1}{2}}\,\hat{p}_{\gamma,\mu}^2\,\mu(\hat{x}_\gamma)^{-\frac{1}{2}}+\frac{i\hbar}{2m}\frac{\gamma''(\hat{x}_\gamma)}{\gamma'(\hat{x}_\gamma)^3}\, \mu(\hat{x}_\gamma)^{\frac{1}{2}}\,\hat{p}_{\gamma,\mu}\,\mu(\hat{x}_\gamma)^{-\frac{1}{2}} + V\!\!\:\big[\gamma(\hat{x}_\gamma)\big] \nonumber\\
    &= \phantom{+}\frac{1}{2m}\frac{1}{\gamma'(\hat{x}_\gamma)^2}\,\Bigg\{\hat{p}_{\gamma,\mu}^2-2i\hbar \,\mu(\hat{x}_\gamma)^{-\frac{1}{2}} \frac{\mathrm{d}\mu(\hat{x}_\gamma)^{\frac{1}{2}}}{\mathrm{d}\hat{x}_\gamma}\,\hat{p}_{\gamma,\mu}-\hbar^2 \,\mu(\hat{x}_\gamma)^{-\frac{1}{2}} \frac{\mathrm{d}^2\mu(\hat{x}_\gamma)^{\frac{1}{2}}}{\mathrm{d}\hat{x}_\gamma^2}\Bigg\} \nonumber \\ 
    &\phantom{=\,} +\frac{i\hbar}{2m}\frac{\gamma''(\hat{x}_\gamma)}{\gamma'(\hat{x}_\gamma)^3}\, \Bigg\{\hat{p}_{\gamma,\mu}-i\hbar \,\mu(\hat{x}_\gamma)^{-\frac{1}{2}} \frac{\mathrm{d}\mu(\hat{x}_\gamma)^{\frac{1}{2}}}{\mathrm{d}\hat{x}_\gamma} \Bigg\} + V\!\!\:\big[\gamma(\hat{x}_\gamma)\big] \nonumber \\ 
    &= \phantom{+}\frac{1}{2m}\frac{1}{\gamma'(\hat{x}_\gamma)^2}\,\hat{p}_{\gamma,\mu}^2 + \frac{i\hbar}{2m}\frac{1}{\gamma'(\hat{x}_\gamma)^2}\bigg\{\frac{\gamma''(\hat{x}_\gamma)}{\gamma'(\hat{x}_\gamma)}+\frac{\mu'(\hat{x}_\gamma)}{\mu(\hat{x}_\gamma)}\bigg\}\,\hat{p}_{\gamma,\mu} \nonumber \\ 
    &\phantom{=\,} - \frac{\hbar^2}{2m}\frac{1}{\gamma'(\hat{x}_\gamma)^2}\Bigg\{\frac{3\mu'(\hat{x}_\gamma)^2-2\mu(\hat{x}_\gamma) \mu''(\hat{x}_\gamma)}{4\mu(\hat{x}_\gamma)^2}+\frac{\gamma''(\hat{x}_\gamma)}{\gamma'(\hat{x}_\gamma)}\frac{\mu'(\hat{x}_\gamma)}{2\mu(\hat{x}_\gamma)}\Bigg\} + V\!\!\:\big[\gamma(\hat{x}_\gamma)\big] \, .
	\label{eq:NonStandard_Hamiltonian_Mu}
\end{align}
Inserting this Hamiltonian into the full results~\eqref{eq:SchematicForm_InfPropagator},~\eqref{eq:CoefficientFunctionsComplete1} and~\eqref{eq:CoefficientFunctionsComplete2} for the infinitesimal propagator and remembering the additional factors entering through the altered projection relation~\eqref{eq:Projection_Altered} yields 
\begin{align}
    K_{\gamma,\mu,\theta}\big(s-\Delta s,s\:\!;\!\!\;\Delta t\big)\,&\doteq\, \underbrace{\mu(s)^{-\frac{1}{2}}\mu(s-\Delta s)^{-\frac{1}{2}}}_{\displaystyle{\begin{matrix}\text{factors from }\braket{s\!\!\:|p}_{\!\!\:\mu}\\ \text{and }\braket[\rbig]{p\!\!\:|s-\Delta s}_{\!\!\:\mu} \end{matrix}}} \hspace{-4.35cm}\underbrace{\frac{\mu(s-\Delta s)}{\gamma'(s-\Delta s)}}_{\displaystyle{\qquad\qquad\qquad\qquad\qquad\qquad\text{factors due to the definition of } K_{\gamma,\mu,\theta}}} \\ 
    &\quad\; \times\cbigg[\frac{m\gamma'(s)^2}{2\pi\hbar ie^{-i\theta} \Delta t}\cbigg]^{\!\!\:\frac{1}{2}} \, \exp\!\!\:\Bigg\{\!\!\;\frac{ie^{i\theta}m}{2\hbar \Delta t}\:\!\cbig[\gamma'(s) \Delta s\cbig]^2-\:\!\frac{ie^{-i\theta}\Delta t}{\hbar}\:\!V\!\!\:\big[\gamma(s)\big]\!\!\;\Bigg\} \nonumber \\ 
    &\quad\; \times \Bigg\{1-\frac{\gamma''(s)}{\gamma'(s)}\,\Delta s+\frac{\gamma'''(s)}{2\gamma'(s)}\,(\Delta s)^{2}-\frac{ie^{i\theta}m}{2\hbar \Delta t}\,\gamma'(s_\ell)\gamma''(s)(\Delta s)^{3} \nonumber \\
    &\quad\; \phantom{\times\Bigg\{1\;} + \frac{ie^{i\theta}m}{2\hbar \Delta t}\frac{15\gamma''(s)^2+4\gamma'(s)\gamma'''(s)}{12}\,(\Delta s)^{4}-\frac{e^{2i\theta}m^2}{8\hbar^2\Delta t^2} \,\gamma'(s)^2\gamma''(s)^2(\Delta s)^{6}\Bigg\} \nonumber \\ 
    &\quad\; \times\Bigg\{1+\frac{\mu'(s)}{2\mu(s)}\,\Delta s+\frac{3\mu'(s)^2-2\mu(s)\mu''(s)}{8\mu(s)^2}\,(\Delta s)^2\Bigg\}\, ,
\end{align}
already factoring the result accordingly while dropping subleading terms in $\Delta t\sim \Delta x^2$. Note that the first four lines are, up to the additional $\mu(s)$ factors, virtually identical to the previous result~\eqref{eq:ShortTimePropContour} for $\mu(s)=1$, whereas the second factor reduces to
\begin{align}
    \bigg[\frac{\mu(s)}{\mu(s-\Delta s)}\bigg]^{\frac{1}{2}}= 1+\frac{\mu'(s)}{2\mu(s)}\,\Delta s+\frac{3\mu'(s)^2-2\mu(s)\mu''(s)}{8\mu(s)^2}\,(\Delta s)^2 + \mathcal{O}\cbig[(\Delta s)^3\cbig]\, ,
\end{align}
with higher-order corrections vanishing when taking the continuum limit $N\to\infty$ of the discretized path integral. Thus, we indeed obtain the short-time propagator as
\begin{align}
    K_{\gamma,\mu,\theta}\big(s-\Delta s,s\:\!;\Delta t\big)=K_{\gamma,\mu=1,\theta}\big(s-\Delta s,s\:\!;\Delta t\big) \, ,
    \label{eq:ShortTimeProp_Final_Mu}
\end{align}
with all factors of $\mu$ canceling fully. The same can be seen when investigating the complete propagator, satisfying
\begin{align}
	K_{\gamma,\mu,\theta}\big(s_\mathrm{i},s_\mathrm{f}\:\!;T\big) &= \frac{\mu(s_\mathrm{i})}{\gamma'(s_\mathrm{i})}\lim_{N \to \infty} \mathlarger{\mathlarger{\int}}_{-\infty}^{\infty} \, \mathlarger{\prod}_{k=1}^{N-1} \:\!\cbig[\mu(s_k)\,\mathrm{d}s_{k}\cbig] \Bigg\{\prod_{\ell=1}^N \braket[\cbigg]{s_\ell|\: \exp\!\!\:\bigg[-\frac{i e^{-i\theta} T}{N\hbar}\, \widehat{H}_{\gamma,\mu}\bigg]|\,s_{\ell-1}\!}\Bigg\}_{\!\substack{\,s_0\:\!=\:\!s_\mathrm{i}\;\;\;\, \\ s_N\:\!=\:\!s_\mathrm{f}\;\;\;\,}}\nonumber \\ 
    &= \frac{\mu(s_\mathrm{i})}{\gamma'(s_\mathrm{i})}\lim_{N \to \infty} \mathlarger{\mathlarger{\int}}_{-\infty}^{\infty} \, \mathlarger{\prod}_{k=1}^{N-1} \:\!\cbig[\mu(s_k)\,\mathrm{d}s_{k}\cbig] \Bigg\{\prod_{\ell=1}^N \bigg[\frac{\gamma'(s_{\ell-1})}{\mu(s_{\ell-1})}\,K_{\gamma,\mu,\theta}\big(s_{\ell-1},s_{\ell}\:\! ; T/N\big)\bigg]\Bigg\}_{\!\substack{\,s_0\:\!=\:\!s_\mathrm{i}\;\;\;\, \\ s_N\:\!=\:\!s_\mathrm{f}\;\;\;\,}} \nonumber \\ 
    &= \lim_{N \to \infty} \mathlarger{\mathlarger{\int}}_{-\infty}^{\infty} \, \mathlarger{\prod}_{k=1}^{N-1} \:\!\cbig[\gamma'(s_{k})\,\mathrm{d}s_{k}\cbig] \,\Bigg\{\prod_{\ell=1}^N K_{\gamma,\mu=1,\theta}\big(s_{\ell-1},s_{\ell}\:\! ; T/N\big)\Bigg\}_{\!\substack{\,s_0\:\!=\:\!s_\mathrm{i}\;\;\;\, \\ s_N\:\!=\:\!s_\mathrm{f}\;\;\;\,}} \nonumber \\ 
    &= K_{\gamma,\mu=1,\theta}\big(s_\mathrm{i},s_\mathrm{f}\:\!;T\big) \, .
	\label{eq:DiscretizedPI_Start_Mu} 
\end{align}
All $N-1$ factors of $\mu(s_k)$ emerging from the additional measure entailed in the completeness relations for the intermediate $s_k$-integrals together with the sole overall $\mu(s_\mathrm{i})$ factor cancel against the $N$ factors $\mu(s_{\ell-1})^{-1}$ emerging from the intermediate matrix elements. This digression explicitly demonstrates the earlier assertion that the choice of the measure $\mu(s)$ is completely arbitrary, as all options eventually lead to identical final results~\eqref{eq:Final_Relation_Propagator} and~\eqref{eq:Final_Relation_Trace}.

\section{Constraints on admissible contours $\Gamma$}
\label{sec:C_Restrictions_Contour}

Whereas the derivation of the spectral representation~\eqref{eq:Propagator_SpectralRepresentation} was mathematically well-defined without inquiring further information on the contour $\Gamma$ apart from the demand of $\Gamma$ to terminate in the Stokes sectors $S_\pm$ for which the eigenfunctions $\Psi_\ell(z)$ are subdominant, the path integral derivation puts two additional restrictions on the contour $\Gamma$. We will briefly discuss these supplementary constraints and argue that the desired requirements can always be met in case $\theta$ is chosen accordingly. Let us first consider the local constraint on the contour, which has to be met at each individual point, while subsequently turning to the more interesting global restriction, asserting in which angular sector the contour is required to terminate.

\subsection{Local restriction}

\begin{figure}[H]
    \centering
    \includegraphics[width=0.95\textwidth]{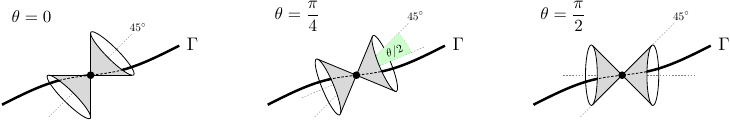}
    \caption{Depiction of the local restriction on the contour $\Gamma$ from demanding the intermediate momentum integrals~\eqref{eq:ShortTimePropagator} to be well-defined. The constraint~\eqref{eq:LocalConstraint} is equivalent to requiring the slope of the contour to always be contained in a tilted light cone. The situation is shown for three different angles $\theta$, with the cone rotating clockwise by $\theta/2$ for increasing $\theta$.}
    \label{fig:LightConeSchematic}
\end{figure}

\begin{figure}[H]
    \centering
    \hspace{0.25cm}
    \begin{subfigure}[c]{0.42\textwidth}
    \includegraphics[width=\textwidth]{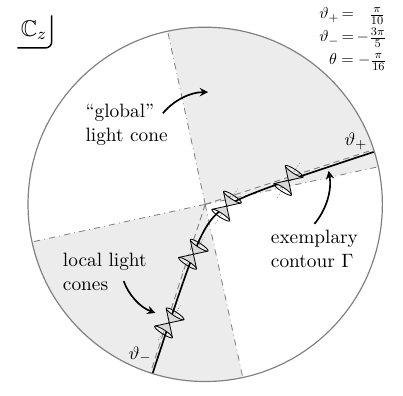}
    \end{subfigure}
    \hfill
    \begin{subfigure}[c]{0.44\textwidth}
    \includegraphics[width=\textwidth]{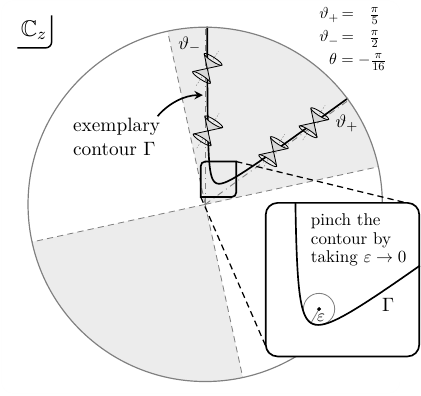}
    \end{subfigure}
    \hspace{0.5cm}
    \caption{Illustration of the arising situation when trying to connect the asymptotic angular regions $\vartheta_\pm$ by a contour $\Gamma$ respecting the local constraint~\eqref{eq:LocalConstraint}. (Left) For $\big\lvert \vartheta_+-\vartheta_-\big\rvert > \pi/2$, this feat is easily achieved by rotating the global light cone such that $\vartheta_+$ and $\vartheta_-$ lie in opposite half-cones. (Right) Situation for which $\big\lvert \vartheta_+-\vartheta_-\big\rvert < \pi/2$. In that case, one chooses both $\vartheta_\pm$ to lie inside the identical half-cone, then pinches the contour such that the violation of~\eqref{eq:LocalConstraint} only appears at a single point.}
    \label{fig:SpecialContour_Cones1}
\end{figure}

In order for the Gaussian momentum integrals~\eqref{eq:ShortTimePropagator} to be well-defined, we previously arrived at the local property 
\begin{align}
    \mathrm{arg}\cbig[\gamma'(s)^2\cbig]\in \cbig(-\!\:\theta,\pi-\theta\cbig) \qquad\forall s\in \mathbb{R}\, ,
    \label{eq:LocalConstraint}
\end{align}
demanding that the slope of the contour $\Gamma$ at each point has to lie in the range specified above. This constraint can be cast into a very natural form by demanding the contour to locally be contained in a tilted light cone, illustratively shown in figure~\ref{fig:LightConeSchematic} for three values of the Wick-rotation angle $\theta$. While this condition prohibits the presence of loops or sharp U-turns in the contour $\Gamma$ conceptualized for the path integral derivation, let us see that up to violating the condition at a singular point, the existence of a contour $\Gamma$ satisfying~\eqref{eq:LocalConstraint} is guaranteed.\footnote{Note that for the entire path integral derivation to go through, we only require one representative $\Gamma$ for which the local property is mandated to hold. Once one arrives at the discretized expression~\eqref{eq:DiscretizedPI_End}, one can indeed freely deform $\Gamma$, with the local restriction~\eqref{eq:LocalConstraint} becoming irrelevant, as previously disallowed choices of $\Gamma$ become admissible.}
\noindent
Assume we desire to construct a contour $\Gamma$ that satisfies 
\begin{align}
    \quad\;\lim_{s\to -\infty} \mathrm{arg}\big[\gamma(s)\big] = \vartheta_- \qquad \mathrm{and} \qquad \lim_{s\to \infty} \mathrm{arg}\big[\gamma(s)\big] = \vartheta_+
    \label{eq:AsymptoticsContour}
\end{align}
for arbitrary, fixed angles $\vartheta_\pm\in (-\pi,\pi]$. We can then ask for which angles $\theta$, if any, an admissible contour $\Gamma$ exists such that condition~\eqref{eq:LocalConstraint} can be met. Excluding pathological cases for which the phase of $\gamma'(s)$ is strongly oscillatory for large $\lvert s\rvert$, relation~\eqref{eq:AsymptoticsContour} implies 
\begin{align}
    \lim_{s\to -\infty} \mathrm{arg}\big[\gamma'(s)\big] \equiv \vartheta_-+\pi \;\;\text{mod } 2\pi \qquad \text{as well as} \qquad \lim_{s\to \infty} \mathrm{arg}\big[\gamma'(s)\big] = \vartheta_+\, .\quad\;
\end{align}
Note that the contour is set to asymptotically emerge from the angle $\vartheta_-$, thus inheriting an additional phase $\pi$ due to the direction of traversal. Since $\gamma(s)$ is assumed to be smooth, by the intermediate value theorem the slope of any such contour $\Gamma$ then necessarily sweeps through all angles lying in between $\vartheta_-+\pi$ and $\vartheta_+$. This effectively means that $\theta$ has to be chosen such that both angles $\vartheta_\pm$ lie inside the extended light cone emanating from the origin, albeit we have to add that $\vartheta_+$ and $\vartheta_-$ are required to lie in opposite halves of this ``global'' light cone, as depicted in the left panel of figure~\ref{fig:SpecialContour_Cones1}, otherwise a sharp U-turn would be immanent. In the case $\big\lvert \vartheta_+-\vartheta_-\big\rvert > \pi/2$, there always exists a non-empty angular sector of selectable $\theta$ enabling the desired construction of an admissible contour $\Gamma$. However, problems will arise in the complementary case $\big\lvert \vartheta_+-\vartheta_-\big\rvert \leq \pi/2$, as one can then only arrange for both $\vartheta_\pm$ to be contained in a single half-cone, necessarily entailing a violation of the condition~\eqref{eq:LocalConstraint} in some region. As shown in the right panel of figure~\ref{fig:SpecialContour_Cones1}, this region can be made arbitrarily small by shrinking the necessary U-turn further and further, to the point where the condition is only violated at a single point. When taking this point as a discretization point for the broken line path entering the discretized path integral, the violation is virtually absent and all steps go through unobstructed. Since the prescribed limiting procedure of pinching the contour clashes with the continuum limit $N\to\infty$, for which the whole U-turn is fully resolved no matter how narrow it is, the above reasoning only provides an ad hoc argument to reason that the construction should in principle apply equally to the case in which one desires to connect Stokes sectors that lie completely within one $90\degree$ sector. 

\noindent 
Note that in all cases at hand, the local constraint~\eqref{eq:LocalConstraint} will restrict the choices of $\theta$. While the provided discussion cannot rule out arising difficulties for the case when the eigenvalue problem is defined in Stokes sectors that are fully contained withing a $90\degree$ sector of the complex plane, we conjecture that the initial path integral derivation can be rigorously extended to circumvent the local constraint~\eqref{eq:LocalConstraint}, at the presumed cost of rendering the illustrated derivation more laborious. To this end, the techniques introduced by McLaughlin~\cite{McLaughlinComplexTimeMethod}---namely being able to freely deform the time contour when performing the path integral derivation---might come in handy. A more thorough analysis is left open for further consideration.

\subsection{Global restriction}

\begin{figure}[H]
    \centering
    \begin{subfigure}[c]{0.32\textwidth}
    \includegraphics[width=\textwidth]{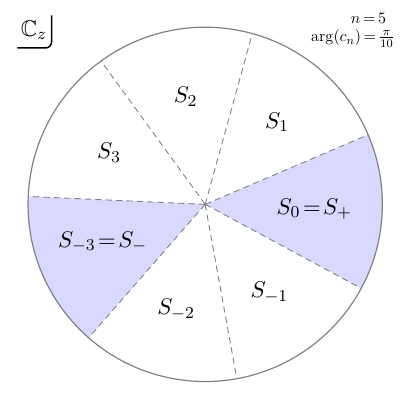}
    \end{subfigure}
    \hfill
    \begin{subfigure}[c]{0.32\textwidth}
    \includegraphics[width=\textwidth]{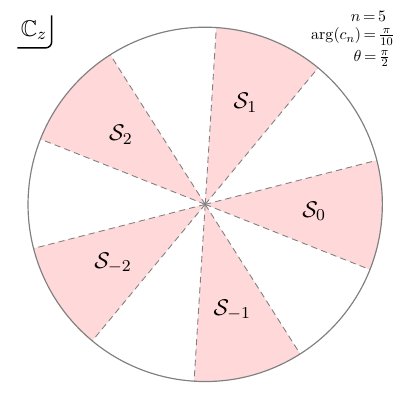}
    \end{subfigure}
    \hfill
    \begin{subfigure}[c]{0.32\textwidth}
    \includegraphics[width=\textwidth]{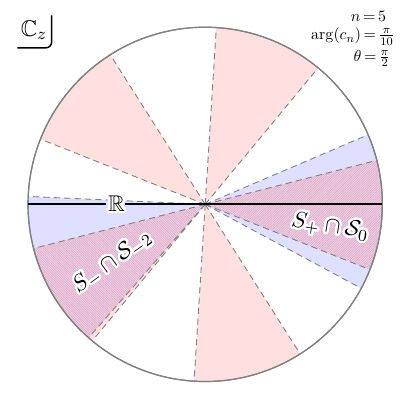}
    \end{subfigure}
    \caption{Exemplary case of the combined constraints inflicted by demanding the contour $\Gamma$ to asymptotically end inside the Stokes sectors $S_\pm$ as well as $\mathcal{S}_k$. (Left) Usual Stokes sector picture portraying the wedges $S_k$, with the eigenvalue problem defined in the two highlighted sectors $S_\pm$. (Center) Illustration of the sectors $\mathcal{S}_k$ constraining the contour $\Gamma$ further, required by the demand that the discretized position-space path integral expression is well-defined. (Right) Synthesis of both figures, with only the hatched overlap region being admissible for $\Gamma$ to end in. Whereas $\Gamma=\mathbb{R}$ constitutes an admissible contour if one solely demands convergence inside $S_\pm$, the combined constraint disallows the real axis to be an admissible contour for the derivation of the path integral.}
    \label{fig:StokesWedgeExamples_GlobalConstraints}
\end{figure}

The global restriction of $\Gamma$ terminating in $S_\pm \cap \big(\mathcal{S}_0 \cup \ldots \cup \mathcal{S}_n\big)$ provides a simpler set of constraints, however there arises one very interesting edge case that could yield interesting relations. As stated previously in section~\ref{sec:3_3_Equating_Representations}, there always exists a non-empty overlap between each Stokes sector $S_\pm$ defined through~\eqref{eq:DefinitionStokesSectors} and the union of all $\mathcal{S}_k$, given by~\eqref{eq:DefinitionStokesSectors_PI}. This is simply a consequence of the Stokes sectors having angular width $2\pi/(n+2)$, whereas the gap between the sectors $\mathcal{S}_k$ and $\mathcal{S}_{k\pm 1}$ only covers a range $\pi/n$. Thus, provided $n>2$, there is always the required overlap, as illustratively shown in figure~\ref{fig:StokesWedgeExamples_GlobalConstraints}. Consequently, the global constraints yield no additional restriction on $\theta$ or other parameters in our theory, their only feature is to narrow the angular sectors in which $\Gamma$ is required to terminate.

\begin{figure}[H]
    \centering
    \hspace{1.5cm}
    \begin{subfigure}[c]{0.32\textwidth}
    \includegraphics[width=\textwidth]{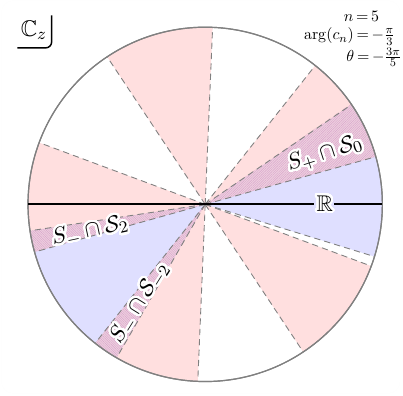}
    \end{subfigure}
    \hfill
    \begin{subfigure}[c]{0.32\textwidth}
    \includegraphics[width=\textwidth]{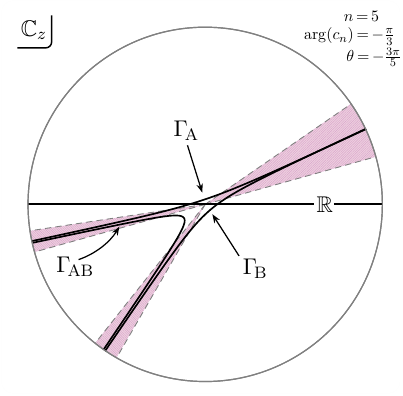}
    \end{subfigure} 
    \hspace{2cm}
    \caption{Case in which a Stokes sector has non-vanishing overlap with two distinct sectors $\mathcal{S}_k$, thus giving rise to disjoint sectors in which one end of $\Gamma$ can be chosen to terminate, yielding seemingly inequivalent functional integrals. However, due to the underlying eigenvalue problem being identical, the path integrals over $\Gamma_{\!\!\:\mathrm{A}}$ and $\Gamma_{\!\!\:\mathrm{B}}$ have to fully agree. This way, the composite functional integration cycle over $\mathcal{C}\big([0,T],\Gamma_{\!\!\:\mathrm{AB}}\big)$ is fully trivialized.}
    \label{fig:StokesWedgeExamples_Split}
\end{figure}

\noindent 
However, an interesting case emerges once the overlap of either $S_+$ or $S_-$ with $\mathcal{S}_0\cup \ldots \cup \mathcal{S}_n$ results in not one, but two disjoint angular sectors, as shown in figure~\ref{fig:StokesWedgeExamples_Split}. In that case, the choices $\Gamma_{\!\!\:\mathrm{A}}$ and $\Gamma_{\!\!\:\mathrm{B}}$ as shown in the right panel of figure~\ref{fig:StokesWedgeExamples_Split} yield inequivalent functional integrals, as both contours end in different regions of convergence of the path integral. Consequently, the path integral defined on the integration cycle $\mathcal{C}\big([0,T],\Gamma_{\!\!\:\mathrm{A}}\big)-\mathcal{C}\big([0,T],\Gamma_{\!\!\:\mathrm{B}}\big)=\mathcal{C}\big([0,T],\Gamma_{\!\!\:\mathrm{AB}}\big)$ necessarily vanishes, which may yield interesting insights into the evaluation of functional integrals. A detailed investigation of this phenomenon is beyond the scope of this work and is left for future research.

\section{Large-order behavior of the eigenvalues associated to a second-order ODE with polynomial coefficient}
\label{sec:D_LargeOrderBehavior_Eigenvalues}

To study the range of validity of the spectral representation~\eqref{eq:Propagator_SpectralRepresentation}, we require the large-order behavior of the eigenvalues $E_\ell$ for $\ell\to\infty$. Since we know the magnitude of the eigenvalues to be unbounded, we strictly speaking would only require information about $\mathrm{arg}(E_\ell)$ for our purposes, however let us state the complete result, as the remaining subject-relevant literature limits itself to presenting special cases.\footnote{Shin's work~\cite{ShinLargeEigenvalues} addresses a sufficiently broad class of cases to allow the general case to be deduced with relative ease. However, we chose to adopt a middle ground by drawing on some of Shin's results without directly appealing to the full conclusion right away.} As before, we are interested in solutions to the ODE
\begin{align}
	\scalebox{1.1}{\bigg\{}\!-\frac{\hbar^2}{2m}\frac{\mathrm{d}^2}{\mathrm{d}z^2} + \sum_{k=1}^n \:\! c_k z^k- E_\ell\scalebox{1.1}{\bigg\}} \,\Psi_\ell(z) = 0\, ,
    \label{eq:InitialODE}
\end{align}
demanding the eigenfunctions $\Psi_\ell(z)$ to decay in the two Stokes sectors $S_\pm$. Recalling the previous definition of the Stokes sectors provided in equation~\eqref{eq:DefinitionStokesSectors}, both sectors are subsequently characterized by their indices $k_\pm$, i.e. we define $S_+\eqqcolon S_{k_+}$ and $S_-\eqqcolon S_{k_-}$. Let us initially bring the eigenvalue problem into a canonical form by inferring the coordinate transformation 
\begin{align}
    z\mapsto \bigg(\frac{\hbar^2}{2mc_n}\bigg)^{\!\!-\frac{1}{n+2}}z\, ,
\end{align}
while defining 
\begin{align}
    \lambda\coloneqq -\frac{2m}{\hbar^2}\bigg(\frac{\hbar^2}{2mc_n}\bigg)^{\!\!\frac{2}{n+2}} E\, ,\qquad a_k\coloneqq \frac{2m}{\hbar^2}\bigg(\frac{\hbar^2}{2mc_n}\bigg)^{\!\!\frac{k+2}{n+2}} c_k\, ,
    \label{eq:RescalingProcedureODE}
\end{align}
granting the rescaled ODE
\begin{align}
	\scalebox{1.1}{\bigg\{}\!-\frac{\mathrm{d}^2}{\mathrm{d}z^2} + z^n+\sum_{k=1}^{n-1} \:\! a_k z^k+ \lambda_\ell\scalebox{1.1}{\bigg\}} \,\widetilde{\Psi}_\ell(z) = 0\, .
    \label{eq:RescaledODE_Sibuya}
\end{align}
This rescaling brings the initial ODE~\eqref{eq:InitialODE} into the canonical form studied by Sibuya~\cite{SibuyaEigenvalueProblems}, with subsequent treatments by Shin~\cite{ShinPTEigenvalues,ShinLargeEigenvalues} adopting Sibuyas notation. The simplified ODE~\eqref{eq:RescaledODE_Sibuya} possesses the new Stokes wedges
\begin{align}
	\widetilde{S}_k\coloneqq\cbigg\{z\in\mathbb{C}: \bigg\lvert\, \mathrm{arg}(z)-\frac{2\pi k}{n+2}\bigg\lvert <\frac{\pi}{n+2}\cbigg\}\, ,
    \label{eq:DefinitionRescaledODE_StokesSectors}
\end{align}
where the indices match the previous definition~\eqref{eq:DefinitionStokesSectors}, thus the eigenfunctions $\widetilde{\Psi}_\ell(z)$ are still demanded to be subdominant in $\widetilde{S}_{k_\pm}$. Following Sibuya, we define the auxiliary quantities
\begin{align}
    \vec{a}\coloneqq \big(a_1,a_2,\ldots , a_{n-1}\big)\in\mathbb{C}^{n-1}\qquad \mathrm{and}\qquad \omega\coloneqq \exp\cBig(\frac{2\pi i}{n+2}\cBig)\, ,
\end{align}
as well as the transformed vector
\begin{align}
    \vec{a}_k\coloneqq \big(\omega^{-k}a_1,\omega^{-2k}a_2,\ldots , \omega^{-(n-1)k}a_{n-1}\big)\, ,\quad k\in\mathbb{Z}\, .
\end{align}
The unique entire function that solves the canonical ODE~\eqref{eq:RescaledODE_Sibuya} for a generic constant term $\lambda$ while being subdominant in the Stokes sector $\widetilde{S}_0$ shall be denoted by $f\big(z,\vec{a},\lambda\big)$. Utilizing a simple scaling argument, Sibuya then shows that the function 
\begin{align}
    f_k\big(z,\vec{a},\lambda\big) \coloneqq f\big(\omega^{-k}z,\vec{a}_k,\omega^{-nk}\lambda\big) = f\big(\omega^{-k}z,\vec{a}_k,\omega^{2k}\lambda\big)
    \label{eq:Scaling_Property_fk}
\end{align}
identically solves the ODE, however being subdominant in the Stokes sector $\widetilde{S}_k$. In the second equality we utilized $\omega^{n+2}=1$, which will also be heavily used subsequently. Having constructed solutions that are subdominant in the different Stokes sectors, finding the eigenvalues $\lambda_\ell$ of our eigenvalue problem associated to the two wedges $\widetilde{S}_{k_\pm}$ amounts to finding the zeros of the Wronskian 
\begin{align}
    W_{k_+,k_-}\big(\vec{a},\lambda\big)\coloneqq f_{k_+}\big(z,\vec{a},\lambda\big)\,\frac{\partial f_{k_-}\big(z,\vec{a},\lambda\big)}{\partial z}-\frac{\partial f_{k_+}\big(z,\vec{a},\lambda\big)}{\partial z}\,f_{k_-}\big(z,\vec{a},\lambda\big)\, .
\end{align}
In case $\lambda$ is an eigenvalue, the two solutions $f_{k_+}$ and $f_{k_-}$ are linearly dependent, leading to the Wronskian vanishing. Utilizing the previous scaling property~\eqref{eq:Scaling_Property_fk} of the $f_k$, one additionally has the useful relation 
\begin{align}
    W_{k_++j,k_-+j}\big(\vec{a},\lambda\big)=\omega^{-j}\,W_{k_+,k_-}\big(\vec{a}_j,\omega^{2j}\lambda\big)\, .
\end{align}
Thus, finding the zeros of $W_{k_+,k_-}\big(\vec{a},\lambda\big)$ is identical to finding the zeros of $W_{k_+-k_-,0}\big(\vec{a}_{k_-},\omega^{2k_-}\lambda\big)$. With these preliminary considerations out of the way, one desires to study the behavior of the Wronskian $W_{k,0}(\vec{a},\lambda)$ for large magnitudes of $\lambda$, allowing us to infer the sector from which zeros can emerge in the limit $\lvert \lambda\rvert \to \infty$.

\subsection{Zeros of the Wronskian $W_{k,0}(\vec{a},\lambda)$ for large $\lvert \lambda\rvert$}

Since Sibuya provided complete results only for the case $W_{2,0}$, we will primarily exploit the insights by Shin on the behavior of $W_{k,0}$ for $k\geq 3$ and $n\geq 3$~\cite[see theorems 4.1 and 4.2]{ShinLargeEigenvalues}. Let us first assert that for large magnitudes of the eigenvalues, their argument $\mathrm{arg}(\lambda_\ell)$ converges toward a specific value $\mathrm{arg}(\lambda_\infty)$, i.e. for any $\varepsilon > 0$ there should be only finitely many eigenvalues for which $\big\lvert \mathrm{arg}(\lambda_\ell)-\mathrm{arg}(\lambda_\infty)\big\rvert >\varepsilon$, thus the eigenvalues condense onto a single ray in the complex plane.\footnote{We will not prove this conjectured as it would require a more in-depth discussion of the Wronskian. To this end, note that the Wronskian $W_{k_+-k_-,0}$ constitutes an entire function of finite order $0<\frac{1}{2}+\frac{1}{n}<1$ (for the relevant case $n\geq 3$), whose zeros fulfill certain regularity conditions provided that its associated indicator attains its minimum in a single direction~\cite{LevinZerosEntireFunctions}. Since this discussion holds limited relevance to the overall paper, we will refrain from exploring it further, rather leaving this open problem unaddressed.} Furthermore, the leading-order result for the eigenvalues in the limit $\ell\to\infty$ does not depend on $\vec{a}$, which only enters next-to-leading corrections due to the Voros symbols entering the quantization condition being dominated by the highest monomial power (and certainly the value of $\lambda$). This allows us to set $\vec{a}=0$ for our purposes, simplifying the intermediate expressions.\footnote{One does indeed not need to infer this simplification, as it would pop out of an explicit computation retaining variable $\vec{a}$. However, let us shorten the discussion by instating this simplification.} Utilizing theorem 4.1 in reference~\cite{ShinLargeEigenvalues}, one finds that for $2\leq k < \lfloor n/2 \rfloor+1$, the Wronskian $W_{k,0}$ admits the asymptotic expansion
\begin{align}
    W_{k,0}(0,\lambda)&= - \,W_{0,k}(0,\lambda) = - \,\omega^{-1}\:\! W_{-1,k-1}(0,\omega^2\lambda) \nonumber \\ 
    &= -\,2\omega^{-1} \omega^{\frac{2-(k-1)}{2}+\frac{n}{4}} \cbig[1+\scalebox{0.65}{$\mathcal{O}$}(1)\cbig] \exp\bigg\{K\big(\omega^{-2}\omega^2\lambda\big)^{\frac{1}{2}+\frac{1}{n}}-K\big(\omega^{2(k-2)}\omega^2\lambda\big)^{\frac{1}{2}+\frac{1}{n}}\bigg\} \nonumber \\ 
    &\phantom{=\;}-2\omega^{-1} \omega^{\frac{2-(k-1)}{2}+\frac{n}{4}} \cbig[1+\scalebox{0.65}{$\mathcal{O}$}(1)\cbig] \exp\bigg\{K\big(\omega^{2(k-1)}\omega^2\lambda\big)^{\frac{1}{2}+\frac{1}{n}}-K\big(\omega^2\lambda\big)^{\frac{1}{2}+\frac{1}{n}}\bigg\} \, ,
    \label{eq:WronskianLargeOrderBehavior}
\end{align}
which is valid for $\lvert\lambda\rvert \to \infty$ in the narrow sector 
\begin{align}
    -\frac{2(k-2)\pi}{n+2} -\delta \leq &\;\mathrm{arg}\big(\omega^2\lambda\big)\leq -\frac{2(k-2)\pi}{n+2} +\delta \qquad\qquad\qquad\qquad \nonumber \\[0.15cm] 
     \Longleftrightarrow \qquad -\frac{2k\pi}{n+2} -\delta \leq &\;\;\;\mathrm{arg}(\lambda)\;\;\leq -\frac{2k\pi}{n+2} +\delta \, .
     \label{eq:AngularSectorLambda}
\end{align}
Hereby, $\delta >0$ is an arbitrarily small positive number, while we furthermore defined the auxiliary constant
\begin{align}
    K\coloneqq \mathlarger{\int}_0^\infty \Big[\sqrt{t^n+1}-t^{\frac{n}{2}}\Big] \mathrm{d}t=\frac{\Gamma\big(\frac{1}{n}\big)\,\Gamma\big(\frac{1}{2}-\frac{1}{n}\big)}{(n+2)\sqrt{\pi}} = \frac{\sqrt{\pi}}{2\cos\!\big(\frac{\pi}{n}\big)}\frac{\Gamma\big(1+\frac{1}{n}\big)}{\Gamma\big(\frac{3}{2}+\frac{1}{n}\big)}\, .
    \label{eq:DefinitionK}
\end{align}
As can be checked explicitly, the Wronskian $W_{k,0}(0,\lambda)$ indeed possesses zeros in that particular strip, whose leading-order behavior can be obtained by solving 
\begin{align}
    \exp\bigg\{K\big(\omega^k\lambda\big)^{\frac{1}{2}+\frac{1}{n}}\cBig(\omega^{-\frac{k(n+2)}{2n}}-\omega^{\frac{(k-2)(n+2)}{2n}}\cBig) \nonumber \qquad\qquad\qquad\qquad\qquad\qquad\qquad\qquad\\ -K\big(\omega^k\lambda\big)^{\frac{1}{2}+\frac{1}{n}}\omega^{\frac{n+2}{n}}\cBig(\omega^{\frac{(k-2)(n+2)}{2n}}-\omega^{-\frac{k(n+2)}{2n}}\cBig)+\scalebox{0.65}{$\mathcal{O}$}(1)\bigg\} = -1\, .
\end{align}
Rearranging the terms suitably, one arrives at the quantization condition
\begin{align}
    K\big(\omega^{-k}\lambda_\ell\big)^{\!\!\:\frac{n+2}{2n}}\Big(\omega^{\frac{n+2}{2n}}+\omega^{-\frac{n+2}{2n}}\Big)\Big[\omega^{\frac{(k-1)(n+2)}{2n}}-\omega^{-\frac{(k-1)(n+2)}{2n}}\Big] +\scalebox{0.65}{$\mathcal{O}$}(1)= (2\ell+1) i\pi
\end{align}
with $\ell\in \mathbb{Z}$. Representing the bracketed terms as sine and cosine, one arrives at the desired result 
\begin{align}
    \lambda_\ell &\xrightarrow{\ell\to\infty} \omega^{-k}\,\Bigg\{\frac{(2\ell+1) \pi}{4\cos\!\big(\frac{\pi}{n}\big)\sin\cbig(\frac{(k-1)\pi}{n}\cbig) K}\Bigg\}^{\!\frac{2n}{n+2}} \cbig[1+\scalebox{0.65}{$\mathcal{O}$}(1)\cbig] \, .
    \label{eq:LargeOrderBehaviorLambda}
\end{align}
Given $n\geq 3$ and $k\geq 2$, for non-negative $\ell\geq 0$ the bracketed expression is positive, such that one indeed finds these zeros of the Wronskian $W_{k,0}(0,\lambda)$ to lie in the correct angular sector~\eqref{eq:AngularSectorLambda}, for which the above derivation was valid. Negative $\ell$ are prohibited by the fact that the so-found $\lambda_\ell$ would lie outside the range of applicability of equation~\eqref{eq:WronskianLargeOrderBehavior}. As we asserted previously, there should not be a second ray where zeros of the Wronskian accumulate, thus the given solutions indeed capture all zeros with sufficiently large magnitude. Note that while relation~\eqref{eq:LargeOrderBehaviorLambda} looks slightly different to equation (29.12) in Sibuyas work~\cite{SibuyaEigenvalueProblems}, they are identical upon using trigonometric identities. \\

\noindent 
While our previous considerations applied to $2\leq k < \lfloor n/2\rfloor +1$, we also require the case of even $n$ with $k=n/2 +1$. This special case was similarly considered by Shin in theorem 4.2, asserting that the Wronskian in that case admits the expansion
\begin{align}
    W_{\frac{n}{2}+1,0}(0,\lambda)&= - \,W_{0,\frac{n}{2}+1}(0,\lambda) = - \,\omega^{-1}\:\! W_{-1,\frac{n}{2}}(0,\omega^2\lambda) \nonumber \\ 
    &= \phantom{+}\,2\omega^{-1} \omega^{2+\frac{n}{2}} \cbig[1+\scalebox{0.65}{$\mathcal{O}$}(1)\cbig] \exp\bigg\{\!-2K\big(\omega^2\lambda\big)^{\frac{1}{2}+\frac{1}{n}}\bigg\} \nonumber \\ 
    &\phantom{=\;}+2\omega^{-1} \omega^{2+\frac{n}{2}} \cbig[1+\scalebox{0.65}{$\mathcal{O}$}(1)\cbig] \exp\bigg\{\!-2K\big(\omega^{n-2}\omega^2\lambda\big)^{\frac{1}{2}+\frac{1}{n}}\bigg\} \, ,
\end{align}
being valid in the angular sector 
\begin{align}
    -\pi+\frac{4\pi}{n+2} -\delta \leq &\;\mathrm{arg}\big(\omega^2\lambda\big)\leq -\pi+\frac{4\pi}{n+2} +\delta \quad\Longleftrightarrow \quad -\pi -\delta \leq \mathrm{arg}(\lambda)\leq -\pi +\delta \, .
    \label{eq:AngularSectorLambda2}
\end{align}
Repeating the identical procedure as above leads to the condition 
\begin{align}
    \exp\bigg\{2K \big(\omega^{\frac{n}{2}+1}\lambda\big)^{\!\!\:\frac{n+2}{2n}}\Big[\big(\omega^{1-\frac{n}{2}}\big)^{\!\!\:\frac{n+2}{2n}}-\big(\omega^{\frac{n}{2}-1}\big)^{\!\!\:\frac{n+2}{2n}}\Big]+\scalebox{0.65}{$\mathcal{O}$}(1)\bigg\} = -1\, .
\end{align}
such that the zeros are found to possess the asymptotic behavior
\begin{align}
    \lambda_\ell &= -\cbigg\{\frac{(2\ell+1) \pi}{4\sin\!\big(\frac{\pi}{2}-\frac{\pi}{n}\big)K}\cbigg\}^{\!\frac{2n}{n+2}} \cbig[1+\scalebox{0.65}{$\mathcal{O}$}(1)\cbig]\, .
    \label{eq:LargeOrderBehaviorLambda2}
\end{align}
For non-negative $\ell$, the solutions to $W_{n/2+1,0}(0,\lambda)$ again lie in the correct sector~\eqref{eq:AngularSectorLambda2} for the derivation to be valid, thus constitute the generic behavior of all zeros with large magnitude. Note that the previous result~\eqref{eq:LargeOrderBehaviorLambda} simply extends to the present case, as the use of $\sin(\pi/2-\pi/n)=\cos(\pi/n)$ proves the equality between~\eqref{eq:LargeOrderBehaviorLambda} and~\eqref{eq:LargeOrderBehaviorLambda2}.

\subsection{Extracting the generic large-order behavior of the eigenvalues $E_\ell$}

Knowing the zeros of the Wronskian $W_{k,0}$ allows us to infer the eigenvalues $\lambda_\ell$ of our (rescaled) eigenvalue problem~\eqref{eq:RescaledODE_Sibuya}, as we formerly argued that the eigenvalues are simply the zeros of $W_{k_+-k_-,0}\big(\vec{a}_{k_-},\omega^{2k_-}\lambda\big)$. Take note of the important fact that we can always label two (non-adjacent) Stokes sectors $S_{k_\pm}$ such that the difference $k_+-k_-$ fulfills $2\leq k_+-k_- \leq \lfloor n/2\rfloor +1$. Except for the case $k_+-k_- = \lfloor n/2\rfloor +1$ this assignment is unambiguous, fully fixing both $k_+$ and $k_-$. Simply invoking the previous results~\eqref{eq:LargeOrderBehaviorLambda} and~\eqref{eq:LargeOrderBehaviorLambda2} yields 
\begin{align}
    \lambda_\ell &= \omega^{-(k_++k_-)}\,\cBigg\{\frac{\sqrt{\pi}\;\Gamma\big(\frac{3}{2}+\frac{1}{n}\big)}{\sin\cbig(\frac{\pi(k_+-k_--1)}{n}\cbig)\:\!\Gamma\big(1+\frac{1}{n}\big)}\cBig(\ell+\frac{1}{2}\cBig)\!\!\:\cBigg\}^{\!\frac{2n}{n+2}} \cbig[1+\scalebox{0.65}{$\mathcal{O}$}(1)\cbig] \, .
    \label{eq:EigenvalueAsymptoticsLambda}
\end{align}
where we additionally utilized expression~\eqref{eq:DefinitionK} for the constant $K$. The given expression~\eqref{eq:EigenvalueAsymptoticsLambda} generalizes the leading-order result stated by Shin in his theorem 1.4. Lastly, we recover the original eigenvalues by inverting relation~\eqref{eq:RescalingProcedureODE}, arriving at the final result\footnote{There might be much simpler constructions on how to arrive at the given result, e.g. by utilizing a leading-order WKB estimate in guise of the typical Bohr--Sommerfeld quantization condition, see e.g.~\cite{BenderPT1}.}  
\begin{align}
    \scalebox{0.99}{$\displaystyle{\!\!E_\ell \sim \frac{\hbar^2}{2m} \bigg(\frac{2mc_n}{\hbar^2}\bigg)^{\!\!\!\:\frac{2}{n+2}}\exp\bigg\{\frac{2i\pi}{n+2}\cBig(\frac{n+2}{2}-k_+-k_-\cBig)\!\!\;\bigg\}\cBigg\{\frac{\sqrt{\pi}\;\Gamma\big(\frac{3}{2}+\frac{1}{n}\big)\big(\ell+\frac{1}{2}\big)}{\sin\cbig(\frac{\pi(k_+-k_--1)}{n}\cbig)\:\!\Gamma\big(1+\frac{1}{n}\big)}\!\!\:\cBigg\}^{\!\!\!\:\frac{2n}{n+2}} .}$}
    \label{eq:EigenvalueAsymptoticsE}
\end{align}

\bibliographystyle{CustomBibliography}  
\addcontentsline{toc}{section}{\protect\numberline{}References}
\bibliography{Literature}

\end{document}